\documentclass[9pt,twocolumn,twoside]{pnas-new}
\usepackage{amssymb}
\usepackage{amsmath}
\usepackage{cancel}
\usepackage{algorithm}
\usepackage{textcomp}
\usepackage{algpseudocode}

\usepackage{amsthm}

\newtheorem{theorem}{Theorem}[section]
\newtheorem{lemma}[theorem]{Lemma}
\newtheorem{corollary}[theorem]{Corollary}
\newtheorem{proposition}[theorem]{Proposition}
\newtheorem{definition}[theorem]{Definition}

\newtheorem{remark}{Remark}

\newtheorem{strategy}{Strategy}

\usepackage{adjustbox}
\usepackage{array}
\newcolumntype{R}[2]{%
	>{\adjustbox{angle=#1,lap=\width-(#2)}\bgroup}%
	l%
	<{\egroup}%
}
\usepackage{multirow} 
\usepackage{booktabs}
\usepackage{relsize}
\usepackage{graphicx}

\usepackage{lscape}

\usepackage{epsfig,subfig}
\usepackage{graphicx}
\usepackage{makeidx}
\usepackage{multicol}

\usepackage{algpseudocode}% http://ctan.org/pkg/algorithms

\usepackage{tikz}
\usetikzlibrary{plotmarks}
\usepackage{pgfplots}
\pgfplotsset{plot coordinates/math parser=false}
\newlength\figureheight
\newlength\figurewidth
\usetikzlibrary{fit}
\tikzset{%
	highlight/.style={rectangle,rounded corners,fill=none,
		thick,inner sep=0pt}
}

\usepackage{smartdiagram}

\makeatletter
\def\bbordermatrix#1{\begingroup \m@th
	\@tempdima 4.75\p@
	\setbox\z@\vbox{%
		\def\cr{\crcr\noalign{\kern2\p@\global\let\cr\endline}}%
		\ialign{$##$\hfil\kern2\p@\kern\@tempdima&\thinspace\hfil$##$\hfil
			&&\quad\hfil$##$\hfil\crcr
			\omit\strut\hfil\crcr\noalign{\kern-\baselineskip}%
			#1\crcr\omit\strut\cr}}%
	\setbox\tw@\vbox{\unvcopy\z@\global\setbox\@ne\lastbox}%
	\setbox\tw@\hbox{\unhbox\@ne\unskip\global\setbox\@ne\lastbox}%
	\setbox\tw@\hbox{$\kern\wd\@ne\kern-\@tempdima\left[\kern-\wd\@ne
		\global\setbox\@ne\vbox{\box\@ne\kern2\p@}%
		\vcenter{\kern-\ht\@ne\unvbox\z@\kern-\baselineskip}\,\right]$}%
	\null\;\vbox{\kern\ht\@ne\box\tw@}\endgroup}
\makeatother

\usepackage{multirow} 
\usepackage{booktabs}
\usetikzlibrary{positioning}
\usetikzlibrary{arrows}
\usetikzlibrary{patterns}

\usepackage{tablefootnote}

\setboolean{displaywatermark}{false}

\usepackage{multirow} 
\usepackage{booktabs}
\usepackage{relsize}

\newcolumntype{d}[1]{D{.}{.}{#1}}

\def\VR{\kern-\arraycolsep\strut\vrule &\kern-\arraycolsep}
\def\VVR{\kern-\arraycolsep\strut\vrule\hspace{0.05em}\vrule &\kern-\arraycolsep}
\def\vr{\kern-\arraycolsep & \kern-\arraycolsep}

\usepackage{arydshln}

\usepackage{url}

\renewcommand{\algorithmicrequire}{\textbf{Input:}}
\renewcommand{\algorithmicensure}{\textbf{Output:}}

\templatetype{pnasresearcharticle} % Choose template 

\title{Random Surfing Revisited: Generalizing PageRank's Teleportation Model}

\author[a,$\star$]{Athanasios N. Nikolakopoulos}

\affil[a]{University of Minnesota, Minneapolis, MN, 55455, USA}

\leadauthor{Athanasios N. Nikolakopoulos}

\correspondingauthor{\textsuperscript{$\star$}To whom correspondence should be addressed. E-mail: (check \href{www.nikolako.net}{www.nikolako.net})}

\keywords{Complex Networks $|$ Random Walks $|$ Markov Chains $|$ Near Decomposability $|$ Higher-order Organization} 

\begin{abstract}
We revisit the \textit{Random Surfer} model, focusing on its---often overlooked---\textit{Teleportation} component, and we introduce \textbf{NCDawareRank}; a novel ranking framework designed to exploit network meta-information as well as aspects of its higher-order structural organization in a way that preserves the mathematical structure and the attractive computational characteristics of PageRank. A rigorous theoretical exploration of the proposed model reveals a wealth of mathematical  properties that entail tangible benefits in terms of robustness, computability, as well as modeling flexibility and expressiveness. 
A set of experiments on real-work networks verify the theoretically predicted properties of NCDawareRank, and showcase its effectiveness as a network centrality measure. 
\end{abstract}

\doi{\textsc{NCDawareRank}}

\begin{document}

\maketitle
\thispagestyle{firststyle}
\ifthenelse{\boolean{shortarticle}}{\ifthenelse{\boolean{singlecolumn}}{\abscontentformatted}{\abscontent}}{}

	\section{Random Surfer Model: A Tale of Two ``Remedies''}
\label{Ch:Intro:Sec:RandomSurfing}
 \dropcap{T}he basic idea behind PageRank's approach to calculating the importance of individual nodes in a network is very intuitive. In their seminal paper Page \textit{et al.}~\cite{pagerank} imagined of a \textit{Random Surfer} of the network that jumps forever from node to node, and then, following this intuitive metaphor, they defined the overall importance of a node to be equal to \textit{the fraction of time this random surfer spends on it, in the long run}.
	Underlying the definition of PageRank is the assumption that the existence of 
	a link from a  node $u$ to a node $v$ testifies the importance of node $v$. 
	Furthermore, the amount of importance conferred to node $v$ is proportional to 
	the importance of node $u$ and inversely proportional to the number of nodes 
	$u$ links to.
	To formulate PageRank’s basic idea with a mathematical model, 	we can construct a \textit{row-normalized adjacency matrix} 
	$\mathbf{H}$, whose element $H_{uv}$ is one over the outdegree of $u$ if there 
	is a link from $u$ to $v$, or zero otherwise. 	
	Notice that in the general case the matrix $\mathbf{H}$ defined in this way, cannot be used as the transition probability matrix of a well-defined random surfing process. In particular, the definition of the final random surfer model needed to overcome two significant problems: 
	 \begin{enumerate}
	 \item The row-normalization of the adjacency matrix of the underlying network does not always yield a valid transition probability matrix. Indeed, not all nodes have outgoing links; in the general case there exist \textit{dangling nodes} that correspond to zero rows of the adjacency matrix---which makes matrix $\mathbf{H}$ strictly substochastic. To deal with this problem, in the random surfer model these zero rows are replaced with a valid probability distribution over the nodes of the network, thereby transforming the initial matrix $\mathbf{H}$, to a stochastic matrix. This intervention  became known as a \textit{stochasticity adjustment}~\cite{LangvilleMeyer06}.
	\item The second problem is slightly more subtle, and arises from the need to ensure that the final ranking vector produced by this random surfing process is well-defined; i.e. that the corresponding  Markov chain that governs the behavior of the random surfer possesses a \textit{unique positive limiting  
	distribution}. Notice that the stochasticity adjustment, is not enough to ensure this. Indeed, the corresponding chain could easily contain more than one closed communicating classes that can trap the random surfer, making the limiting distribution, dependent on the starting vector, and hence, not unique. 	To address this issue Page \textit{et al.}~\cite{pagerank} introduced a \textit{damping factor} $\alpha$ and a convenient rank-one \textit{teleportation} matrix $\mathbf{E}$, formally defined as 
	\begin{displaymath}
	\mathbf{E} \triangleq \mathbf{1}\mathbf{v^\intercal},
	\end{displaymath}
	for some probability vector $\mathbf{v}$, usually chosen to be $\mathbf{v} = \frac{1}{n}\mathbf{1}$. This second adjustment is sometimes referred to as a \textit{primitivity adjustment}~\cite{LangvilleMeyer06}, since it ensures the irreducibility and the aperiodicity of the final stochastic matrix that captures the transition probabilities of the random surfer. 
	\end{enumerate}

	After incorporating the above ``remedies'' the resulting stochastic matrix that corresponds to the final random surfer model---sometimes called the \textit{Google matrix}---can be expressed as 
		\begin{equation}
		\label{PageRank_Matrix_G}
		\mathbf{G} \triangleq \alpha \mathbf{H} + (1-\alpha)\mathbf{E}, 
		\end{equation}
		with the \textit{damping factor} $\alpha$, chosen to fall between 0 and 1. The PageRank vector is then well-defined as the unique stationary distribution of the Markov chain with transition probability matrix $\mathbf{G}$. 
		
		Returning to the random surfing metaphor, the final model translates to a random surfer who at each step,
		\begin{itemize}
			\item[-] with probability $\alpha$ follows the outgoing links of the node she is currently visiting,  uniformly at random; and,
			\item[-] with probability $1-\alpha$ \textit{teleports} to a different node of the network according to distribution $\mathbf{v}$.
		\end{itemize}
		
		\noindent We are now ready to dive deeper into the mathematical properties as well as the  modeling implications of  these two adjustments.  
		\subsection{Stochasticity Adjustment}	
			\textit{Dangling nodes} are a reality in many real networks. Take for example the   Web-graph. Its dangling nodes include Web documents of potentially high quality, as well as pages that could not be crawled because they are protected. Eiron \textit{et al.}~\cite{Eiron:2004:RWF:988672.988714} reported that many of the highest ranked Web-pages are in fact dangling. These include certain types of naturally dangling URLs such as downloadable files, PDFs, images, MP3s, movies, and so on.  Furthermore, due to the existence of sites that produce dynamic content, the potential number of pages in the Web is practically infinite; and crawling this infinite Web will necessarily produce a large number of dangling pages that reside in the boundaries of its crawled portion (these nodes are sometimes referred to as the ``\textit{web frontier}''\cite{Eiron:2004:RWF:988672.988714}.)

			From a mathematical point of view dangling nodes correspond to zero rows of the adjacency matrix. The row-normalized version of this matrix therefore, needs some sort of ``patching'' in order to produce a well-defined transition probability matrix. In PageRank, this patching is typically done in a homogeneous manner. The standard approach is to select a \textit{dangling-node distribution}---usually defined to be the same as the preference distribution used for the teleportation matrix---and then use it to replace all the zero rows of the substochastic normalized adjacency matrix, thereby transforming it to stochastic. 
			
		 Formally, the final stochastic matrix will be given by
			\begin{eqnarray}
			\mathbf{G} & = & \alpha\mathbf{H} + (1-\alpha)\mathbf{E} \nonumber \\
			& = & \alpha(\mathbf{H}_{\text{ND}} + \mathbf{a}\mathbf{f}^\intercal) + (1-\alpha)\mathbf{1}\mathbf{v}^\intercal,
			\end{eqnarray}
			where $\mathbf{H}_{\text{ND}}$ denotes the originally substochastic matrix (with zero rows for the dangling nodes), $\mathbf{a}$ is a vector that indicates the zero rows of matrix $\mathbf{H}_{\text{ND}}$ (i.e. its $i^{\textit{th}}$ element is 1 if and only if $[\mathbf{H}_{\text{ND}}]_{ij} = 0$ for every $j$), and $\mathbf{f}^\intercal$ is the patching distribution.	
			
			The patching distribution $\mathbf{f}^\intercal$, may or may not be the same with the teleportation distribution $\mathbf{v}^\intercal$ used for the definition of the teleportation matrix $\mathbf{E}$. When these two distributions are the same, the PageRank model is referred to as \textit{strongly preferential}; when they are different it is referred to as \textit{weakly preferential}. 
			 In the vast majority of applications of PageRank in the literature the strongly preferential model is preferred, with both teleportation and  patching distribution usually chosen to be the standard uniform distribution, $\tfrac{1}{n}\mathbf{1}^\intercal$.
			
			The above approach translates to a model for which: when the random surfer finds herself visiting a dangling node, in the next step she jumps to a different node of the network with a standard probability, \textit{irrespectively from the origin node she currently occupies}. We believe that for many applications this approach might be suboptimal. For example, in Web ranking this approach certainly deviates from  common experience and does not capture true Web-surfing behavior. Indeed, if we imagine a user surfing a dangling page, then it is much more likely that she uses the ``back button'' to follow a different link, or type-in a different website address to jump to a related page. Intuitively, the fact alone that she is currently visiting this specific ``dangling page'', says something about where she is more likely to go next. So, treating dangling nodes in a homogeneous manner, is somewhat simplistic and unrealistic. Unfortunately, due to the size of many network centrality problems, the full-extend insertion of features like ``back jumps,'' or ``jumps based on visiting history'' to the standard PageRank model could obscure its mathematical simplicity and compromise its applicability in large-scale problems. On the other end, choosing to completely ignore the dangling nodes could lead to the complete oversight of potentially important nodes, and it could also neglect their effect on the rankings of other nodes of the network. This makes the inclusion of dangling nodes in the final ranking model necessary, and their intuitive and computationally efficient handling, an important and subtle issue.

		\subsection{Primitivity Adjustment}
		 The teleportation matrix is an artificial addition to the random surfer model, that ensures the ergodicity of the random surfing process. It can be thought as a regularization component that warrants a well-defined solution to an ill-posed problem~\cite{PageRankBeyondWeb}. Its involvement in the final model is controlled by the strictly positive damping factor $\alpha$, the value of which determines how often the random surfer follows the actual network connections rather than jumping at a random node. 
		The choice of the damping factor is very important and has received great research attention \cite{Avrachenkov:2007:DPM:1777879.1777881,Boldi:2009:PFD:1629096.1629097,LangvilleMeyer06}. Picking a very small damping factor ignores the link structure of the network in favor of the artificial teleportation matrix, and thus, results in uninformative ranking vectors. On the other hand, as $\alpha$ gets closer to 1, the network component becomes increasingly important. While this may seem intuitively preferable, picking an $\alpha$ very close to 1, also results in a number of significant problems. From a computational point of view, the number of iterations till convergence required by the Power Method and its variants (commonly used for the extraction of the PageRank vector) grows with $\alpha$ and the computation of the ranking vector becomes numerically ill-conditioned \cite{kamvar2003condition}. Furthermore, from a qualitative prospective, various studies also indicate that damping factors close to 1 result into counterintuitive ranking vectors where the network’s core component is assigned null rank and all the PageRank gets concentrated mostly in irrelevant nodes \cite{Avrachenkov:2007:DPM:1777879.1777881,DBLP:conf/dagstuhl/BoldiSV07,Boldi:2009:PFD:1629096.1629097}; whereas keeping $\alpha$ away from 1, could help protect the final centrality measure against outliers in the network~\cite{PageRankBeyondWeb}, and produce a better ranking.
			
		We see that, both qualitative and computational reasons, suggest that $\alpha$ should be chosen neither too big, nor too small, which means that the final ranking vector will always be affected to a certain degree by the teleportation model. In the presence of sparsity, this effect becomes even more significant; and if we take into account that under strongly preferential patching the teleportation vector is followed with probability 1 whenever the random surfer is in a dangling node, we see that this artificially introduced component dictates a large part of the random surfing behavior.  
		
		Most formal generalizations of PageRank proposed in the literature, focus on the  damping issue. Horn and Serra-Capizzano~\cite{horn2006} consider the use of a complex-valued\footnote{Notice that despite deviating from the intuitive random surfing paradigm this is a reasonable  generalization since mathematically the PageRank vector is a rational function of $\alpha$~\cite{Boldi:2005:PFD:1060745.1060827,DBLP:conf/dagstuhl/BoldiSV07}.} $\alpha$ and later Gleich and Rossi~\cite{GleichAndRossiTimeDependentTeleportation} showed that complex values of the damping factor $\alpha$ arise in the solution of a time-dependent generalization of PageRank, as well. Boldi~\cite{Boldi:2005:TRW:1062745.1062787}, in an attempt to eliminate PageRank's dependency on the arbitrarily chosen parameter $\alpha$,  proposed TotalRank; an algorithm that integrates the ranking vector over the entire range of possible damping factors. Constantine and Gleich~\cite{constantine2009random} proposed Random-$\alpha$ PageRank, a ranking method that considers the influence of a population of random surfers, each choosing its own damping factor according to a given distribution. 
				
		Baeza-Yates \textit{et al.}~\cite{Baeza-Yates:2006:GPD:1148170.1148225} provide maybe the most general setting for the above ideas. Their formulation arises from an alternative characterization of the PageRank vector as the normalized solution of the linear system,  
				\begin{eqnarray}
				(\mathbf{I}-\alpha\mathbf{H^\intercal})\boldsymbol{\pi} & = & (1-\alpha)\mathbf{v}. \nonumber 
				\end{eqnarray} 
	    The stochasticity of matrix $\mathbf{H}$ implies that the spectral radius $\rho(\alpha\mathbf{H})$ is less than one thereby allowing the PageRank vector to be expressed by means of the convergent Neumann series,
		\begin{displaymath}\boldsymbol{\pi} = (1-\alpha) \sum_{k=0}^{\infty} \alpha^k(\mathbf{H^\intercal})^k \mathbf{v}. 				\end{displaymath} The above relation expresses the PageRank vector as a weighted sum of powers of $\mathbf{H^\intercal}$, with the weights, $\alpha^k$, decaying geometrically. Baeza-Yates \textit{et al.}~\cite{Baeza-Yates:2006:GPD:1148170.1148225} generalize this series representation as 
				\begin{equation}
				\boldsymbol{\pi} = \sum_{k=0}^{\infty} \psi(k) (\mathbf{H^\intercal})^k \mathbf{v},
				\label{eq:dampingFunctions}
				\end{equation} where $\psi(k)$ is a suitably selected \textit{damping function}. Equation~(\ref{eq:dampingFunctions}) models how the quantities in the teleportation vector $\mathbf{v}$, probabilistically diffuse through the network, with the probability of a path of length $k$, being damped by $\psi(k)$. 
				Kollias and Gallopoulos~\cite{Gallopoulos32,Gallopoulos} ``bridge the gap'' between these functional rankings and the traditional random surfer model, proposing a fruitful \textit{multidamping} reformulation 
				that allows intuitive interpretations of the damping functions in terms of random surfing habits. Furthermore, their framework facilitates fast approximation algorithms for finding the highest ranked nodes and it also lends itself naturally to computation of the rankings in massively parallel/distributed environments~\cite{Gallopoulos}.

			However, little have been done towards a generalization of the teleportation matrix itself. The vast majority of applications of PageRank in the literature adopt the traditional rank-one teleportation matrix that is defined using, either the standard uniform vector proposed by Page \textit{et al.}~\cite{pagerank} or, in some cases, an application-specific teleportation vector~\cite{AdersenFOCS06,ProteinPageRank,Tong:2006:FRW:1193207.1193363}.
			While mathematically the introduction of some sort of teleportation is necessary to ensure that the final Markov chain becomes irreducible and aperiodic, the standard teleportation matrix does not cease to be an artificial addition to the Random Surfer model, the homogeneous approach of which can be restrictive and sometimes even counterintuitive. Furthermore, the very existence of the standard teleportation matrix gives incentive for direct manipulation of the ranking score through link-spamming~\cite{constantine2009random,Eiron:2004:RWF:988672.988714} and is also known to impose fundamental limitations to the quality of the ranking vectors (sensitivity to the effects of sparsity, biased ranking of newly added nodes etc.~\cite{Xue:2005:EHS:1076034.1076068}).	Moreover, from a purely computational perspective, choosing a teleportation model that is completely ``blind'' to the spectral characteristics of the underlying network, could result in unnecessary burden for the extraction of the ranking vector that could be alleviated through a smarter teleportation selection.

\subsection{Random Surfing Model Redux} 
	The basic idea behind PageRank is very intuitive and generic, a fact that has helped the method to be applied with significant success in many application areas arising from diverse disciplines including Biology~\cite{ProteinRank,jiang2009gene,Morrison2005,Singh2007,PageRankAdenocarcinoma}, Chemistry~\cite{mooney2012molecularnetworks}, Neuroscience~\cite{Zuo01082012}, Literature~\cite{kontopoulou2012graph,meng2009computing}, Bibliometrics~\cite{bollen2006journal,Chen20078,Liu20051462,1742-5468-2007-06-P06010}, Sports~\cite{govan2008generalizing,PageRankTennis}, etc (see also~\cite{PageRankBeyondWeb}). 
 	PageRank has also been used as the fundamental building block, for many recently proposed methods for learning-over-graphs, targeting applications such as community detection~\cite{li2019optimizing,he2016local,he2015detecting}, semi-supervised classification~\cite{berberidis2019adaptive,klicpera2018predict,berberidis2018adadif,berberidis2018random}, and recommendation~\cite{Gori:2007:IRB:1625275.1625720,nikolakopoulos2019personalized,nikolakopoulos2020boosting,nikolakopoulos2019recwalk} to name a few. However, the majority of approaches to generalize PageRank, as well as its numerous applications in the literature, silently adopt the traditional teleportation model as a given. We believe that by doing so, they open the door to a number of unintentional qualitative consequences that arise from the underlying properties of such restrictive primitivity adjustment strategy---many of which primarily manifest in practical settings. More importantly, they are missing out on exploiting this component for the incorporation of available node meta-information, or for capturing aspects of the higher-order organization of the underlying network. Motivated by this, here we revisit the random surfer model focusing on the teleportation component, which we try to enrich in a flexible, versatile, and computationally efficient way. 
 	
 	From a conceptual point of view, our approach draws inspiration from the theory of \textit{Decomposable} systems by Simon~\cite{Simon:1996:SA:237774}: 
 	\begin{quote}
 		\textit{“To a Platonic mind, everything in the world is connected to everything
 			else---and perhaps it is. Everything is connected, but some things are more
 			connected than others. The world is a large matrix of interactions in
 			which most of the entries are close to zero, and in which, by ordering those
 			entries according to their orders of magnitude, a distinct hierarchic structure can be discerned.”}
 		\\
 		\phantom{dsfsdfdsfsdfsdfsdfsdfsdfdsfdsd} – {\upshape \textsc{Herbert A. Simon}}
 	\end{quote} 
 	In his seminal work on the architecture of complexity~\cite{Simon:1996:SA:237774}, Simon argued that the majority of sparse hierarchically structured systems share the property of having a \textit{Nearly Completely Decomposable} (NCD) architecture: they can be seen as comprised of a hierarchy of \textit{interconnected blocks}, sub-blocks and so on, in such a way that elements within any particular such block relate much more vigorously with each other than do elements belonging to different blocks, and this property holds between any two levels of the hierarchy. Simon's powerful ideas lay the framework for our approach; they are exploited conceptually, qualitatively, as well as computationally throughout our work\footnote{For an introduction the theory of NCD stochastic systems, see Appendix~\ref{Ch2:Preliminaries}.}.

\subsection*{Overview and Summary of Contributions}	 	
	The main contribution of this work\footnote{Preliminary results related to this  work can be found in~\cite{nikolakopoulos2015random,nikolakopoulos2013ncdawarerank,nikolakopoulos2016ranking}. Here we focus on general directed networks. For specific applications of the framework that exploit special properties of bipartite and multi-partite networks see also~\cite{RSMG,nikolakopoulos2020boosting,nikolakopoulos2015top,nikolakopoulos2014ncdrec}.} is the proposal of \textbf{NCDawareRank}; 
	a novel ranking framework that 
		generalizes the teleportation part of the random surfer model in an intuitive and computationally efficient way. 	
	\begin{itemize}
		\item 	We decompose the network into \textit{NCD blocks}, introducing a new level of abstraction which we proceed to exploit without creating coarser-level graph models that may obscure the direct link structure of the network and hide valuable information. At the heart of our approach lies the idea that the existence of a single link from a node $u$ to a node $v$ suggests \textit{multiple implied connections} of $u$ with other nodes that are considered related to the target node $v$ under the prism of the chosen decomposition.  \textit{In other words, in our model the existence of an outgoing link, except for justifying the importance of the node it points to, also ``hints'' about the importance of the block that contains this node.} 
			
			 To formulate mathematically the above idea, we introduce the notion of \textit{Proximal Sets} and we define a novel \textit{Inter-level Proximity Component} that quantifies these indirect inter-node relations in a way that inherits the attractive mathematical characteristics of PageRank’s traditional teleportation matrix. Our novel inter-level proximity stochastic matrix is low-rank, and we show the way it can be expressed as a product of two extremely sparse components, eliminating the need to be explicitly computed and stored, thereby ensuring that the final model can be handled  efficiently. In our model, in the general case, we have two levels of teleportation, which are translated in terms of random surfing behavior as follows:
			\begin{itemize}
				\item 
				Given that the random surfer is in node $u$, in the next step:
				\begin{enumerate}
					\item With probability $\eta$ goes to one of the outgoing links of $u$, i.e. follows the link structure of the network.
					\item With probability $\mu$ goes to the NCD proximal sets of $u$, i.e. the union of the NCD blocks that contain $u$ and the nodes it links to.
					\item With probability $1-\eta-\mu$ \textit{teleports} to a different node according to a given distribution.
				\end{enumerate}
			\end{itemize} 
			In other words, in our model a fraction $\mu$ of the importance that would be scattered throughout the network in a uniform manner, is propagated instead, to nodes that are considered ``close'' to the one currently visited by the random surfer.
		
			\item Based on the notions of NCD blocks and the related proximal sets, we propose an alternative approach to handling the \textit{dangling nodes} of the network, and we show that it implies no additional computational burden with respect to the traditional strongly- and weakly-preferential patching strategies. Our approach provides \textit{heterogeneous handling} and achieves \textit{more fair importance propagation} from the dangling nodes to their affiliated nodes. At the same time,  our approach has the advantage of \textit{lowering the incentive for link-spamming} and---under realistic assumptions about the decompositions---the advantage of \textit{highlighting useful properties of the underlying network’s structure that can result to tangible computational benefits}.			
		\item Albeit reducing its involvement to the final model, NCDawareRank in the general case also includes a standard rank-one teleportation component as a purely mathematical necessity for achieving primitivity. \textit{But, is it always necessary to include such component?} Interestingly the answer is not.
		In particular, we study theoretically the structure of our inter-level proximity model and we derive necessary and sufficient conditions, under which \textit{the underlying decomposition alone could result in a well-defined ranking vector}---eliminating the need for uniform teleportation. Furthermore, we examine the case where the underlying network is decomposed subject to more than one criteria simultaneously, and we show that \textit{primitivity can be achieved by their superposition even if none of the decompositions can ensure it by itself}. Our approach here is based on the theory of Non-Negative Matrices, and our proofs ensure that the primitivity criteria of the final stochastic matrix can be checked very efficiently; solely in terms of properties of the proposed decompositions.  
		\item We propose an efficient algorithm for computing the NCDawareRank vector in the general case. In particular, we show that our approach enables an exploitation of the coarser-level reducibility of the network that can lead to fundamentally faster computation of the ranking vector. We derive the conditions under which the final Markov chain becomes \textit{Nearly Completely Decomposable}, and \textit{Lumpable} with respect to the same coarse-level decomposition, and then, using an approach based on Stochastic Complementation, \textit{we predict analytically the aggregate-level limiting distribution, and we express the final ranking vector in terms of solutions of structurally identical, lower-dimensional ranking problems that can be solved in parallel}. Finally, we show that this approach could be applied for the computation of the standard PageRank problem as well---as long as the chosen handling strategy of the dangling nodes does not ``interfere'' with the connectivity properties of the actual network.
		\item We conduct a comprehensive set of experiments using real snapshots of the  Web-graph, and we show that our model alleviates the negative effects of the uniform teleportation matrix, and it produces ranking vectors that display \textit{low sensitivity to the effects of sparsity} and, at the same time, exhibit resistance to \textit{direct manipulation through link-spamming}. NCDawareRank outperforms several link-analysis generalizations of PageRank, in every experimental setting considered; both when we follow the traditional strongly preferential patching of the dangling nodes, and in the case we exploit our alternative dangling node handling strategy.  
	\end{itemize} 
		
	\section{NCDawareRank}
	\label{Ch4:NCDawareRank:Sec:Model}

		Before we proceed further, we need to define the parameters of our problem.
	
	\subsection{NCDawareRank Model Definitions}
	\label{Sec_Model}
		\begin{description}
			\item[Underlying Network.]  Let $\mathcal{G} = \{\mathcal{V,E}\}$ be a directed graph and denote $n=|\mathcal{V}|$. Consider a node $u$ in $\mathcal{V}$, and let  $\mathcal{G}_u$ denote the set of nodes that can be visited in a single step from $u$. Clearly,  $d_u\triangleq|\mathcal{G}_u|$ is the out-degree of $u$, i.e. the number of  outgoing edges of $u$. 
				\item[Decomposition.\index{NCDawareRank!General Definition!Decomposition}] Our underlying space is assumed to be decomposable, subject to given set of criteria into possibly overlapping blocks of related nodes. 
				For example the set of Web-pages can be decomposed into blocks that depict sites, domains, languages, topics of the content of the page etc. The first two decompositions are partitions, whereas in the third and fourth the blocks may be overlapping, since a Web-page may contain material written in more than one languages or covering more than one topics. 
				Formally, a decomposition is defined to be an indexed family of non-empty sets,
				\begin{equation}
				\mathcal{M} \triangleq \{\mathcal{D}_1,\dots,\mathcal{D}_{K}\}
				\end{equation} 
				that collectively cover the underlying space, i.e. 
				\begin{equation}
				\mathcal{V}=\bigcup_{k=1}^{K}\mathcal{D}_k.
				\end{equation}
				Each set of nodes $\mathcal{D}_I$ is referred to as an \textit{NCD block}. 
				\item[Proximal Sets.\index{NCDawareRank!General Definition!Proximal Sets}]  We define $\mathcal{M}_u$ to be the set of the \textit{proximal} nodes  of $u$, subject to the decomposition $\mathcal{M}$, i.e. the union of the blocks that contain $u$ and the nodes it links to. Formally, the set  $\mathcal{M}_u$ is defined by 
				\begin{equation}
				\mathcal{M}_u \triangleq \bigcup_{{w \in (u\cup\mathcal{G}_u),w \in \mathcal{D}_k}}\mathcal{D}_k,
				\label{def:proximal_ovelapping}
				\end{equation}  
				and we use $N_u$ to denote the number of different blocks in $\mathcal{M}_u$. 

		\item[Normalized Adjacency Matrix $\mathbf{H}$.\index{NCDawareRank!General Definition!Normalized Adjacency Matrix $\mathbf{H}$}] As in the traditional PageRank model, this matrix depicts the relations between the nodes as they arise directly from the data. In particular, matrix $\mathbf{H}$ is defined to be the row-normalized version of the adjacency matrix of the graph. 
		Formally, its $uv^{\textit{th}}$ element is defined as follows:
		\begin{equation}
		H_{uv} \triangleq \left\{
		\begin{array}{r l}
		\frac{1}{d_u}, & \quad \mbox{if $v \in \mathcal{G}_u$},\\
		0, & \quad \mbox{otherwise}.\\
		\end{array} \right. 
		\end{equation}	
		Matrix $\mathbf{H}$ is assumed to be a row-stochastic matrix. The matter of dangling nodes (i.e. nodes with no outgoing edges) is considered fixed through some sort of stochasticity adjustment. For reasons of better presentation we postpone further discussion of this matter to  Section~\ref{Ch4:NCDawareRank:Sec:Dangling}. 
			\item[Inter-Level Proximity Matrix $\mathbf{M}$.] The Inter-Level Proximity matrix is created to depict the inter-level connections between the nodes in the network, that arise from the decomposition. In particular, each row of matrix $\mathbf{M}$ denotes a probability vector $\mathbf{m}^\intercal_u$, that distributes evenly its mass between the $N_u$ blocks of $\mathcal{M}_u$, and then, uniformly to the included nodes of each block.
			Formally, the $uv^{\textit{th}}$ element of matrix $\mathbf{M}$, that relates the node $u$ with node $v$, is defined as
			\begin{equation}
			M_{uv}\triangleq \sum_{\mathcal{D}_k \in \mathcal{M}_{u}, v \in \mathcal{D}_k}\frac{1}{N_{u}\lvert \mathcal{D}_k\rvert}.
			\label{def:M_overlapping}
			\end{equation}
			When the blocks define a partition of the underlying space the above definition is simplified to 
			\begin{displaymath}
			M_{uv}\triangleq \left\{
			\begin{array}{r l}
			\frac{1}{N_u|\mathcal{D}_{(v)}|}, & \quad \mbox{if $v \in \mathcal{M}_u$},\\
			0, & \quad \mbox{otherwise},\\
			\end{array} \right. 
			\end{displaymath} where we used $\mathcal{D}_{(v)}$ to denote the unique (in this case) NCD block that contains node $v$.

				\begin{figure}
				\centering
				\includegraphics[height = 4.5cm]{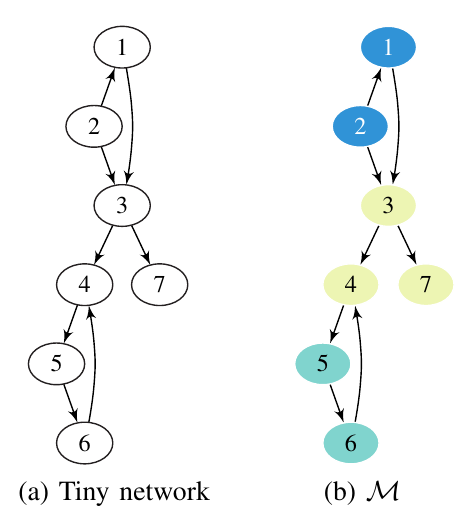}
				\caption{Example network and its associated decomposition.}
				\label{tiny_NCD_web}
			\end{figure}

				\textbf{Factorization of Matrix} $\mathbf{M}$. Matrix $\mathbf{M}$ is by definition a sparse matrix. For large-scale applications though, it might not be  sparse enough. Fortunately, $\mathbf{M}$ has a very special structure that can be exploited in order to achieve efficient storage and computability as well as other advantages (to be explored in the upcoming sections of this work). 
				In particular, from the definition of the NCD blocks and the proximal sets, it is clear that whenever the number of blocks is smaller than the number of nodes in the network, i.e. $K<n$, the corresponding matrix $\mathbf{M}$ is necessarily low-rank; in fact, a closer look at the definitions~(\ref{def:proximal_ovelapping}) and~(\ref{def:M_overlapping}) above, suggests that matrix $\mathbf{M}$ admits a very useful factorization, which ensures the tractability of the resulting model. In particular, every matrix 
				$\mathbf{M}$ can be expressed as a product of 2 extremely sparse matrices, 
				$\mathbf{R}$ and $\mathbf{A}$, defined below. 
				
				We first define a matrix $\mathbf{X}\in \mathfrak{R}^{n \times K}$, whose $ik^{\textit{th}}$ element is 1, if $\mathcal{D}_k \in \mathcal{M}_i$ and zero otherwise, and a matrix $\mathbf{Y}\in \mathfrak{R}^{K\times n}$, whose $kj^{\textit{th}}$ element is 1 if $v_j \in \mathcal{D}_k$ and zero otherwise. Then, if $\mathbf{R}$ and $\mathbf{A}$, denote the row-normalized versions $\mathbf{X}$ and $\mathbf{Y}$ respectively, matrix $\mathbf{M}$ can be expressed as: 
				\begin{equation}
				\mathbf{M}  \equiv \mathbf{R} \mathbf{A}, \quad \mathbf{R} \in \mathfrak{R}^{n\times K}, \quad \mathbf{A} \in  \mathfrak{R}^{K\times n}.
				\end{equation}
				For the sake of example, these matrices for the tiny network of Fig.~\ref{tiny_NCD_web}  where we see 3 NCD blocks, are: 
					\begin{align}
				\mathbf{R} & = 
				{\small 	\begin{pmatrix}
					1/2 & 1/2 & 0 \\
					1/2 & 1/2 & 0 \\
					0 & 1 & 0 \\
					0 & 1/2 & 1/2 \\
					0 & 0 & 1  \\
					0 & 1/2 & 1/2  \\
					0 & 1 & 0    
					\end{pmatrix}},
				\nonumber \\
				\mathbf{A} & = 
				{\small 	\begin{pmatrix} 
					1/2 & 1/2 & 0 & 0 & 0 & 0 & 0 \\
					0 & 0 & 1/3 & 1/3 & 0 & 0 & 1/3 \\ 
					0 & 0 & 0 & 0 & 1/2 & 1/2 & 0 
					\end{pmatrix} },\nonumber
				\end{align}
%					\begin{align}
%					\mathbf{R} & = 
%					{\small \begin{pmatrix}
%					1 & 0 & 0 \\
%					1 & 0 & 0 \\
%					1/3 & 1/3 & 1/3 \\
%					0 & 1 & 0 \\
%					0 & 1 & 0 \\
%					0 & 1 & 0 \\
%					0 & 0 & 1    
%					\end{pmatrix}}, \nonumber \\
%					\mathbf{A} & = 
%				{\small 	\begin{pmatrix} 
%					1/3 & 1/3 & 1/3 & 0 & 0 & 0 & 0 \\
%					0   & 0   & 0   & 1/3 & 1/3 & 1/3 & 0 \\ 
%					0   & 0   & 0   & 0 & 0 & 0 & 1 
%					\end{pmatrix}}	\nonumber
%					\end{align}
				and the related matrix $\mathbf{M}$  is  
%					\begin{displaymath}
%					\mathbf{M}= 
%					{\small \begin{pmatrix}
%					1/3 & 1/3 & 1/3 & 0   & 0   & 0   & 0 \\
%					1/3 & 1/3 & 1/3 & 0   & 0   & 0   & 0 \\
%					1/9 & 1/9 & 1/9 & 1/9 & 1/9 & 1/9 & 1/3 \\
%					0   & 0   & 0   & 1/3 & 1/3 & 1/3 & 0 \\
%					0   & 0   & 0   & 1/3 & 1/3 & 1/3 & 0 \\
%					0   & 0   & 0   & 1/3 & 1/3 & 1/3 & 0 \\
%					0   & 0   & 0   & 0   & 0   & 0   & 1   
%					\end{pmatrix}}.
%					\end{displaymath}
				\begin{equation}
				\mathbf{M} = {\small 	\begin{pmatrix} 
					{1}/{4} & {1}/{4} & {1}/{6} & {1}/{6} & 0 & 0 & {1}/{6} \\
					{1}/{4} & {1}/{4} & {1}/{6} & {1}/{6} & 0 & 0 & {1}/{6} \\
					0 & 0 & {1}/{3} & {1}/{3} & 0 & 0 & {1}/{3} \\
					0 & 0 & {1}/{6} & {1}/{6} & {1}/{4} & {1}/{4} & {1}/{6} \\
					0 & 0 & 0 & 0 & {1}/{2} & {1}/{2} & 0 \\
					0 & 0 & {1}/{6} & {1}/{6} & {1}/{4} & {1}/{4} & {1}/{6} \\
					0 & 0 & {1}/{3} & {1}/{3} & 0 & 0 & {1}/{3} \\
					\end{pmatrix} }. \nonumber
				\end{equation}
				Notice that the inter-level proximity matrices are well-defined stochastic matrices, for every possible decomposition. Their stochasticity can arise immediately from the row normalization of matrices $\mathbf{R}, \mathbf{A}$, together with the fact that neither matrix $\mathbf{X}$ nor matrix $\mathbf{Y}$  have zero rows\footnote{Indeed, the existence of a zero row in matrix $\mathbf{X}$ implies 
					\begin{math}
					\mathcal{V} \neq \cup_{k=1}^{K}\mathcal{D}_k,
					\end{math}
					which contradicts the definition of $\mathcal{M}$; similarly the existence of a zero row in matrix $\mathbf{Y}$ contradicts the definition of the NCD blocks $\mathcal{D}$ which are defined to be non-empty.}.

				In Table~\ref{storage}, we see the storing requirements of these matrices for some of the networks we experimented on\footnote{More information about these networks can be found in Appendix~\ref{App:Datasets}.}. These networks denote snapshots of the Web and the criterion behind the decomposition in this case is assumed to be the partition of the Web into websites.  
				\begin{table}[h!]
					\centering
					\caption[Storage Savings due to the Factorization of Matrix $\mathbf{M}$]{Storage Savings due to the Factorization of Matrix $\mathbf{M}$}
					\resizebox{0.99\linewidth}{!}{
						\begin{tabular}{c c c c c c}
							\hline
							Network &\# of nodes &\# of blocks & $\mathbf{M}$ & $\mathbf{R}+\mathbf{A}$ & $\mathbf{H}$ \\ \hline
							\texttt{cnr-2000}  & 326K & 0.7K & 42.57 GB    & \textbf{8.209 MB}  & 51.56 MB     \\ 
							\texttt{eu-2005}  &  863K & 0.4K &  588.8 GB    & \textbf{34.42 MB}  &  300.1 MB   \\
							\texttt{india-2004} & 1.38M & 4.3K & 131.6 GB    &\textbf{38.82 MB}  & 268.7 MB     \\ 
							\texttt{uk-2002}  & 18.5M & 97.4K & 1.323 TB    & \textbf{604.5 MB}  & 4.580 GB     \\
							\hline
						\end{tabular}
					}
					\label{storage}
				\end{table}
				
				Notice that after the factorization, the combined storage needs for matrices $\mathbf{R}$ and $\mathbf{A}$ are significantly lower, even when compared to the storage requirements of adjacency matrix $\mathbf{H}$.
			\item[Standard Teleportation Matrix $\mathbf{E}$.] 	Finally, the NCDawareRank model also includes a rank-one teleportation matrix $\mathbf{E}$. One very simple and convenient way to define $\mathbf{E}$ is the following: 
			\begin{displaymath}
			\mathbf{E} \triangleq \mathbf{1}\mathbf{v}^{\intercal} ,
			\end{displaymath}
			where vector $\mathbf{v}^{\intercal}$ is a probability vector that first distributes the rank evenly between the NCD blocks and then, in each block, evenly to the included nodes. Of course, as in PageRank, one can construct a \textit{personalized} version of the teleportation matrix, using a different stochastic vector instead of $\mathbf{v}^{\intercal}$. The introduction of this matrix, 
%			can be seen as a remedy to 
			ensures that the underlying Markov chain, 
%			 corresponding to the final matrix, 
			becomes irreducible and aperiodic, and therefore possesses a unique positive stationary probability distribution.
			\item[NCDawareRank Vector.] The ranking vector produced by our model is defined as the normalized left Perron-Frobenius eigenvector of the stochastic matrix that brings together the normalized adjacency matrix $\mathbf{H}$, the inter-level proximity matrix $\mathbf{M}$, and the standard teleportation matrix $\mathbf{E}$. Concretely, the final stochastic matrix, which we denote $\mathbf{P}$, is given by
			\begin{equation}
			\label{NCDawareRank}
			\mathbf{P} \triangleq \eta \mathbf{H} + \mu \mathbf{M} + (1-\eta - \mu) \mathbf{E}  ,
			\end{equation}
			with $\eta,\mu > 0$ such that $\eta+\mu \leq 1$. Parameter $\eta$ controls the fraction of importance delivered to the outgoing links and parameter $\mu$  controls the fraction of importance that will be propagated to the proximal nodes. 
		\end{description}

In order to ensure the irreducibility and aperiodicity of the final stochastic matrix in the general case, $\eta + \mu$ must be less than $1$. However, when the  inter-level proximity matrix is enough to ensure the ergodicity of the final Markov chain, $\eta + \mu = 1$, leads to a well-defined NCDawareRank vector also. The conditions for achieving primitivity without resorting to the standard teleportation are explored in depth in % the Section to follow. 
Section \ref{Ch4:NCDawareRank:Sec:PrimitivityAnalysis}.

\subsection{Handling the Dangling Nodes}
	\label{Ch:Dangling:Sec:NCDawareDanglingNodeStrategy}
	\label{Ch4:NCDawareRank:Sec:Dangling}
	Our goal is to exploit the partition of the network into NCD blocks, in order to propose a richer dangling node handling strategy without undermining the efficiency of the overall approach. Having this in mind, in this section we build on the underlying idea of the NCDawareRank approach and we describe a simple dangling node patching strategy involves \textit{different patching behavior depending on the origin NCD block of the dangling node}, and its formal definition is given below: 
	\begin{strategy}
		\label{strategy}
		For every dangling node $d$, the corresponding $\boldsymbol{0}^{\intercal}$ row of the original adjacency matrix is replaced with a probability distribution $\mathbf{f}^\intercal$  that distributes evenly its mass between the $N_d$ blocks of $\mathcal{M}_d$, and then, uniformly to the included nodes of each block. Concretely, the $v^{\textit{th}}$ element of the vector $\mathbf{f}^\intercal$ is defined to be
		\begin{equation}
		[\mathbf{f}]_{v}\triangleq \sum_{\mathcal{D}_k \in \mathcal{M}_{d}, v \in \mathcal{D}_k}\frac{1}{N_{d}\lvert \mathcal{D}_k\rvert},
		\label{def:fd_overlapping}
		\end{equation}
		for the general case where $d$ might belong to more than one blocks. When the blocks are defined to be non-overlapping, this simplifies to
		\begin{equation}
		[\mathbf{f}]_{v} \triangleq \left\{\begin{array}{rl}
		\dfrac{1}{\lvert \mathcal{D}_{k}\rvert}, & \textrm{ if } d,v \in \mathcal{D}_{k}, \\
		0, & \textrm{otherwise} .
		\end{array} \right.
		\end{equation}
		\end{strategy}

	        Notice that the strategy we propose can be handled very efficiently exploiting the factorization of matrix $\mathbf{M}$ we introduced in the previous section. Concretely, if we write the final matrices $\mathbf{H,M}$ as a sum of two matrices
	        \begin{eqnarray}
	        \mathbf{H} & = & \mathbf{H}_{\text{ND}} + \mathbf{H}_{\text{D}}, \\
	        \mathbf{M} & = & \mathbf{M}_{\text{ND}} + \mathbf{M}_{\text{D}},
	        \end{eqnarray} 
	        with matrices $\mathbf{H}_{\text{ND}},\mathbf{M}_{\text{ND}}$ containing the non-dangling nodes of the network (and zero rows for the dangling ones) and $\mathbf{H}_{\text{D}},\mathbf{M}_{\text{D}}$ containing the dangling nodes of the network (and zero rows for the non-dangling ones), we have 
	        \begin{eqnarray}
	        \mathbf{P} & = & \eta\mathbf{H} + \mu\mathbf{M} + (1-\eta-\mu)\mathbf{E} \nonumber \\
	        & = & \eta(\mathbf{H}_{\text{ND}} + \mathbf{H}_{\text{D}}) + \mu(\mathbf{M}_{\text{ND}} + \mathbf{M}_{\text{D}}) + (1-\eta-\mu)\mathbf{E} \nonumber \\
	        & = & \eta\mathbf{H}_{\text{ND}} + \eta \mathbf{H}_{\text{D}} + \mu \mathbf{M}_{\text{D}} + \mu \mathbf{M}_{\text{ND}} + (1-\eta-\mu)\mathbf{E}.
	        \end{eqnarray}
	        However, from the definition of the NCD blocks and the proximal sets (see Section \ref{Sec_Model}) we get that under the above handling strategy, it holds:
	        \begin{equation}
	        \mathbf{H}_{\text{D}} \equiv \mathbf{M}_{\text{D}}.
	        \end{equation} 
	        Therefore, we have
	        \begin{eqnarray}
	        \mathbf{P} & = & \eta\mathbf{H}_{\text{ND}} + \eta \mathbf{H}_{\text{D}} + \mu \mathbf{M}_{\text{D}} + \mu \mathbf{M}_{\text{ND}} + (1-\eta-\mu)\mathbf{E} \nonumber \\
	        & = & \eta\mathbf{H}_{\text{ND}} + (\eta+\mu) \mathbf{M}_{\text{D}} + \mu \mathbf{M}_{\text{ND}} + (1-\eta-\mu)\mathbf{E},
	        \end{eqnarray} which implies no additional computational or storage burden with respect to the standard NCDawareRank model.  In particular, the number of non-zero entries that arise from the application of Strategy 1 is strictly less than the number of non-zeros of any weakly preferential patching strategy applied to the NCDawareRank model. 
	
	The simple strategy we propose has a number of useful characteristics that make it a reasonable candidate for dealing with the dangling node problem. In Appendix~\ref{SI:dangling} we discuss this matter briefly focusing primarily on what this strategy implies in terms of importance propagation to the nodes and also on how it translates in terms of random surfing behavior.

	\section{Primitivity Analysis of NCDawareRank’s Teleportation Model}
	\label{Ch4:NCDawareRank:Sec:PrimitivityAnalysis}

	Intuitively, NCDawareRank tries to alleviate the negative effects of uniform teleportation by introducing an intermediate level of proximity between the one that comes directly from the actual network topology (matrix $\mathbf{H}$) and the one that relates naively all the elements with each other for purely mathematical reasons. NCDawareRank’s matrix $\mathbf{M}$, informally, ``augments'' the scarce internode connections of the actual network interpreting them in a ``synecdochical manner'' that permits a single link to relate many more nodes at once under the prism of the chosen decomposition, and then reduces the involvement of the rank-one teleportation component in the final model in favor of matrix $\mathbf{M}$. However, albeit alleviating some of its negative effects, NCDawareRank model also includes a traditional rank-one teleportation matrix as a purely mathematical necessity. But, is it?  
	
	The main question we try to address in this section is: \textit{Is it possible to discard the uniform teleportation altogether?} 	Thankfully, the answer is yes. In particular, we show that, the definition of the NCD blocks, can be enough to ensure the production of well-defined ranking vectors without resorting to rank-one teleportation. The criterion for this to be true is expressed solely in term of properties of the proposed decomposition, which makes it very easy to check and at the same time gives insight that can lead to better decompositions for the particular ranking problems under consideration.

	Our approach is based on non-negative matrix theory~\cite{seneta2006non} which simplifies greatly the derivation of our results, and improves the presentation. Before we proceed to the proof of our main results we present here some necessary preliminary definitions and terminology as well as the Perron-Frobenius theorem.
	
	\subsection{Preliminaries}

	\begin{definition}[Irreducibility] 	
		An $n\times n$ non-negative matrix $\mathbf{P}$ is called \textit{irreducible} if for every pair of indices $i,j \in [1,2,\dots,n]$, there exists a positive integer $m \equiv m(i,j)$ such that $[\mathbf{P}^m]_{ij}>0$. The class of all non-negative irreducible matrices is denoted $\mathfrak{I}$.
	\end{definition}		

	\begin{definition}[Period]	
		The \textit{period} of an index $i\in[1,2,\dots,n]$ is defined to be the greatest common divisor of all positive integers $m$ such that $[\mathbf{P}^m]_{ii}>0$.
	\end{definition}
	
	\begin{proposition}[Periodicity is a Matrix Property]
		For an irreducible matrix, the period of every index is the same and is referred to as the period of the matrix.
	\end{proposition}

	\begin{definition}[Primitivity]	
		An irreducible matrix with period $d=1$, is called \textit{primitive}. The important subclass of all primitive matrices will be denoted $\mathfrak{P}$.
	\end{definition}

	\begin{definition}[Allowability]	\index{Allowability!Definition}
		A non-negative matrix $\mathbf{A}$ is said to be row-allowable if it has at least one positive entry in each row. It is said to be column-allowable if $\mathbf{A}^\intercal$ is row-allowable. It is said to be allowable if it is both row- and column-allowable.
		\label{def:Allowability}
	\end{definition}
	
	Finally, we also state here the Perron-Frobenius theorem for irreducible matrices which will be used throughout this section. For a proof of the theorem as well as detailed treatment of the theory of non-negative matrices, the interested reader may refer to~\cite{seneta2006non}.

	\begin{theorem}[Perron-Frobenius Theorem for Irreducible Matrices~\cite{Frobenius-1908-theorem,Perron-1907-theorem}]
		Let $\mathbf{T}$ be an $n\times n$ irreducible non-negative matrix. Then, there exists an eigenvalue $r$ such that:
		\begin{enumerate}
			\item $r$ is real and positive.
			\item With $r$ can be associated strictly positive left and right eigenvectors.
			\item $r \ge \lvert\lambda_i \rvert$ for any eigenvalue $\lambda_i \neq r$. Furthermore, when $\mathbf{T}$ is cyclic with period $d>1$ there are present precisely $d$ distinct eigenvalues $\lambda_i$ with $\lvert\lambda_i \rvert = r$. These eigenvalues are the complex roots of the equation $\lambda^d - r^d = 0$, i.e. 
			\begin{displaymath} \lambda_1=r\omega^0, \lambda_2 = r\omega^1,\dots,\lambda_d = r\omega^{d-1} , 
			\end{displaymath} 
			where $\omega = e^{2 \pi i /d}$.
			\item The eigenvectors associated with $r$ are unique to constant multiples.
			\item If $0\leq \mathbf{B} \leq \mathbf{T}$ and $\beta$ is an eigenvalue of $\mathbf{B}$, then $\lvert \beta \rvert \leq r$. Moreover, 
			\begin{displaymath}
			\lvert \beta \rvert = r \quad \Longrightarrow \quad \mathbf{B}=\mathbf{T}.
			\end{displaymath}
			\item $r$ is a simple root of the characteristic equation of $\mathbf{T}$.
		\end{enumerate}
	\end{theorem}

	\subsection{Primitivity Criterion for the Single Decomposition Case}
	\label{Ch:PrimitivityAnalysis:Sec:PrimitivitySingleDecomposition}
	As we discussed in Section \ref{Ch:Intro:Sec:RandomSurfing}, mathematically, in the standard PageRank model the introduction of the teleportation matrix can be seen as a \textit{primitivity adjustment} of the final stochastic matrix. Indeed, the adjacency matrix of many directed networks is typically reducible, so if the teleportation matrix had not existed the PageRank vector would not be well-defined~\cite{LangvilleMeyer06,pagerank}. 
	
	In the general case, the same holds for NCDawareRank, as well. However, for suitable decompositions of the underlying network, matrix $\mathbf{M}$ opens the door for achieving primitivity without resorting to the uninformative teleportation matrix.  Here, we show that this ``suitability'' of the decompositions can, in fact, be reflected on the properties of a low-dimensional  \textbf{Indicator Matrix} defined below:

	\begin{definition}[Indicator Matrix]
		For every decomposition $\mathcal{M}$, we define an Indicator Matrix $\mathbf{W}\in \mathfrak{R}^{K \times K}$ designed to capture the existence of inter-block relations in the underlying network. Concretely, matrix $\mathbf{W}$ is defined as follows:
		\begin{displaymath}
		\mathbf{W} \triangleq \mathbf{A}\mathbf{R},  
		\end{displaymath} 
		where $\mathbf{A,R}$ are the factors of the inter-level proximity matrix $\mathbf{M}$.
	\end{definition}
	
	Clearly, whenever $W_{IJ}$ is positive, there exists a node $u \in \mathcal{D}_I$ such that $\mathcal{D}_J \in \mathcal{M}_u$. Intuitively, one can see that a positive element in matrix $\mathbf{W}$ implies the existence of possible inter-level ``random surfing paths'' between the nodes belonging to the corresponding blocks. Thus, if the indicator matrix $\mathbf{W}$ is irreducible, these paths exist between every pair of nodes in the network, which makes the stochastic matrix $\mathbf{M}$ also irreducible. 
	
	In fact, in the following theorem we show that the irreducibility of matrix $\mathbf{W}$ is enough to certify the primitivity of the final NCDawareRank matrix, $\mathbf{P}$. Then, just choosing positive numbers $\eta,\mu$ that sum to one, leads to a well-defined ranking vector produced by an NCDawareRank model without a traditional rank-one teleportation component.
	
	\begin{theorem}[Primitivity Criterion for the Single Decomposition Case]
		Matrix  $\mathbf{P} = \eta\mathbf{H} + \mu\mathbf{M}$, with $\eta$ and $\mu$ positive real numbers such that $\eta + \mu = 1$, is primitive if and only if the indicator matrix $\mathbf{W}$ is irreducible. Concretely, $\mathbf{P} \in \mathfrak{P} \iff \mathbf{W} \in \mathfrak{I}$.
		\label{thm:PrimitivitySinleDecomposition}
	\end{theorem}
	
	\begin{proof}
		We will first prove that 
		\begin{equation}
		\mathbf{W} \in \mathfrak{I} \implies \mathbf{P} \in \mathfrak{P}.
		\end{equation} 
		
		First notice that whenever matrix $\mathbf{W}$ is irreducible then it is also primitive. In particular, it is known that when a non-negative irreducible matrix has at least one positive diagonal element, then it is also primitive~\cite{meyer2000matrix}. In case of matrix $\mathbf{W}$, notice that by the definition of the proximal sets and matrices $\mathbf{A,R}$, we get that $W_{II}>0$ for every $1\leq I \leq K$. Thus, the irreducibility of the indicator matrix ensures its primitivity also. Formally, we have 
		\begin{equation}
		\mathbf{W} \in \mathfrak{I} \implies \mathbf{W} \in \mathfrak{P}.
		\end{equation}

		Now if the indicator matrix $\mathbf{W}$ is primitive, the same is true for the inter-level proximity matrix $\mathbf{M}$. 
		Before we prove this we will prove the following useful lemma.
		\begin{lemma}
			If a positive matrix $\mathbf{X}$ is multiplied by a row-allowable matrix from the left, or a column-allowable matrix from the right, the product matrix $\mathbf{Y}$ remains positive.
			\label{lemma:allowability}
		\end{lemma}	
		\begin{proof}
			Let us consider the first case. Let $\mathbf{L}$ be a row-allowable matrix. By definition  there is at least one $k$ such that $L_{ik}$ is positive. Thus,
			\begin{displaymath}
			Y_{ij} = \sum_{k} L_{ik}X_{kj} >0, \qquad \text{for all } i,j.
			\end{displaymath}
			Therefore, $\mathbf{Y}$ is positive as well. Following exactly the same argument one can prove that  multiplication from the right with a column-allowable matrix produces a positive product also.
		\end{proof}
		
		\begin{lemma} The primitivity of the indicator matrix $\mathbf{W}$ implies the primitivity of the inter-level proximity matrix $\mathbf{M}$, defined over the same decomposition, i.e
			\begin{equation}
			\mathbf{W} \in \mathfrak{P} \implies \mathbf{M} \in \mathfrak{P}.		
			\end{equation}
			\label{lemma:W2M}
		\end{lemma}
		\begin{proof}
			It suffices to show that there exists a number $m$, such that for every pair of indices $i,j$, $[\mathbf{M}^m]_{ij}>0$ holds. Or equivalently, there exists a positive integer $m$ such that $\mathbf{M}^m$ is a positive matrix (see \cite{seneta2006non}).
			
			This can be seen easily using the factorization of matrix $\mathbf{M}$ given above. In particular, since $\mathbf{W}\in\mathfrak{P}$, there exists a positive integer $k$ such that $\mathbf{W}^k>0$. Now, if we choose $m = k+1$, we get:
			\begin{eqnarray}
			\mathbf{M}^m & = & (\mathbf{RA})^{k+1} \nonumber \\
			& = & \underbrace{\mathbf{(RA)(RA)\cdots (RA)}}_{k+1\text{ times}} \nonumber \\
			& = & \mathbf{R} \underbrace{\mathbf{(AR)(AR)\cdots (AR)}}_{k\text{ times}}\mathbf{A} \nonumber \\
			& = & \mathbf{R} \mathbf{W}^k \mathbf{A}.
			\label{rel:M_Prim}
			\end{eqnarray}
			
			However, matrix $\mathbf{W}^k$ is positive and since both matrices $\mathbf{R}$ and $\mathbf{A}$ are---by definition---allowable, by Lemma~\ref{lemma:allowability} we get that matrix $\mathbf{M}^m$, is also positive. Therefore, $\mathbf{M} \in \mathfrak{P}$, and the proof is complete. 
		\end{proof}
		
		Now, in order to get the primitivity of the final stochastic matrix $\mathbf{P}$, we use the following useful lemma which shows that any convex combination of stochastic matrices that contains at least one primitive matrix, is also primitive.
		\begin{lemma}
			Let $\mathbf{T}$ be a primitive stochastic matrix and $\mathbf{B_1,B_2,\dots,B_n}$ stochastic matrices, then matrix
			\begin{displaymath}
			\mathbf{C} = \alpha \mathbf{T}+\beta_1\mathbf{B_1}+\dots+\beta_n\mathbf{B_n},
			\end{displaymath} where $\alpha>0$ and $\beta_1,\dots,\beta_n\geq0$ such that $\alpha+\beta_1+\dots+\beta_n=1$ is a primitive stochastic matrix.
			\label{lemma:ConvexPrimitivity}
		\end{lemma}

		\begin{proof}
			Clearly matrix $\mathbf{C}$ is stochastic as a convex combination of stochastic matrices (see~\cite{horn2012matrix}).
			For the primitivity part it suffices to show that there exists a natural number, $m$, such that $\mathbf{C}^m>0$. This can be seen very easily. In particular, since matrix $\mathbf{T} \in \mathfrak{P}$, there exists a number $k$ such that every element in $\mathbf{T}^{k}$ is positive.
			
			Consider the matrix $\mathbf{C}^m$:
			\begin{eqnarray}
			\mathbf{C}^m & = & (\alpha\mathbf{T}+\beta_1\mathbf{B_1}+\dots+\beta_n\mathbf{B_n})^m \nonumber \\
			& = & \alpha^m\mathbf{T}^m + (\text{sum of non-negative matrices}).
			\end{eqnarray}
			Now letting $m=k$, we get that every element of matrix $\mathbf{C}^{m}$ is strictly positive, which completes the proof.   
		\end{proof}

		As we have seen, when $\mathbf{W}\in \mathfrak{I}$, matrix $\mathbf{M}$ is primitive. Furthermore, $\mathbf{M}$ and $\mathbf{H}$ are by definition stochastic. Thus, Lemma~\ref{lemma:ConvexPrimitivity} applies and we get that the NCDawareRank  matrix $\mathbf{P}$, is also primitive. 
		In conclusion, we have shown that: 
		\begin{equation}
		\mathbf{W}\in\mathfrak{I} 
		\implies\mathbf{W}\in\mathfrak{P}\implies\mathbf{M}\in\mathfrak{P}\implies\mathbf{P}\in\mathfrak{P},
		\label{ReverseProof}
		\end{equation}
		which proves the reverse direction of the theorem.
		
		To prove the forward direction (i.e. $\mathbf{P} \in \mathfrak{P} \implies \mathbf{W} \in \mathfrak{I}$) it suffices to show that whenever matrix $\mathbf{W}$ is reducible, matrix $\mathbf{P}$ is also reducible (and thus, not primitive \cite{seneta2006non}). 	First observe that when matrix $\mathbf{W}$ is reducible the same holds for matrix $\mathbf{M}$. 
		
		\begin{lemma}
			The reducibility of the indicator matrix $\mathbf{W}$ implies the reducibility of the inter-level proximity matrix $\mathbf{M}$. Concretely,
			\begin{equation}
			\mathbf{W} \notin \mathfrak{I} \implies \mathbf{M} \notin \mathfrak{I}.
			\end{equation}
		\end{lemma}
		
		\begin{proof}
			Assume that matrix $\mathbf{W}$ is reducible. Then, there exists a permutation matrix ${\mathbf{\Pi}}$ such that $\mathbf{\Pi W \Pi^\intercal}$ has the form 
			\begin{equation}
			\begin{pmatrix}
			\mathbf{X} & \mathbf{Z} \\
			\mathbf{0} & \mathbf{Y}
			\end{pmatrix} ,
			\label{rel:BlockUpperDiagonal}
			\end{equation}
			where $\mathbf{X,Y}$ are square matrices~\cite{seneta2006non}. Notice that a similar block upper triangular form can be then achieved for matrix $\mathbf{M}$. In particular, the existence of the block zero matrix in~(\ref{rel:BlockUpperDiagonal}), together with the definition of matrices $\mathbf{A,R}$, ensures the existence of a set of blocks that have the property none of their including nodes to have outgoing edges to the rest of the nodes in the network\footnote{Notice that if this was not the case, there would be a nonzero element in the block below the diagonal necessarily.}. Thus, organizing the rows and columns of matrix $\mathbf{M}$ such that these nodes are assigned the last indices, results in a matrix $\mathbf{M}$ that has a similarly block upper triangular form. This makes $\mathbf{M}$ reducible too.  
		\end{proof}
		
		The only remaining thing we need to show is that the reducibility of matrix $\mathbf{M}$ implies the reducibility of matrix $\mathbf{P}$ also. This can arise from the fact that by definition 
		\begin{equation}
		M_{ij}=0 \implies H_{ij}=0.
		\end{equation} So, the permutation matrix that brings $\mathbf{M}$ to a block upper triangular form, has exactly the same effect on matrix $\mathbf{H}$. Similarly, the  final stochastic matrix $\mathbf{P}$ has the same block upper triangular form as a sum of matrices $\mathbf{H}$ and $\mathbf{M}$. This makes matrix $\mathbf{P}$ reducible and hence, non-primitive.
		
		Therefore, we have shown that $\mathbf{W} \notin \mathfrak{I} \implies \mathbf{P} \notin \mathfrak{P}$, which is equivalent to  
		\begin{equation}
		\mathbf{P} \in \mathfrak{P} \implies \mathbf{W} \in \mathfrak{I}.
		\label{ForwardProof}
		\end{equation}
		Putting everything together, we see that both directions of our theorem have been established. Thus we get,
		\begin{equation}
		\mathbf{P} \in \mathfrak{P} \iff \mathbf{W} \in \mathfrak{I},
		\end{equation} and our proof is complete. 
	\end{proof}

	Now, when the stochastic matrix $\mathbf{P}$ is primitive, from the Perron-Frobenius theorem it follows that its largest eigenvalue---which is equal to 1---is unique and it can be associated with strictly positive left and right eigenvectors. Therefore, under the conditions of Theorem~\ref{thm:PrimitivitySinleDecomposition}, the ranking vector produced by the NCDawareRank model---which is defined to be the stationary distribution of the stochastic matrix $\mathbf{P}$: (a) is uniquely determined as the (normalized) left eigenvector of $\mathbf{P}$ that corresponds to the eigenvalue 1 and, (b) its support includes every node in the underlying network. The following corollary summarizes the result.
	
	\begin{corollary}
		When the indicator matrix $\mathbf{W}$ is irreducible, the ranking vector produced by NCDawareRank with $\mathbf{P} = \eta\mathbf{H} + \mu\mathbf{M}$, where $\eta,\mu$ positive real numbers such that $\eta + \mu = 1$ holds, denotes a well-defined distribution that assigns positive ranking to every node in the network.
	\end{corollary}

	\subsection{Primitivity Criterion for the Multiple Decompositions Case}
	\label{Ch:PrimitivityAnalysis:Sec:PrimitivityMultipleDecomposition}
	In our discussion so far, we assumed that there has been defined only one decomposition of the underlying space. However, clearly one can decompose the underlying space in more than one ways and incorporate these decompositions into the model simply by introducing new inter-level proximity matrices $\mathbf{M_1,M_2,\dots,\mathbf{M_S}}$ and associated parameters $\mu_1,\mu_2,\dots,\mu_S$. In this case, the final stochastic matrix $\mathbf{P}$, will be given by
	\begin{equation}
	\label{eq:GeneralRankingMatrix}
	\mathbf{P} = \eta \mathbf{H} + \mu_1 \mathbf{M_1} + \cdots +  \mu_S \mathbf{M_S} + (1-\eta - \sum_i \mu_i) \mathbf{E} ,
	\end{equation}
	with $\eta,\mu_1,\mu_2,\dots,\mu_S>0$ such that $\eta+\sum_i\mu_i \leq 1$. 	Besides the possible qualitative benefits that come from this straightforward generalization, here we  will prove that multiple inter-level proximity matrices make it possible to achieve primitivity even when each of the inter-level components are incapable of doing so by themselves. In particular, we get the following theorem:

	\begin{theorem}[Primitivity Criterion for the Multiple Decompositions Case]
		Matrix  $\mathbf{P} =  \eta \mathbf{H} + \mu_1 \mathbf{M_1} + \cdots +  \mu_S \mathbf{M_S}$, with $\eta$ and $\mu_i$ positive real numbers such that $\eta + \sum_i \mu_i = 1$, is primitive if and only if the matrix 	\index{Indicator Matrix!Multiple Decompositions Case}
		\begin{displaymath}
		\mathbf{W'} = \begin{pmatrix}
		\mathbf{A_1} \\ \mathbf{A_2} \\ \vdots \\ \mathbf{A_S}
		\end{pmatrix}\begin{pmatrix}
		\mathbf{R_1} & \mathbf{R_2} & \cdots & \mathbf{R_S}
		\end{pmatrix}
		\end{displaymath} is irreducible. Concretely, $\mathbf{P} \in \mathfrak{P} \iff \mathbf{W'} \in \mathfrak{I}$.
		\label{thm:PrimitivityMultipleDecomposition}
	\end{theorem}	
	\begin{proof}
		Notice that matrix $\mathbf{P}$ can be written as follows:
		\begin{eqnarray}
		\mathbf{P}  & = &  \eta \mathbf{H} + \mu_1 \mathbf{M_1} + \cdots +  \mu_S \mathbf{M_S} \nonumber \\
		& = &  \eta \mathbf{H} + \mu \left(\dfrac{\mu_1}{\mu} \mathbf{M_1} + \cdots +  \dfrac{\mu_S}{\mu} \mathbf{M_S}\right) \nonumber \\
		& = &  \eta \mathbf{H} + \mu \left(\dfrac{\mu_1}{\mu} \mathbf{R_1A_1} + \cdots +  \dfrac{\mu_S}{\mu} \mathbf{R_SA_S}\right) \nonumber \\
		& = & \eta\mathbf{H} + \mu \begin{pmatrix}\frac{\mu_1}{\mu}
		\mathbf{R_1} & \frac{\mu_2}{\mu}\mathbf{R_2} & \cdots & \frac{\mu_S}{\mu}\mathbf{R_S}
		\end{pmatrix}\begin{pmatrix}
		\mathbf{A_1} \\ \mathbf{A_2} \\ \vdots \\ \mathbf{A_S}
		\end{pmatrix} \nonumber \\
		& = &  \eta \mathbf{H} + \mu \mathbf{R'A'},
		\end{eqnarray}
		with $\mu=\sum_{i=1}^{S}\mu_i$.  Hence, we see that the problem of determining the primitivity of matrix $\mathbf{P}$ in case of multiple decompositions is equivalent to the problem of determining the primitivity of an NCDawareRank model admitting a single decomposition with overlapping blocks, and with an inter-level proximity matrix defined such that its rows propagate their importance according to an appropriately weighted distribution to the blocks (the rows of matrix $\mathbf{R'}$) and then uniformly to their included nodes (the rows of matrix $\mathbf{A'}$). 	Therefore, Theorem~\ref{thm:PrimitivitySinleDecomposition} applies with the indicator matrix in this case being 
		\begin{equation}
		\begin{pmatrix}
		\mathbf{A_1} \\ \mathbf{A_2} \\ \vdots \\ \mathbf{A_S}
		\end{pmatrix}\begin{pmatrix}
		\frac{\mu_1}{\mu}\mathbf{R_1} & \frac{\mu_2}{\mu}\mathbf{R_2} & \cdots & \frac{\mu_S}{\mu}\mathbf{R_S}
		\end{pmatrix}.
		\end{equation}  
		Furthermore, since we are interested only in the irreducibility of this matrix we can safely ignore the scalar multiplications in the formation of the second factor and form the indicator 
		\begin{equation}
		\mathbf{W'}  \triangleq \begin{pmatrix}
		\mathbf{A_1} \\ \mathbf{A_2} \\ \vdots \\ \mathbf{A_S}
		\end{pmatrix}\begin{pmatrix}
		\mathbf{R_1} & \mathbf{R_2} & \cdots & \mathbf{R_S}
		\end{pmatrix} .
		%		 		& = & 
		\end{equation}
		Indeed, the particular values of the strictly positive scalars $\frac{\mu_1}{\mu},\frac{\mu_2}{\mu}, \dots, \frac{\mu_S}{\mu}$ have no effect to the position of the zero values in the corresponding inter-level proximity matrices, which is the only thing that can affect the primitivity of the final stochastic matrix $\mathbf{P}$~\cite{meyer2000matrix,seneta2006non}. In conclusion we have,  
		\begin{align}
		\mathbf{W'} \in \mathfrak{I}  & \iff 	\begin{pmatrix}
		\mathbf{A_1} \\ \mathbf{A_2} \\ \vdots \\ \mathbf{A_S}
		\end{pmatrix}\begin{pmatrix}
		\frac{\mu_1}{\mu}\mathbf{R_1} & \frac{\mu_2}{\mu}\mathbf{R_2} & \cdots & \frac{\mu_S}{\mu}\mathbf{R_S}
		\end{pmatrix} \in \mathfrak{I} \nonumber \\
		& \iff   \mathbf{P} \in \mathfrak{P}, \nonumber
		\end{align}
		which completes our proof.
	\end{proof}	
	
	The criterion of Theorem~\ref{thm:PrimitivityMultipleDecomposition} requires testing the irreducibility of matrix $\mathbf{W'}$, which is a square matrix of order $(K_1+K_2+\cdots+K_S)$, where $K_I$ denotes the number of blocks in the decomposition that defines the inter-level proximity matrix $\mathbf{M_I}$. Although for reasonable decompositions $K_I\ll n$ holds for every $I$, and therefore the above criterion can be  tested very efficiently, here we show that in case of multiple decompositions, the primitivity of matrix $\mathbf{P}$ is implied by certain properties of the factor matrices that are even easier to check. The following theorem gives two sufficient conditions for primitivity.

	\begin{theorem}[Sufficient Conditions for Primitivity for the Multiple Decompositions Case]
		If at least one of the following is true:
		\begin{enumerate}
			\item there exists an $I$ such that the indicator matrix $\mathbf{W_I} \triangleq \mathbf{A_IR_I}$ is irreducible, 
			\item there exist $1 \leq I \neq J \leq S$ such that $\mathbf{A_IR_J}>\mathbf{0},$
		\end{enumerate}
		then matrix  
		\begin{displaymath}
		\mathbf{P} = \eta \mathbf{H} + \mu_1 \mathbf{M_1} + \cdots +  \mu_S \mathbf{M_S},
		\end{displaymath} with $\eta$ and $\mu_i$ positive real numbers such that $\eta + \sum\mu_i = 1$, is primitive. 
		\label{Theorem:PrimitivityConditions}
	\end{theorem}
	\begin{proof}
		We will prove that if either one of the conditions (i) or (ii) hold, then $\mathbf{P}\in\mathfrak{P}$. Let us start with condition (i):
		\begin{lemma} The irreducibility of any matrix $\mathbf{W_I}, 1\leq I \leq S$ implies the primitivity of $\mathbf{P}$. Concretely, $\mathbf{W_I}\in \mathfrak{I} \implies \mathbf{P} \in \mathfrak{P}$. 
		\end{lemma}	
		\begin{proof}
			First remember that, as we argued in the proof of Theorem~\ref{thm:PrimitivitySinleDecomposition}, every indicator matrix $\mathbf{W_I}$ has by definition positive diagonal elements and thus, whenever matrix $\mathbf{W_I}$ is irreducible it will also be primitive. Then, from Lemma~\ref{lemma:W2M} it immediately follows that $\mathbf{M_I}$ will also be primitive. Therefore, using Lemma~\ref{lemma:ConvexPrimitivity}, we get the primitivity of matrix $\mathbf{P}$ which can be expressed as a linear combination of primitive and stochastic matrices, and our proof is complete.
		\end{proof}	
		
		Let us now consider the case of condition (ii). 
		
		\begin{lemma} The positivity of any matrix $\mathbf{A_IR_J}, 1\leq I,J \leq S$ implies the primitivity of the final stochastic matrix $\mathbf{P}$. Concretely,
			$\mathbf{A_IR_J}>\mathbf{0} \implies \mathbf{P} \in \mathfrak{P}$.
		\end{lemma}	
		
		\begin{proof}
			Consider the matrix $\left(\mu_1 \mathbf{M_1} + \cdots +  \mu_S \mathbf{M_S}\right)^S$. Its multinomial expansion gives:
			\begin{align}
			& \left(\mu_1 \mathbf{M_1} + \cdots +  \mu_S \mathbf{M_S}\right)^S 
			 =  \nonumber \\
			& = \sum_{i=1}^{S!}\left(\prod_{j=1}^{S} C \mathbf{M}_{\sigma_i(j)}\right) + \cdots \nonumber \\
			& =  \sum_{i=1}^{S!}\left(\prod_{j=1}^{S} C \mathbf{R}_{\sigma_i(j)}\mathbf{A}_{\sigma_i(j)}\right) + \cdots \nonumber \\
			& =  \sum_{i=1}^{S!}C \left( \mathbf{R}_{\sigma_i(1)}\mathbf{A}_{\sigma_i(1)}
			\cdots
			\mathbf{R}_{\sigma_i(S)}\mathbf{A}_{\sigma_i(S)}\right) + \cdots , \nonumber
			\end{align} 
			where $C$ is the constant $\prod \mu_i$ and $\sigma_i$ a permutation of the set of integers $[1,\dots,S]$. 
			
			Let us consider one of the $(S-1)!$ permutations that contain the product $\mathbf{A_IR_J}$. We will denote this permutation $\sigma_{i^\star}$. By definition our factor matrices are allowable, i.e. for every valid decomposition, matrices $\mathbf{A}$ and $\mathbf{R}$ are both row- and column-allowable. Therefore, since by assumption $\mathbf{A_IR_J}>\mathbf{0}$, and this matrix is multiplied only by column-allowable matrices from the right, and by row-allowable matrices from the left, from successive applications of Lemma~\ref{lemma:allowability} we get that the final matrix
			\begin{displaymath}
			\mathbf{R}_{\sigma_{i^\star}(1)}\mathbf{A}_{\sigma_{i^\star}(1)}\mathbf{R}_{\sigma_{i^\star}(2)}\mathbf{A}_{\sigma_{i^\star}(2)}\cdots\mathbf{R}_{\sigma_{i^\star}(S)}\mathbf{A}_{\sigma_{i^\star}(S)},
			\end{displaymath}
			will be strictly positive, and therefore matrix 
			\begin{displaymath}
			\left(\mu_1 \mathbf{M_1} + \cdots +  \mu_S \mathbf{M_S}\right)^S
			\end{displaymath}
			will be positive too, as a sum of a positive and non-negative matrices. This means that matrix $\mu_1 \mathbf{M_1} + \cdots +  \mu_S \mathbf{M_S}$ is primitive, and from lemma~\ref{lemma:ConvexPrimitivity} it follows that $\mathbf{P}$ is also primitive.
		\end{proof}	
		
		In conclusion we have showed that if either one of the conditions (i) and (ii) is true, matrix $\mathbf{P}$ is primitive.
	\end{proof}
	
	For an illustrative example of the primitivity critetia presented above, see also Appendix~\ref{App:ExampleInter-Level}.

	\section{Computing NCDawareRank: Algorithm and Theoretical Analysis}
	\label{Ch4:NCDawareRank:Sec:Algorithm}
	 
	Let us now proceed to the computation of the NCDawareRank vector. Of course in the literature there have been proposed a plethora of methods for the computation of the steady state distribution of a Markov chain~\cite{stewart:1994introduction}. Furthermore, because of the similarity in the definitions between PageRank and NCDawareRank, every algorithm or approach proposed for PageRank could be applied to the calculation of NCDawareRank as well\footnote{Notice that the general NCDawareRank model can easily be brought in a form similar to PageRank (we will use this in the proof of Theorem~\ref{thm:convergence}).}. In fact, even \textit{Power Method} which is one of the most simple---and for good reasons frequently used~\cite{LangvilleMeyer06}---approaches for computing the PageRank vector, can be used successfully for the computation of NCDawareRank. Indeed,  the proposed factorization of matrix $\mathbf{M}$ ensures that the final stochastic matrix can be expressed as a sum of sparse and low-rank components, which makes the use of matrix-free approaches as the Power Method, particularly well-suited for large-scale applications.   
		
		Here, however, we propose an approach that takes advantage of the particular structural properties of our model in order to compute the final ranking vector fundamentally faster, in the general case. Our approach is motivated by the observation that under reasonable decompositions the NCDaware dangling strategy respects the structure of the network and avoids artificially connecting parts of the network that are in reality disconnected (see also the related discussion in Appendix~\ref{SI:dangling}). We will see that this is a very interesting property that can lead to mathematically elegant implications that shed new light to the potential of heterogeneous handling of the dangling nodes in the classic PageRank model as well. 

In the following section we will discuss the conditions under which the Markov chain that corresponds to NCDawareRank enjoys the properties of \textit{nearly complete decomposability} and \textit{lumpability} with respect to the same coarse-level decomposition.

		\subsection{Block-Level Decomposable NCDawareRank Models}
		\label{Ch4:NCDawareRank:SubSec:AggregableTheory}
		
		\begin{definition}[Block-Level Separability]
			When there exists a partition of the NCD blocks, 
			\begin{math}
			\{\mathcal{D}_1,\mathcal{D}_2,\dotsc,
			\mathcal{D}_K\},
			\end{math}
			into $L$ super-blocks called \textbf{Aggregates},
			\begin{equation}
			\mathcal{B} \triangleq \{\mathcal{B}_1,\mathcal{B}_2,\dotsc,
			\mathcal{B}_L\},
			\end{equation}
			such that there exist no pair of nodes $u\in\mathcal{B}_{I},v\in\mathcal{B}_{J}, I\neq J$ for which $[\mathbf{H}]_{uv}>0$ holds, the corresponding NCDawareRank model is called \textbf{Block-Level Separable} with respect to partition $\mathcal{B}$. 
		\end{definition}

		\begin{theorem}[Decomposability of the NCDawareRank Chain]
		For every, block-level separable NCDawareRank model with respect to a partition $\mathcal{B}$, when the value of the teleportation probability $(1-\eta-\mu)$ is small enough, 	the Markov chain that corresponds to matrix $\mathbf{P} = \eta\mathbf{H}+\mu\mathbf{M}+(1-\eta-\mu)\mathbf{E}$, will be Nearly Completely Decomposable subject to the partition of the nodes of the initial network, into the $L$  aggregates $\{\mathcal{B}_1,\mathcal{B}_2,\dotsc,\mathcal{B}_L\}$. 
			\label{thm:NCDawareRankNCD}
		\end{theorem}
		\begin{proof}
			The crucial observation is that when the model is block-level separable, nodes belonging to different aggregates are communicating in the final Markov chain only through the rank-one teleportation matrix $\mathbf{E}$. 	Now taking into account the fact that the teleportation probability is typically chosen to be small, the final stochastic matrix will be nearly completely decomposable with respect to partition $\mathcal{B}$.  
			
			Concretely, let us assume that the rows and columns of matrix $\mathbf{P}$, are organized such that, nodes within the same aggregate occupy consecutive rows and columns in $\mathbf{P}$. It suffices to show that the maximum degree of coupling $\varepsilon$ with respect to the proposed partition, will be upper bounded by $(1-\eta-\mu)$ (see Section \ref{Subsec:NCDMarkovChains} for details).
			By definition we have  
			\begin{displaymath}
			\varepsilon = \max_{m_I}\left(\sum_{J\ne I}\sum_{l=1}^{n(J)}P_{m_Il_J}\right), 
			\end{displaymath}
			with the RHS denoting the maximum probability with which the random surfer leaves a set $\mathcal{B}_{I}$ for another. Of course, when the model is block-level separable, this can only happen through the teleportation matrix, which by definition is followed by the random surfer with probability $1-\eta-\mu$.
			
			\noindent Therefore, we have 
			\begin{eqnarray}
			\varepsilon & = & \max_{m_I}\left(\sum_{J\ne I}\sum_{l=1}^{n(J)}P_{m_Il_J}\right) \nonumber \\
			& = & \max_{m_I}\left(\sum_{J\ne I}\sum_{l=1}^{n(J)}(1-\eta-\mu)E_{m_Il_J}\right) \nonumber \\ 
			& = & (1-\eta-\mu)\max_{m_I}\left(\sum_{J\ne I}\sum_{l=1}^{n(J)}E_{m_Il_J}\right) \nonumber \\
			& \leq & (1-\eta-\mu)   \lVert \mathbf{E} \rVert_\infty \nonumber \\
			& = & (1-\eta-\mu),
			\end{eqnarray}
			which means that the maximum value degree of coupling between the $L$ aggregates will always be upper bounded by $1-\eta-\mu$. %(see Fig.~\ref{example} for a small example).
			Therefore, for small enough values of $1-\eta-\mu$, the maximum degree of coupling will be small and the corresponding Markov chain will be nearly completely decomposable, which completes the proof. 
		\end{proof}

		Notice that this effectively makes the overall model, \textit{multilevel nearly completely decomposable}; with the decomposability of the outer-level being controlled directly by the parameters of our model, whereas the decomposability at the lower-level reflecting the topological characteristics that spontaneously occur~\cite{blockrank} in the network. 
		As a matter of fact, the ``conformed'' and symmetric way this outer-level decomposability manifests itself, implies another property that is particularly useful when combined with decomposability; the property of \textit{lumpability}. Before we proceed further, we briefly outline the definition of Lumpable Markov chains.%In the following theorem we state this  we prove this rigorously.
		
		\subsubsection{Lumpability}
		\index{Lumpability!Definition|(}
		Let $\mathbf{P}$  be a transition matrix of a first-order homogeneous Markov chain, and initial vector ${\boldsymbol{\pi}}_{(0)}^{\intercal}$. Let
		\begin{displaymath}
		\mathcal{A}=\{\mathcal{A}_1,\mathcal{A}_2,\ldots ,\mathcal{A}_R\}
		\end{displaymath}  
		be a partition of the set of states. Each subset ${\mathcal{A}_I},I=1,\ldots ,R$  can be considered a state of a new process. If we use $S_t$, to denote the state occupied by this new process at time $t$, the probability of a transition occurring at time $t$ from state $\mathcal{A}_I$  to $\mathcal{A}_J$, subject to the initial distribution being ${\boldsymbol{\pi}}_{(0)}^{\intercal}$, can be denoted
		\begin{equation} 
		P_{\mathcal{A}_I\mathcal{A}_J}(t)=\Pr \{S_t=\mathcal{A}_J|S_{t-1}=\mathcal{A}_I\wedge \cdots \wedge S_0=\mathcal{A}_M\} .
		\label{def:Ch4:Lumped}
		\end{equation}
		The new process is called a \textit{lumped process}~\cite{kemeny1983finite}. Notice that the above probability in the general case depends on the choice of the initial state.
		\begin{definition}[Lumpable Markov Chain]
			A Markov chain is called lumpable with respect to a partition $$\mathcal{A}=\{\mathcal{A}_1,\mathcal{A}_2,\ldots ,\mathcal{A}_R\},$$  if for every starting vector ${\boldsymbol{\pi}}_{(0)}^{\intercal}$ the lumped process defined by~(\ref{def:Ch4:Lumped}) is a first-order homogeneous Markov chain which does not depend on the choice of the initial state.
		\end{definition}
		
		We are now ready to prove the following theorem.

		\begin{theorem}[Lumpability of the NCDawareRank Chain]
			In every  block-level separable NCDawareRank model 	with respect to partition $\mathcal{B}$, the corresponding Markov Chain is lumpable with respect to the same partition $\mathcal{B}$.
		\end{theorem}
		
		\begin{proof}
			It suffices to show that the probability of moving from a state $i\in\mathcal{B}_k$ to the set $\mathcal{B}_\ell$, i.e. 	
			\begin{equation}
			\Pr\{i\rightarrow\mathcal{B}_\ell\}=\sum\limits_{j\in \mathcal{B}_\ell}{P_{ij}}
			\label{lumpProof:1stPart}
			\end{equation}
			has the same value, for every $i\in\mathcal{B}_k$, and that this holds for every $k, \ell$ in $[1,\dots,L]$ (see Kemeny and Snell~\cite{kemeny1983finite} for a proof). For $\ell\neq k$ we have: 
			\begin{eqnarray}
			& & \Pr\{i\rightarrow\mathcal{B}_\ell\} = \nonumber\\
			& & = \sum\limits_{j\in \mathcal{B}_\ell}{P_{ij}} \nonumber\\
			& & =  \sum\limits_{j\in \mathcal{B}_\ell}\left({\eta H_{ij} + \mu M_{ij} + (1-\eta -\mu)E_{ij}}\right) \nonumber\\
			& & =  \cancelto{0}{\eta \sum\limits_{j\in \mathcal{B}_\ell}H_{ij}} + \cancelto{0}{\mu \sum\limits_{j\in \mathcal{B}_\ell}M_{ij}} + (1-\eta-\mu) \sum\limits_{j\in \mathcal{B}_\ell}E_{ij} \label{Ch:Dangling:Eq:lump1}\\
			& & =  (1-\eta-\mu) \sum_{j \in \mathcal{B}_\ell} v_j, \qquad \text{for all }i\in\mathcal{B}_k, 
			\label{eq:LumpedMatrixOffDiagonal}
			\end{eqnarray}
			with the cancellation of the first two terms in~(\ref{Ch:Dangling:Eq:lump1}), coming directly from the definition of block-level separability together with the definition of proximal sets (see Section \ref{Ch4:NCDawareRank:Sec:Model}).   For $\ell=k$ we have: 
			\begin{eqnarray}
			& & \Pr\{i\rightarrow\mathcal{B}_k\} = \nonumber\\
			& & = \sum\limits_{j\in \mathcal{B}_k}{P_{ij}} \nonumber\\
			& & =  \sum\limits_{j\in \mathcal{B}_k}\left({\eta H_{ij} + \mu M_{ij} + (1-\eta -\mu)E_{ij}}\right) \nonumber\\
			& & =  \cancelto{\eta}{\eta \sum\limits_{j\in \mathcal{B}_k}H_{ij}} + \cancelto{\mu}{\mu \sum\limits_{j\in \mathcal{B}_k}M_{ij}} + (1-\eta-\mu) \sum\limits_{j\in \mathcal{B}_k}E_{ij} \nonumber\\
			& & =  \eta + \mu + (1-\eta-\mu) \sum_{j \in \mathcal{B}_k} v_j, \quad \text{for all }i\in\mathcal{B}_k . 
			\label{eq:LumpedMatrixDiagonal}
			\end{eqnarray}
			Thus, the criterion of lumpability is verified, and the proof is complete. 
		\end{proof}

		\begin{remark}
				Notice that the block-level decomposability of an NCDawareRank model applied to realistic networks is not restrictive. Especially if the NCD blocks correspond to nodes of the network forming weakly connected subgraphs. Then, using the NCDaware handling strategy we proposed earlier (or, in fact, any other strategy chosen such that the support of the patching distribution of each dangling node includes only  nodes of the same weakly connected component,) results in a Block-Level Separable NCDawareRank model with respect to the partition of the network into different weakly connected components. 
		\end{remark}
		
		\bigskip

	In the sections to follow we will show that the particular structure and the symmetries of the NCD Markov chain that corresponds to a block-level separable NCDawareRank model, enables a useful analysis of our model into structurally identical submodels which correspond to the block diagonal submatrices of $\mathbf{P}$, that can be studied in complete isolation and solved in parallel. In what follows we will consider the general case where the model is block-level separable into $L$ aggregates. However, the algorithm we propose  trivially covers the case where $L=1$, which is the expected outcome of using a strongly preferential dangling node patching approach.

			\subsection{Exploiting the Block-Level Decomposability}		
			\label{Ch4:NCDawareRank:Sec:AggregableAlgorithm}
			Generally, the decomposability of a system into nearly uncoupled subsystems, gives us the theoretical grounds to study each subsystem in isolation and then to bring together the independent solutions in order to get a good approximation of the overall system’s behavior (see Appendix~\ref{Ch2:Preliminaries}).  As before, we assume that the rows and columns of matrix $\mathbf{P}$ are  arranged such that nodes belonging to the same aggregate $\mathcal{B}_I$ are together. The first thing we have to do is to define rigorously the exact way the strictly substochastic diagonal block matrices of $\mathbf{P}$ will be made stochastic. In particular, the off block-diagonal elements of  each row of $\mathbf{P}$, have to be added to each row of the diagonal blocks, in order to transform them to stochastic matrices. This can be done in several ways, and it is known to have an effect to the degree of the approximation one gets by analyzing each block separately~\cite{stewart:1994introduction} (see our discussion in Appendix~\ref{Ch2:Preliminaries}). 		
			
			Our approach here is based on the theory of \textit{Stochastic Complementation}~\cite{Meyer:1989:SCU:75568.75571,meyer2000matrix}, which can provide \textit{exact} results at the cost of a  computationally expensive construction of the appropriate stochastic matrices for the subsystems. While in the general case, such approach is known to be more costly than
%			offer no computational advantages with respect to 
			other decompositional methods like Iterative Aggregation/Disaggregation Algorithms~\cite{IADHaviv,KMS,IADConvergence,IADMultiLevelRAMA,stewart:1994introduction}\index{Iterative Aggregation/Disaggregation}, in the sections to follow we will show that for our particular case, it is not. And this is because \textit{we can express analytically the stochastic matrices of the subsystems in terms of smaller NCDawareRank models applied to the corresponding subgraphs, and also predict \textit{a priori} the solution of the coupling matrix, eliminating the need to (implicitly or explicitly) form it and compute its stationary distribution.} For a brief overview of Meyer’s stochastic complementation the reader can see our discussion in Appendix~\ref{subsec:PreliminariesStochasticComplementation} (for detailed treatment the interested reader is referred to~\cite{Meyer:1989:SCU:75568.75571,stewart:1994introduction}).
			
			\subsection{Stochastic Complementation of the Aggregates}
			\label{subsec:NCDawareRankComplementation}
			Let us consider the matrix $\mathbf{P}$ arranged so that nodes corresponding to the same aggregate are in consecutive rows and columns:
			\begin{equation}
			\mathbf{P} = \begin{pmatrix}
			\mathbf{P_{11}} & \mathbf{P_{12}} & \dots & \mathbf{P_{1L}} \\
			\mathbf{P_{21}} & \mathbf{P_{22}} & \dots & \mathbf{P_{2L}} \\
			\vdots	& \vdots & \ddots & \vdots \\
			\mathbf{P_{L1}} & \mathbf{P_{L2}} & \dots & \mathbf{P_{LL}} 
			\end{pmatrix}.
			\label{eq:GeneralPwithCC}
			\end{equation}
			We use $\mathbf{P_{\star i}}$ to denote the $i^{\textit{th}}$ column of blocks from which $\mathbf{P_{ii}}$ is excluded, and $\mathbf{P_{i \star}}$ to denote the $i^{\textit{th}}$  row of blocks from which $\mathbf{P_{ii}}$ is excluded. Furthermore, we use $\mathbf{P^\star_i}$ to denote the principal block submatrix of $\mathbf{P}$ obtained by deleting the $i^{\textit{th}}$ row and $i^{\textit{th}}$ column of blocks from $\mathbf{P}$. The stochastic complements $\mathbf{S_i}$ of $\mathbf{P}$ are equal to 
			\begin{eqnarray}
			\mathbf{S_i} \triangleq \mathbf{P_{ii}} + \mathbf{P_{i\star}}(\mathbf{I} - \mathbf{P^\star_i})^{-1}\mathbf{P_{\star i}}.	
			\end{eqnarray}
			Finally, the coupling matrix with respect to the decomposition $\{\mathcal{B}_1,\mathcal{B}_2,\dots,\mathcal{B}_L\}$ is given by
			\begin{equation}
			\mathbf{C} \triangleq \begin{pmatrix}
			\mathbf{s^\intercal_1}\mathbf{P_{11}}\mathbf{1} & \mathbf{s^\intercal_1}\mathbf{P_{12}}\mathbf{1} & \dots & \mathbf{s^\intercal_1}\mathbf{P_{1L}}\mathbf{1} \\[0.3em]
			\mathbf{s^\intercal_2}\mathbf{P_{21}}\mathbf{1} & \mathbf{s^\intercal_2}\mathbf{P_{22}}\mathbf{1} & \dots & \mathbf{s^\intercal_2}\mathbf{P_{2L}}\mathbf{1} \\
			\vdots	& \vdots & \ddots & \vdots \\
			\mathbf{s^\intercal_L}\mathbf{P_{L1}}\mathbf{1} & \mathbf{s^\intercal_L}\mathbf{P_{L2}}\mathbf{1} & \dots & \mathbf{s^\intercal_L}\mathbf{P_{LL}} \mathbf{1}
			\end{pmatrix},
			\end{equation}
			where $\mathbf{s^\intercal_i}$ is the stationary distribution of the stochastic complement $\mathbf{S_i}$. Let $\mathbf{\xi}^\intercal$, be the stationary distribution of the Markov chain with transition probability matrix $\mathbf{C}$. Then, from the disaggregation Theorem~\ref{Theorem:Dissagregation} (see Appendix~\ref{Ch2:Preliminaries}) the stationary distribution ${\boldsymbol{\pi}}^\intercal$ of $\mathbf{P}$ is given by
			\begin{equation}
			{\boldsymbol{\pi}}^\intercal = \begin{pmatrix}
			\xi_1 \mathbf{s^\intercal_1} & \xi_2 \mathbf{s^\intercal_2} & \cdots & \xi_L \mathbf{s^\intercal_L}
			\end{pmatrix}.
			\end{equation}

			While in the general case the computation of the stochastic complements $\mathbf{S_i}$ is a computationally intensive task~\cite{meyer2000matrix,meyer2012stochastic}---which makes the above approach for computing the stationary distribution $\boldsymbol{\pi}$ impractical---in our case we can express them directly as isolated NCDawareRank submodels. We prove this in the following theorem:  
			\begin{theorem}[Stochastic Complements as NCDawareRank Submodels]
				Each stochastic complement $\mathbf{S_i}$ coincides with the final matrix of a smaller NCDawareRank model, with the same parameters $\eta,\mu$, applied to the aggregate $\mathcal{B}_i$, and  using as teleportation vector the normalized version of the corresponding subvector of $\mathbf{v}$.
				\label{thm:stochasticcomplements}
			\end{theorem}
			\begin{proof}
				Let us consider a block-level separable NCDawareRank model with final matrix
				\begin{displaymath}
				\mathbf{P} = \eta\mathbf{H}+\mu\mathbf{M}+(1-\eta-\mu)\mathbf{E},
				\end{displaymath}
				organized as in equation~(\ref{eq:GeneralPwithCC}). Its $i^{\textit{th}}$ stochastic complement is given by
				\begin{displaymath}
				\mathbf{S_i} = \mathbf{P_{ii}} + \mathbf{P_{i\star}}(\mathbf{I} - \mathbf{P^\star_i})^{-1}\mathbf{P_{\star i}}.	
				\end{displaymath}
				By the definition of the property of block-level separability, together with the definition of the NCD proximal sets we get,
				\begin{eqnarray}
				\mathbf{P_{ii}} & = & \eta\mathbf{H_{ii}} + \mu\mathbf{M_{ii}} + (1-\eta-\mu)\mathbf{1}\mathbf{v^\intercal_i} ,\\
				\mathbf{P_{i\star}} & = & (1-\eta-\mu)\mathbf{1}\mathbf{v^\intercal_{i\star}}, \\
				\mathbf{P_{\star i}} & = & (1-\eta-\mu)\mathbf{1}\mathbf{v^\intercal_{i}},
				\end{eqnarray}
				where we consider the teleportation vector $\mathbf{v}$ organized and partitioned according to matrix $\mathbf{P}$:
				\begin{displaymath}
					\mathbf{v^\intercal} = \begin{pmatrix}
					\mathbf{v_1^\intercal} & \mathbf{v_2^\intercal} & \cdots & \mathbf{v_L^\intercal}
					\end{pmatrix} 
				\end{displaymath} and we use $\mathbf{v_{i\star}}$ to denote the vector arising from $\mathbf{v}$, after deleting $\mathbf{v_i}$:
				\begin{displaymath}
				\mathbf{v_{i\star}^\intercal} = \begin{pmatrix}
				\mathbf{v_1^\intercal} & \cdots & \mathbf{v_{i-1}^\intercal} & \mathbf{v_{i+1}^\intercal} & \cdots &  \mathbf{v_L^\intercal}
				\end{pmatrix} .
				\end{displaymath}
				From the stochasticity of $\mathbf{P}$ we get
				\begin{eqnarray}
				(\mathbf{P_{\star i}}+\mathbf{P^\star_{i}})\mathbf{1}  &=& \mathbf{1} \nonumber \\ 
				\mathbf{P_{\star i}}\mathbf{1}+\mathbf{P^\star_{i}}\mathbf{1}  &=& \mathbf{1} \nonumber \\  
				\mathbf{P_{\star i}}\mathbf{1} & = & (\mathbf{I} - \mathbf{P^\star_{i}})\mathbf{1} \nonumber \\
				\Rightarrow(\mathbf{I} - \mathbf{P^\star_{i}})^{-1}
				\mathbf{P_{\star i}}\mathbf{1} & = & \mathbf{1}	\nonumber \\
				\Rightarrow(\mathbf{I} - \mathbf{P^\star_{i}})^{-1}
				(1-\eta-\mu)\mathbf{1}\mathbf{v^\intercal_{i}}\mathbf{1} & = & \mathbf{1}, \nonumber 
				\end{eqnarray}
				and if we take into account that in a well-defined NCDawareRank (or PageRank) model, $\mathbf{v_i}\neq\mathbf{0}$ holds (otherwise, the final stochastic matrix would be reducible), we can finally get
				\begin{eqnarray}
				(\mathbf{I} - \mathbf{P^\star_{i}})^{-1}
				\mathbf{1} & = & \dfrac{1}{\mathbf{v^\intercal_{i}}\mathbf{1}(1-\eta-\mu)}\mathbf{1}.
				\end{eqnarray}
				Thus, returning to the stochastic complement $\mathbf{S_i}$ and substituting we get
				\begin{eqnarray}
				\mathbf{S_i} &=& \mathbf{P_{ii}} + \mathbf{P_{i\star}}(\mathbf{I} - \mathbf{P^\star_i})^{-1}\mathbf{P_{\star i}} \nonumber \\
				&=& \eta\mathbf{H_{ii}} + \mu\mathbf{M_{ii}} + (1-\eta-\mu)\mathbf{1}\mathbf{v^\intercal_i} + \nonumber \\
				& & + (1-\eta-\mu)\mathbf{1}\mathbf{v^\intercal_{i\star}}(\mathbf{I} - \mathbf{P^\star_i})^{-1}(1-\eta-\mu)\mathbf{1}\mathbf{v^\intercal_{i}}\nonumber \\
				&=& \eta\mathbf{H_{ii}} + \mu\mathbf{M_{ii}} + (1-\eta-\mu)\mathbf{1}\mathbf{v^\intercal_i} + \nonumber \\
				& & + (1-\eta-\mu)^2\mathbf{1}\mathbf{v^\intercal_{i\star}}\dfrac{1}{\mathbf{v^\intercal_{i}}\mathbf{1}(1-\eta-\mu)}\mathbf{1}\mathbf{v^\intercal_{i}} \nonumber \\
				&=& \eta\mathbf{H_{ii}} + \mu\mathbf{M_{ii}} + (1-\eta-\mu)\mathbf{1}\mathbf{v^\intercal_i} + \dfrac{(1-\eta-\mu)}{\mathbf{v^\intercal_{i}}\mathbf{1}}\mathbf{1}\mathbf{v^\intercal_{i\star}}\mathbf{1}\mathbf{v^\intercal_{i}} \nonumber \\
				&=& \eta\mathbf{H_{ii}} + \mu\mathbf{M_{ii}} + (1-\eta-\mu)\mathbf{1}\mathbf{v^\intercal_i} + \nonumber \\
				& & + (1-\eta-\mu)\dfrac{1-\mathbf{v^\intercal_{i}}\mathbf{1}}{\mathbf{v^\intercal_{i}}\mathbf{1}}\mathbf{1}\mathbf{v^\intercal_{i}} \nonumber \\
				&=& \eta\mathbf{H_{ii}} + \mu\mathbf{M_{ii}} +(1-\eta-\mu) \dfrac{1}{\mathbf{v^\intercal_{i}}\mathbf{1}}\mathbf{1}\mathbf{v^\intercal_{i}} \nonumber \\
				& = & \eta\mathbf{H_{ii}} + \mu\mathbf{M_{ii}} +(1-\eta-\mu)\mathbf{E_{ii}}. 
				\end{eqnarray}
				Therefore, we can see that $\mathbf{S_i}$ is the final stochastic matrix of an NCDawareRank model over the subgraph $\mathcal{B}_i$, considered in isolation:
				\begin{displaymath}
				\mathbf{S_i} = \eta\mathbf{H_{ii}} + \mu\mathbf{M_{ii}} +(1-\eta-\mu)\mathbf{E_{ii}},
				\end{displaymath}  
				and our proof is complete. 
			\end{proof}

			\subsection{Solving the NCDawareRank Submodels}

				 For simplicity we suggest using the Power Method for the computation of the steady state probability distribution. Another reasonable option would be to use a decompositional approach such as Iterative Aggregation/Disaggregation which is ideal to exploit the lower-level decomposability of submodels into the NCD blocks. However, because of the sparseness of the involving components and for simplicity of exposition, we will proceed with the Power Method, which in fact, is known to be one of the most popular approaches to compute PageRank as well. The algorithm is given below:
				 
				 \begin{algorithm}
				 	\caption{NCDawareRank for Subnets ($\mathbf{H},\mathbf{A},\mathbf{R},{\boldsymbol{\pi}}^\intercal_{(0)},\epsilon)$}
				 	\label{s-NCDawareRank}
				 	\begin{algorithmic}
				 		\State \textbf{1}. Let the initial approximation be ${\boldsymbol{\pi}}^\intercal_{(0)}$. Set $k=0$.
				 		\State \textbf{2}. Compute
				 		{\small \begin{eqnarray}
				 			{\boldsymbol{\pi}}_{(k+1)}^{\intercal}& = & {\boldsymbol{\pi}}_{(k)}^{\intercal}\mathbf{P} \nonumber \\
				 			& = & \eta {\boldsymbol{\pi}}_{(k)}^{\intercal} \mathbf{H} + \mu{\boldsymbol{\pi}}_{(k)}^{\intercal} \mathbf{RA} +(1-\eta - \mu){\boldsymbol{\pi}}_{(k)}^{\intercal} \mathbf{E}.  \nonumber
				 			\end{eqnarray}}
				 		\State \textbf{3}. Normalize $ {\boldsymbol{\pi}}_{(k+1)}^{\intercal} $ and compute  
				 		\begin{displaymath}
				 		r = \lVert {\boldsymbol{\pi}}_{(k+1)}^{\intercal} - {\boldsymbol{\pi}}_{(k)}^{\intercal}\rVert_1.
				 		\end{displaymath}
				 		\State \quad  If $ r < \epsilon$, quit with $ {\boldsymbol{\pi}}_{(k+1)}^{\intercal} $, otherwise set $k=k+1$ and go to step 2.
				 	\end{algorithmic}
				 \end{algorithm}

				 Each iteration of the Power Method involves a calculation that looks like this:
				 \begin{eqnarray}
				 \hat{{\boldsymbol{\pi}}}^\intercal& = & {\boldsymbol{\pi}}^{\intercal}\mathbf{P} \nonumber \\
				 & = & \eta \underbrace{{\boldsymbol{\pi}}^{\intercal} \mathbf{H}}_{\Omega_\mathbf{H}} + \mu\underbrace{{\boldsymbol{\pi}}^{\intercal} \mathbf{RA}}_{\Omega_\mathbf{M}} +(1-\eta - \mu)\cancelto{1}{\boldsymbol{\pi}^{\intercal}\mathbf{1}}\mathbf{v^\intercal} . \nonumber
				 \end{eqnarray}
				 The extreme sparsity of the factors of matrix $\mathbf{M}$, together with the fact that their dimension is typically orders of magnitude less than the number of nodes\footnote{This is true for even fine-grained decompositions in Web-ranking applications where the decomposition depicts site-based partitioning.}, suggest that the bottleneck of each iteration is by far the Sparse Matrix$\times$Vector (SpMV) product  of the previous estimate with the normalized adjacency matrix $\mathbf{H}$, i.e. it holds
				 \begin{displaymath}
				 \Omega_\mathbf{H} \gg \Omega_\mathbf{M},
				 \end{displaymath}
				 where we used $\Omega_\mathbf{H}$ (resp. $\Omega_\mathbf{M}$ ) to denote the number of floating point operations needed for the computation $\boldsymbol{\pi}^\intercal \mathbf{H}$ (resp. $(\boldsymbol{\pi}^\intercal \mathbf{R})\mathbf{A}$). 
				   
				Finally, note here that the products $\boldsymbol{\pi}^\intercal \mathbf{H}$ and  $(\boldsymbol{\pi}^\intercal \mathbf{R})\mathbf{A}$ of each iteration can be computed in parallel, with the introduction of very small communication cost. All these observations suggest that the computational overhead per iteration induced by the inter-level proximity matrix, with respect to PageRank is practically very small. 
				
				\textit{But what about the number of iterations till convergence? }	
				It is known that the convergence rate of the Power Method applied to a stochastic matrix depends on the magnitude of the subdominant eigenvalue, $\lvert\lambda_2\rvert$. Precisely, the asymptotic rate of convergence is the rate at which $\lvert\lambda_2(\mathbf{P})\rvert^k\rightarrow 0$. The following theorem bounds the subdominant eigenvalue of the NCDawareRank matrix $\mathbf{P}$. 
				   
				 \begin{theorem}[Subdominant Eigenvalue of the NCDawareRank Matrix]
				 	The subdominant eigenvalue of any NCDawareRank stochastic matrix $\mathbf{P}= \eta\mathbf{H}+\mu\mathbf{M}+(1-\eta-\mu)\mathbf{E}$, is upper bounded by $\eta +\mu$. 
				 	\label{thm:convergence}
				 \end{theorem}
				 \begin{proof}
				 	We define the matrix 
				 	\begin{displaymath}
				 	\mathbf{Z} \triangleq \frac{1}{\eta+\mu}(\eta\mathbf{H}+\mu\mathbf{M}).
				 	\end{displaymath}
				 	Notice that since matrices $\mathbf{H}$ and $\mathbf{M}$ are row-stochastic, matrix $\mathbf{Z}$ is also row-stochastic:

				 	\begin{eqnarray}
				 	\mathbf{Z}\mathbf{1} & = &
				 	\frac{1}{\eta+\mu}(\eta\mathbf{H}+\mu\mathbf{M})\mathbf{1} \nonumber \\
				 	& = & \frac{1}{\eta+\mu}(\eta\mathbf{H}\mathbf{1}+\mu\mathbf{M}\mathbf{1})
				 	\nonumber \\
				 	& = & \frac{1}{\eta+\mu}(\eta\mathbf{1}+\mu\mathbf{1}) = \mathbf{1}. \nonumber
				 	\end{eqnarray}
				 	Now, notice that we can express matrix $\mathbf{P}$ in terms of $\mathbf{Z}$ as follows:
				 	\begin{equation}
				 	\mathbf{P} = \alpha'\mathbf{Z}+(1-\alpha')\mathbf{E},
				 	\label{eq:NCDawareRankasPageRank}
				 	\end{equation}
				 	where $\alpha' \triangleq \eta+\mu<1$ and
				 	$\mathbf{E}=\mathbf{1}\mathbf{v}^{\intercal}$, for some probability vector with non-zero elements $\mathbf{v}^{\intercal}$. 				 	
				 	But from the celebrated Google lemma\footnote{In fact the first proof of this result can be traced back to Brauer~\cite{brauer1952}.} (for a proof see \cite{LangvilleMeyer06}) we know that if the spectrum of the stochastic matrix $\mathbf{S}$ is
				 	\begin{displaymath}
				 	\{1,\lambda_2,\dotsc,\lambda_n\},
				 	\end{displaymath}
				 	then the spectrum of matrix $\mathbf{G} = \alpha\mathbf{S}+(1-\alpha)\mathbf{1}\mathbf{v}^{\intercal}$, where $\mathbf{v}^{\intercal}$ a probability vector with positive elements, is \begin{displaymath}\{1,\alpha\lambda_2,\dotsc,\alpha\lambda_n\}.
				 	\end{displaymath}
				 	Thus, applying this result to matrix $\mathbf{P}$ expressed as in  relation~(\ref{eq:NCDawareRankasPageRank}) implies that its subdominant eigenvalue is 
				 	\begin{displaymath}
				 	\lambda_2(\mathbf{P}) = \alpha'\lambda_2(\mathbf{Z}) =(\eta+\mu)\lambda_2(\mathbf{Z}) .
				 	\end{displaymath} However, for every eigenvalue of a stochastic matrix, $\lvert\lambda\rvert \leq 1$ holds (see \cite{seneta2006non}), and the bound follows.
				 \end{proof}
				 
				  Assuming that the number of operations of a SpMV is twice the number of non-zero elements of the sparse matrix involved and if we use $\mathit{nnz}(\cdot)$ to denote the number of non-zero elements of a matrix, we get that the number of floating point operations needed to satisfy a tolerance criterion $\epsilon$ is approximately
				  \begin{displaymath}
				  \Omega = \frac{\mathrm{\Theta}(\operatorname{nnz}(\mathbf{H}))\log \epsilon}{\log|\eta+\mu|}.
				  \end{displaymath}

				 	\subsubsection{Convergence Tests}
				 	To test the effect of the introduction of the inter-level proximity matrix, we run NCDawareRank introducing increasingly big values of $\mu$ while keeping the teleportation parameter $1-\eta-\mu$ constant to 0.10; the criterion of convergence is taken to be a difference in the $L_1$-norm of successive estimations, lower than $10^{-8}$. The results are presented in Table~\ref{Table:convergence_mu}. We see that even for values of $\mu$ as low as 0.005, we have a drop of the number of iterations till convergence, and this drop continues as $\mu$ increases and then stabilizes. This is true for all the networks we experimented on. These results and the fact that we want to avoid ignoring the direct link structure, suggest that a good choice for parameter $\mu$ is a value near 0.10. 
				 	
				 	\begin{table}[h!] 
				 		\centering
				 		\caption{Iterations till Convergence for Different Values of  $\mu$}
				 		\resizebox{0.99\linewidth}{!}{ 
				 			\begin{tabular}{ccccccccccccccc}
				 				\hline  
				 				$\mu= $        & 0 & 0.005 & 0.01 &  0.05 & 0.10 & 0.15   &  0.20 & 0.25 & 0.30
				 				\\\hline
				 				\texttt{cnr-2000}       & 137  &  131  &   127    & 121 &   \textbf{122} &   122 & 122 & 122 & 122 
				 				\\                    
				 				\texttt{eu-2005}        & 129  &  125  &   123    & 120 &   \textbf{121} &   121 & 121 & 120 & 120
				 				\\                         
				 				\texttt{india-2004}     &  135   &   129 &   125    & 117 &   \textbf{117} &   117 & 117 & 117 & 117 
				 				\\          
				 				\texttt{uk-2002}        &131   &  127  &   124    & 122 &   \textbf{122} &   123 & 123 & 123 & 123 
				 				\\ \hline %      
				 			\end{tabular}
				 		}
				 		\label{Table:convergence_mu}
 					 	\end{table}

 				 	\setlength\figureheight{0.165\textwidth} 
 				 	\setlength\figurewidth{.38\textwidth} 
 				 	\begin{figure*}
							\centering
							\includegraphics[width = 0.92\textwidth]{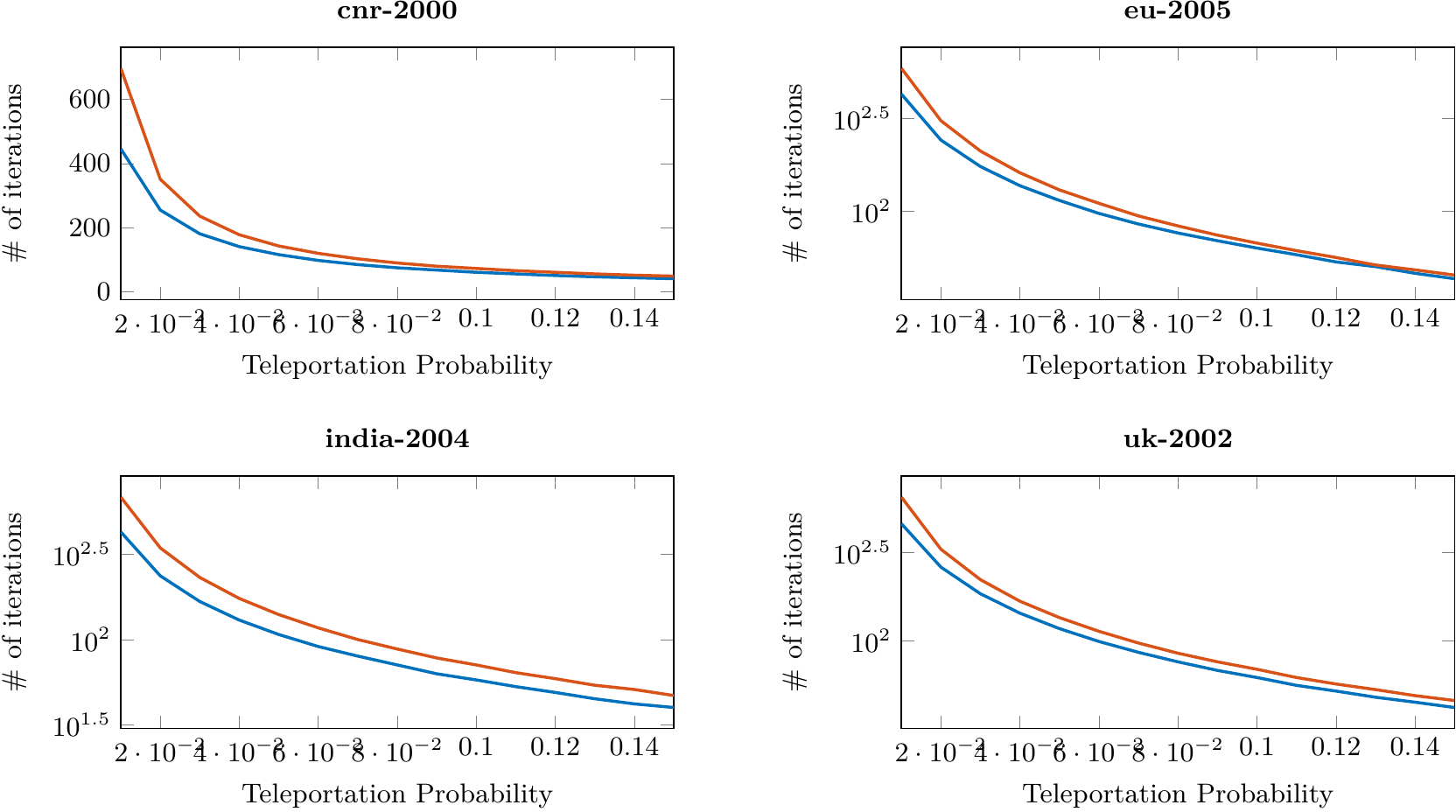}
 				 			\caption{NCDawareRank (\textcolor{blue}{$\bullet$}) vs PageRank(\textcolor{red}{$\bullet$}).  Number of iterations till convergence for different values of the teleportation probability.}
 				 			\label{fig:NCDawareRankIterations}
 				 		\end{figure*}

				 	Furthermore, we fix parameter $\mu$ to the value 0.10 and we test the convergence of our algorithm against PageRank for different values of the teleportation probability. Fig.~\ref{fig:NCDawareRankIterations} reports the results. We see that the number of iterations till convergence for our method is smaller than PageRank. Notice here, that while theoretically the convergence rate upper bound is the same (for a given teleportation probability) for both methods, the introduction of the inter-level proximity matrix makes NCDawareRank converge a bit faster. 
	 	
				 	To highlight their difference we present the percentage of the increase in iterations needed till convergence in Fig.~\ref{fig:PercentageIncrease}. We observe that PageRank generally needs more iterations to converge, with the difference reaching up to 60\% for the smallest teleportation probability tested (0.01).

			\subsection{Analytical Solution of the Coupling Matrix}
			
			The only thing that we need in order to compute the final ranking vector is the stationary distribution of the coupling matrix $\mathbf{C}$. Thankfully, a careful exploitation of the symmetries of our model, reveals that this matrix also has a convenient structure that allows us to predict its unique stationary distribution analytically.
			
			Let us consider the elements of $\mathbf{C}$. For its $C_{ii}$ element it holds
			\begin{eqnarray}
			C_{ii} & = & \mathbf{s^\intercal_i}\mathbf{P_{ii}}\mathbf{1} \nonumber \\
			& = &  \mathbf{s^\intercal_i}\left(\eta\mathbf{H_{ii}} + \mu\mathbf{M_{ii}} + (1-\eta-\mu)\mathbf{1}\mathbf{v^\intercal_i}\right)\mathbf{1} \nonumber \\
			& = & \mathbf{s^\intercal_i} \left(\eta\mathbf{H_{ii}}\mathbf{1} + \mu\mathbf{M_{ii}}\mathbf{1} + (1-\eta-\mu)\mathbf{1}\mathbf{v^\intercal_i}\mathbf{1} \right) \nonumber \\
			& = & \mathbf{s^\intercal_i} \left(\eta\mathbf{1} + \mu\mathbf{1} + (1-\eta-\mu)(\mathbf{v^\intercal_i}\mathbf{1}) \mathbf{1} \right) \nonumber \\
			& = & \mathbf{s^\intercal_i} \big(\underbrace{\eta + \mu + (1-\eta-\mu)(\mathbf{v^\intercal_i}\mathbf{1})}_{\text{scalar}}\big) \mathbf{1}  \nonumber \\
			& = & \big(\eta + \mu + (1-\eta-\mu)(\mathbf{v^\intercal_i}\mathbf{1})\big) \mathbf{s^\intercal_i} \mathbf{1}  \nonumber \\
			& = & \eta + \mu + (1-\eta-\mu)(\mathbf{v^\intercal_i}\mathbf{1}) .
			\end{eqnarray}
			For every $C_{ij}, i\neq j$, we have:
			\begin{eqnarray}
			C_{ij} & = & \mathbf{s^\intercal_i}\mathbf{P_{ij}}\mathbf{1} \nonumber \\
			& = & \mathbf{s^\intercal_i}(1-\eta-\mu)\mathbf{1}\mathbf{v^\intercal_j}\mathbf{1} \nonumber \\
			& = & (1-\eta-\mu)\mathbf{v^\intercal_j}\mathbf{1} .
			\end{eqnarray}
			The final stochastic matrix $\mathbf{C}$ is, therefore, given by
			{\normalsize \begin{eqnarray}
				\mathbf{C} &= & 
				\begin{pmatrix}
				\mathbf{s^\intercal_1}\mathbf{P_{11}}\mathbf{1} & \mathbf{s^\intercal_1}\mathbf{P_{12}}\mathbf{1} & 
				\dots & \mathbf{s^\intercal_1}\mathbf{P_{1L}}\mathbf{1} \\[0.3em]
				\mathbf{s^\intercal_2}\mathbf{P_{21}}\mathbf{1} & \mathbf{s^\intercal_2}\mathbf{P_{22}}\mathbf{1} & \dots & \mathbf{s^\intercal_2}\mathbf{P_{2L}}\mathbf{1} \\
				\vdots	& \vdots & \ddots & \vdots \\
				\mathbf{s^\intercal_L}\mathbf{P_{L1}}\mathbf{1} & \mathbf{s^\intercal_L}\mathbf{P_{L2}}\mathbf{1} & \dots & \mathbf{s^\intercal_L}\mathbf{P_{LL}} \mathbf{1}
				\end{pmatrix} \nonumber \\[.6em]
				&= & \begin{pmatrix}
				\eta + \mu + (1-\eta-\mu)(\mathbf{v^\intercal_1}\mathbf{1}) &
				%		 (1-\eta-\mu)(\mathbf{v^\intercal_2}\mathbf{1}) & 
				\dots & (1-\eta-\mu)(\mathbf{v^\intercal_L}\mathbf{1}) \\
				\vdots & 
				%		\vdots & 
				\ddots & \vdots \\
				(1-\eta-\mu)(\mathbf{v^\intercal_1}\mathbf{1}) & 
				%		 (1-\eta-\mu)(\mathbf{v^\intercal_2}\mathbf{1}) & 
				\dots & \eta + \mu + (1-\eta-\mu)(\mathbf{v^\intercal_L}\mathbf{1})
				\end{pmatrix}. \nonumber
				\end{eqnarray}}

			Notice that the elements of the final coupling matrix are independent of the solutions of the stochastic complements,  $\mathbf{s^\intercal_i}$. This follows directly by the outer-level lumpability of matrix $\mathbf{P}$. Furthermore, notice that matrix $\mathbf{C}$ can be written in the following useful form: 
			\begin{eqnarray}
			\mathbf{C}	& = & (\eta+\mu)\mathbf{I} + (1-\eta-\mu) \boldsymbol{1}\boldsymbol{\xi}^\intercal,
			\end{eqnarray}
			with $\boldsymbol{\xi}^\intercal$ defined to be equal to:
			\begin{eqnarray}
			\boldsymbol{\xi}^\intercal &\triangleq& \begin{pmatrix}
			\mathbf{v^\intercal_1}\mathbf{1} & \mathbf{v^\intercal_2}\mathbf{1} & \dots & \mathbf{v^\intercal_L}\mathbf{1}
			\end{pmatrix}  \\ 
			&=& \begin{pmatrix}
			\sum_{j \in \mathcal{B}_1} v_j & \sum_{j \in \mathcal{B}_2} v_j & \dots &  \sum_{j \in \mathcal{B}_L} v_j
			\end{pmatrix} \nonumber.
			\end{eqnarray}
			
			Clearly matrix $\mathbf{C}$ has a very special structure. This enables us to find its stationary distribution analytically, without ever needing to form the matrix and solve it computationally. The following theorem predicts this distribution and proves its uniqueness.  
			\begin{theorem}[Stationary Distribution of the Coupling Matrix]
				The unique stationary distribution of $\mathbf{C}$  is $\boldsymbol{\xi}^\intercal$:
				\begin{displaymath}
				\boldsymbol{\xi}^\intercal \triangleq \begin{pmatrix}
				\mathbf{v^\intercal_1}\mathbf{1} & \mathbf{v^\intercal_2}\mathbf{1} & \dots & \mathbf{v^\intercal_L}\mathbf{1}
				\end{pmatrix} ,
				\end{displaymath}
				where $\mathbf{v}$ is the teleportation vector of the model, used in the definition of the teleportation matrix $\mathbf{E}$.
				\label{thm:couplingsolution}
			\end{theorem}
			\begin{proof}
				We need to show that 
				\begin{eqnarray}
				\boldsymbol{\xi}^\intercal\mathbf{C} & = &\boldsymbol{\xi}^\intercal, \nonumber \\
				\boldsymbol{\xi}^\intercal\mathbf{1} & = &1, \nonumber
				\end{eqnarray} hold, and that distribution $\mathbf{\xi}$ is the unique distribution that solves the above system. 
				The verification part can be done by straightforward calculations:
				\begin{eqnarray}
				\boldsymbol{\xi}^\intercal\mathbf{1} & = & \begin{pmatrix}
				\sum_{j \in \mathcal{B}_1} v_j & \sum_{j \in \mathcal{B}_2} v_j & \dots &  \sum_{j \in \mathcal{B}_L} v_j
				\end{pmatrix}\mathbf{1} \nonumber \\
				& = & \sum_{j}v_j = 1,
				\label{eq:xi_1}
				\end{eqnarray}	
				and 
				\begin{eqnarray}
				\boldsymbol{\xi}^\intercal\mathbf{C} & = &\boldsymbol{\xi}^\intercal (\eta+\mu)\mathbf{I} + \boldsymbol{\xi}^\intercal(1-\eta-\mu) \boldsymbol{1}\boldsymbol{\xi}^\intercal \nonumber \\
				& = & (\eta+\mu)\boldsymbol{\xi}^\intercal + (1-\eta-\mu) \cancelto{1}{\boldsymbol{\xi}^\intercal\mathbf{1}} \mathbf{\xi}^\intercal \nonumber \\
				& = &\boldsymbol{\xi}^\intercal.
				\label{eq:xi_2}
				\end{eqnarray}	
				Furthermore, since the irreducibility of $\mathbf{P}$ implies the irreducibility of $\mathbf{C}$ (see~\cite{stewart:1994introduction}), the above stationary distribution is the unique stationary distribution of $\mathbf{C}$ and the proof is complete.
			\end{proof}

			\subsection{Putting Everything Together: The Complete NCDawareRank Algorithm }		
			\label{SubSec:ConfinedNCDawareRankAlgorithm}
			In light of Theorems~\ref{thm:stochasticcomplements} and~\ref{thm:couplingsolution}, the final computation of the NCDawareRank vector can be summarized as Algorithm~\ref{alg:AggregableNCDawareRank}. For an illustrative numerical example of the overall algorithm, see Appendix~\ref{App:NCDawareRankComputation}.
			\begin{algorithm}
				\caption{NCDawareRank}
				\begin{algorithmic}[1]
					\renewcommand{\algorithmicrequire}{\textbf{Input:}}
					\renewcommand{\algorithmicensure}{\textbf{Output:}}
					\Require Matrices $\mathbf{H},\mathbf{A},\mathbf{R}$ and teleportation vector $\mathbf{v}$.
					\Ensure Ranking vector ${\boldsymbol{\pi}}$.
					%\Statex
					\State Compute in parallel the stationary distributions of the stochastic complements $\mathbf{S_i}$ 
					\begin{eqnarray}
					\mathbf{s^\intercal_i} & = & \mathbf{s^\intercal_i}\mathbf{S_i} \nonumber \\
					\mathbf{s^\intercal_i} \mathbf{1} & = & 1 \nonumber
					\end{eqnarray}
					using Algorithm~\ref{s-NCDawareRank} (or any other equivalent solver) with input $\mathbf{H_{ii}},\mathbf{M_{ii}}$ (i.e. the corresponding rows of $\mathbf{R}$ and columns of $\mathbf{A}$) and teleportation vector $\tfrac{1}{\mathbf{v^\intercal_i} \mathbf{1}}\mathbf{v_{i}^\intercal}$.
					\State Form the solution of the coupling matrix $\mathbf{\xi}^\intercal$, predicted by Theorem~\ref{thm:couplingsolution}:
					\begin{displaymath}
					\boldsymbol{\xi}^\intercal  = \begin{pmatrix}
					\xi_1 & \xi_2 & \dots & \xi_L
					\end{pmatrix} = \begin{pmatrix}
					\mathbf{v^\intercal_1}\mathbf{1} & \mathbf{v^\intercal_2}\mathbf{1} & \dots & \mathbf{v^\intercal_L}\mathbf{1}
					\end{pmatrix}
					\end{displaymath}
					\State Combine the solutions to create the final ranking vector ${\boldsymbol{\pi}}$:
					\begin{displaymath}
					{\boldsymbol{\pi}}^\intercal = \begin{pmatrix}
					\xi_1\mathbf{s^\intercal_1} & \xi_2 \mathbf{s^\intercal_2} & \dots & \xi_L \mathbf{s^\intercal_L}
					\end{pmatrix}
					\end{displaymath}
					\State \textbf{return} $ {\boldsymbol{\pi}}$
				\end{algorithmic}
				\label{alg:AggregableNCDawareRank}
			\end{algorithm}

	\subsection{Applying the Analysis to the Standard PageRank Model}
	 Let us note here that all the analysis presented in this section holds for the traditional PageRank model too, provided that the rows that correspond to dangling nodes are patched using a distribution that assigns positive probability only to nodes originated from the same weakly connected component of the graph. We call the class of handling strategies that satisfy this property ``\textit{Confined Dangling Strategies}.'' 
	 
	 The aggregates in this case are defined to be the weakly connected components of the network, and the stochastic complements are independent PageRank models defined on the subgraphs, as in the NCDawareRank case. The following theorem---the proof of which is completely analogous to our proofs presented earlier and therefore skipped for the sake of tighter presentation---states this formally.
	 \newpage
	 \begin{theorem}[PageRank Under Confined Dangling Strategies]
	 	Under a confined dangling strategy (or complete absence of dangling nodes), the Markov chain that corresponds to the final stochastic matrix of the PageRank model 
	 	\begin{displaymath}
	 	\mathbf{G} = \alpha\mathbf{H} + (1-\alpha)\mathbf{1}\mathbf{v}^\intercal,
	 	\end{displaymath} 
	 	satisfies the following: 
	 	\begin{itemize}
	 		\item \textbf{Lumpability.} It is lumpable with respect to the partition of the nodes into weakly connected components.
	 		\item \textbf{Decomposability.}  For small enough teleportation probability, the chain is outer-level nearly completely decomposable with respect to the same partition.
	 		\item \textbf{Stochastic Complements as PageRank Models.} The stochastic complements of the accordingly organized matrix $\mathbf{G}$ are PageRank models with the same parameter $\alpha$ and teleportation vector $\tfrac{1}{\mathbf{v^\intercal_i} \mathbf{1}}\mathbf{v_{i}^\intercal}$, defined on each weakly connected component in isolation.
	 		\item \textbf{Coupling}. The corresponding coupling matrix is independent of the stationary distributions of the stochastic complements and it is given by
	 		\begin{eqnarray}
	 		\mathbf{C}	& = & \alpha\mathbf{I} + (1-\alpha) \boldsymbol{1}\boldsymbol{\xi}^\intercal,
	 		\end{eqnarray}  
	 		with the probability vector $\boldsymbol{\xi}$, defined as above.
	 		\item \textbf{PageRank vector}.  The unique stationary distribution of the chain i.e. the PageRank vector $\boldsymbol{\pi}$ satisfies:
	 		\begin{displaymath}
	 		{\boldsymbol{\pi}}^\intercal = \begin{pmatrix}
	 		\xi_1\mathbf{s^\intercal_1} & \xi_2 \mathbf{s^\intercal_2} & \dots & \xi_L \mathbf{s^\intercal_L}
	 		\end{pmatrix}.
	 		\end{displaymath}
	 	\end{itemize}   
	 \end{theorem}  
	
    \section{Experimental Evaluation}
    \label{Sec:experiments}
    \subsection{Dataset Preparation}
		\label{Ch4:NCDawareRank:SubSec:Methodology}
		
		Throughout this work we experiment with several medium and large sized snapshots of the Web obtained from the collection~\cite{UbiCrawler}. In particular, we used the \texttt{cnr-2000}, \texttt{eu-2005}, \texttt{india-2004}  and  \texttt{uk-2002}  Web-graphs. More information about these graphs, as well as links to download them can be found in Appendix~\ref{App:Datasets}. The larger network (\texttt{uk-2002}) have been used, only for the storage need tests as well as for the computational and convergence comparisons of Section \ref{Ch4:NCDawareRank:Sec:Algorithm}, whereas the medium- and the small-sized ones for the sparsity and link-spamming experiments presented in the following sections. During dataset preparation, we sorted the URLs lexicographically, we extracted the lengths of the NCD blocks (which in our experiments correspond to websites) and we created matrices $\mathbf{R}$ and $\mathbf{A}$, as discussed in Section \ref{Ch4:NCDawareRank:Sec:Model}. 
	  	
	  	\subsection{Testing the Effect of the New Teleportation Model on Sparsity and Link Spamming}
		\label{Ch4:NCDawareRank:Sec:Experiments}
			\subsubsection{Competing Methods and Metrics}
		We compare \textbf{NCDawareRank} (with $\eta=0.85$ and $\mu=0.1$) against several other state-of-the-art link-analysis algorithms. In particular:
		\begin{itemize}
			\item \textbf{HyperRank} (Baeza-Yates \textit{et al.}\cite{Baeza-Yates:2006:GPD:1148170.1148225})
			with $\beta=3$,
			\item \textbf{LinearRank} (Baeza-Yates \textit{et al.}\cite{Baeza-Yates:2006:GPD:1148170.1148225})
			with $L=10$,
			\item \textbf{PageRank} (Page \textit{et al.}\cite{pagerank}) using the canonical value for the damping factor, $\alpha=0.85$,
			\item \textbf{RAPr} (Constantine and Gleich \cite{constantine2009random}) with the random variable $A$, following the $\textit{Beta}(1,1,[0,1])$ distribution (the default distribution used by the authors in their publicly available implementation),
			\item \textbf{TotalRank} (Boldi \cite{Boldi:2005:TRW:1062745.1062787}).
		\end{itemize} 
		These were the parameters we used for our experiments, with a few specifically stated exceptions.	Finally, note that for some of the qualitative experiments conducted here, we have used two variants of our methods. The first one is denoted \textbf{NCDawareRank}, and handles the dangling nodes using Strategy 1, defined in the previous section. The second is denoted \textbf{NCDawareRank(Naive)} and uses the traditional strongly preferential handling. The second version is included, in order to isolate and illuminate the effects of the introduction of our novel inter-level proximity model alone.
		
		In our tests we make use of the Kendall’s $\tau$ correlation coefficient \cite{kendall1990rank,KENDALL01061938}; This is an intuitive nonparametric correlation index that has been widely used for ranking comparisons (see e.g.,  \cite{Baeza-Yates:2006:GPD:1148170.1148225,Boldi:2005:TRW:1062745.1062787, blockrank, NikolakopoulosK15,NikolakopoulosWebIntelligence2015,nikolakopoulos2014use}). 
		 The value of $\tau$ is $1$ for perfect match and $-1$ for reversed ordering. All the experiments were performed in Matlab on a 64bit machine with 24GB RAM.

			\subsubsection{Resistance to Manipulation}
			\label{Sec_antimanipulation}
			
			One of the most important problems faced by PageRank as well as other link-analysis algorithms is susceptibility to manipulation through link-spamming. Spam Web-pages represent a significant fraction of the overall pages in most domains. In some of them this fraction is alarming (according to~\cite{Ntoulas:2006:DSW:1135777.1135794}, about 70\% of .biz domain can be characterized as spam).  Taking into account that Web search is the \textit{de facto} control point for e-commerce, and that high ranking in search engines is considered to have high value, the economic incentive behind high ranking increases the need for spam-resistant ranking schemes~\cite{Eiron:2004:RWF:988672.988714,Yang:2007:DPP:1277741.1277815}.  
			
			\begin{figure}
				\centering
				\includegraphics[width = 0.92\columnwidth]{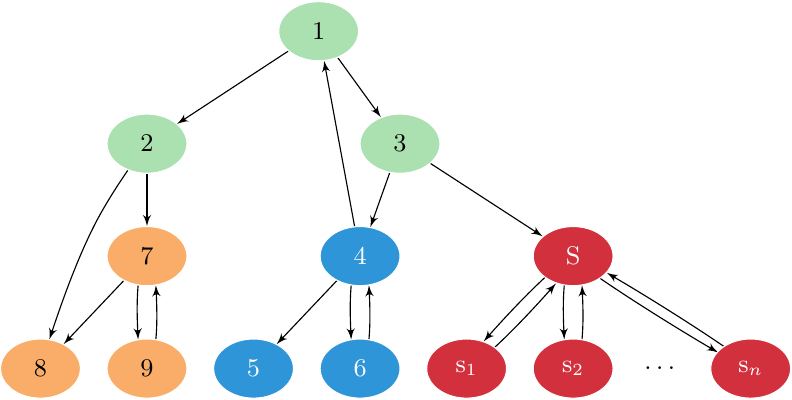}
				\caption[Example graph with spamming node.]{Example Spamgraph: the spamming node $S$ has created $n$ nodes, $s_1,s_2,\dots,s_n$ with their only outgoing link leading to $S$.}
				\label{SpamGraph}
			\end{figure}

			In this section, we show how the intensity of manipulation affects NCDawareRank’s ranking scores. The example graph of Fig.~\ref{SpamGraph} shows our approach to simulate the link-spamming phenomenon. In this example, the spamming node $S$ creates $n$ nodes that funnel all their rank towards it. In fact, this is known to be a standard manipulation technique commonly used in practice~\cite{Eiron:2004:RWF:988672.988714}.

			We follow the same approach in our experiments using the \texttt{cnr-2000} graph instead. In particular, we randomly pick a node with small initial ranking (from now on referred to as the ``spamming node'') and we add a number of $n$ nodes that their only incoming link is from the spamming node and their only outgoing link is towards it, in the same manner as in the small example graph of Figure \ref{SpamGraph}. Then, we run both variants of NCDawareRank and the other algorithms, for several values of $n$.

			\setlength\figureheight{0.42\linewidth} 
			\setlength\figurewidth{0.74\linewidth}
			\begin{figure}
				\centering
				\includegraphics[width = 0.95\columnwidth]{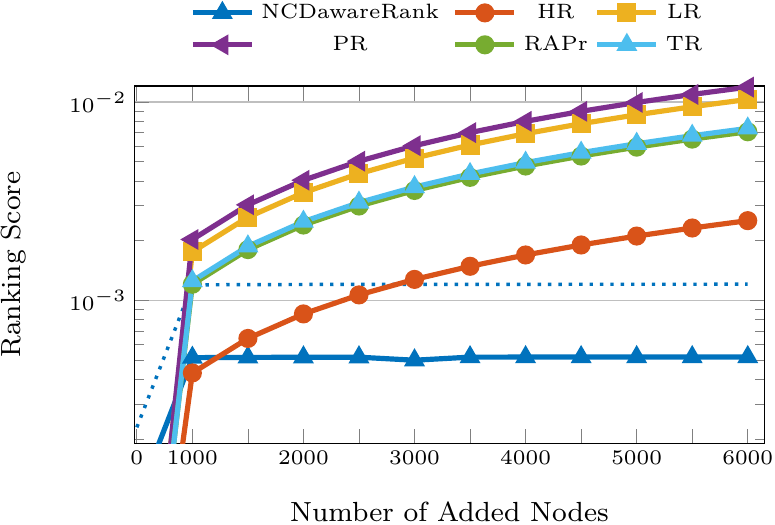}
				\caption[Ranking of the spamming node as a function of added pages, for different algorithms.]{Ranking of the spamming node as $n$ grows, for different algorithms. The dotted blue line represents the ranking score of a naive version of NCDawareRank with strongly preferential dangling node handling.}
				\label{fig:SpammingResults}
			\end{figure}

			In Fig.~\ref{fig:SpammingResults}, we see the ranking score of the spamming node as a function of $n$, for all the different algorithms we experimented with. The anti-manipulation effect of NCDawareRank becomes immediately clear. The introduction of the matrix $\mathbf{M}$ and the NCD-based teleportation of our method, ensure that the ranking increment rate, after a small number of added nodes, becomes very small; with the absolute ranking score in the case of the Naive algorithm being a little bigger than the standard version. In our method in order to gain rank, a node has to have incoming links originated from other NCD blocks as well; the effect of nepotistic links is limited and thus, artificial boosting of the ranking score becomes harder.
			
			\setlength\figureheight{0.44\linewidth} 
			\setlength\figurewidth{0.77\linewidth}
			\begin{figure}
				\centering
				\centering
				\includegraphics[width = 0.85\columnwidth]{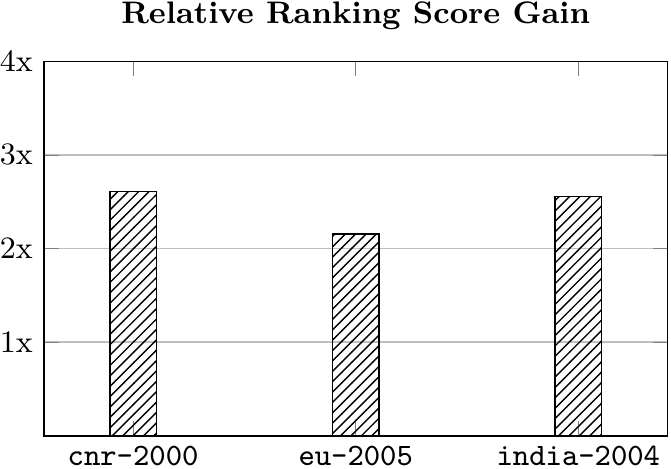}
				\caption[Testing the relative spamming node’s ranking score gain per added node.]{Testing the relative spamming node’s ranking score gain per added node. The figure reports the ratio of this ranking gain between the Naive version of NCDawareRank (strongly-preferential handling) and the Standard version (using the strategy of Section \ref{Ch4:NCDawareRank:Sec:Dangling}).}
				\label{fig:SpammingRelativeGain}
			\end{figure}

			Notice here, that the ranking gain with the addition of more and more nodes for both variants of our method is so small that they appear to be flat when put in the same graph with the competing methods. In fact the spamming node gains a very small amount of ranking, with  the rate of this gain being different for the two versions of our algorithm. To illuminate this and to quantify the relative benefits that come from adopting the NCDaware dangling strategy, we conduct the following experiment: We take the \texttt{cnr-2000,eu-2005,india-2004} graphs,  we randomly  sample 100 of their nodes, and we treat each of them (one at a time) as a spammer  adding $n$ artificially created nodes (for $n$ = 5\textperthousand\ to 30\textperthousand\ of the cardinality of the complete set of pages) that funnel their rank towards it; we then run both the Naive and the standard NCDawareRank versions, and we report the ratio between the mean ranking score gain per added node. The results are presented in Fig.~\ref{fig:SpammingRelativeGain}. We see that the ranking gain when we use the NCDaware strategy are about 2.5 times smaller than the gains of the strongly preferential variant. This was expected and is in accordance with the characteristics of dangling Strategy 1, discussed in Appendix \ref{SI:dangling}.

			\setlength\figureheight{0.44\linewidth} 
			\setlength\figurewidth{0.74\linewidth}	
			\begin{figure}
				\centering
				\includegraphics[width = 0.85\columnwidth]{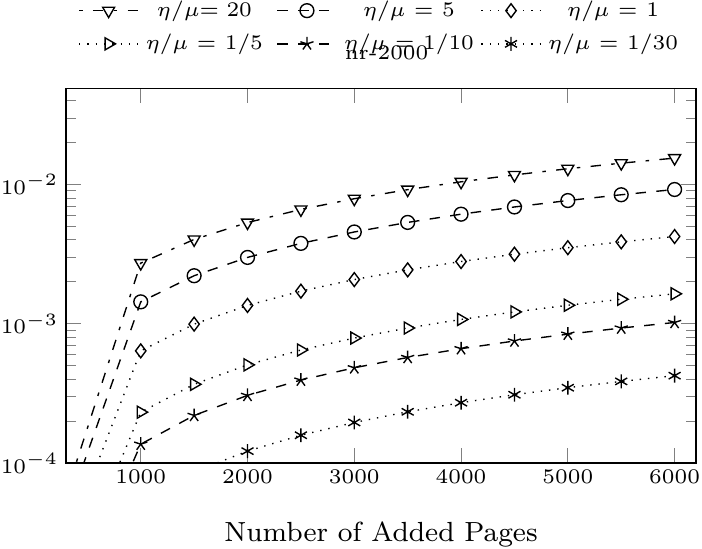}
				\caption[Ranking of the spamming node as a function of added pages, using the standard uniform teleportation and strongly preferential handling.]{Ranking of the spamming node as $n$ grows using the standard uniform teleportation and strongly preferential handling, in order to isolate the effect of the introduction of matrix $\mathbf{M}$.}
				\label{fig:SpammingResults2}
			\end{figure}

			Finally, in order to better isolate the resistance effect induced by the introduction of parameter $\mu$ and the corresponding matrix $\mathbf{M}$ alone, we run NCDawareRank using PageRank’s uniform teleportation, for several values of the ratio $\eta/\mu$, while holding $\eta+\mu$ equal to 0.95, and we plot the results in Figure \ref{fig:SpammingResults2}. We see that even with the introduction of a very small $\mu$,  NCDawareRank starts to exhibit positive resistance properties with the anti-manipulation effect increasing as the ratio $\eta/\mu$ tends to zero, as expected. Of course as  $\eta/\mu $ is getting more close to zero, the direct link structure of the network values less, because the adjacency matrix $\mathbf{H}$ gets increasingly ignored. However, this increased sensitivity of the spamming node’s ranking score for small values of the ratio $\eta/\mu$ could lead to interesting alternative uses of our measure (e.g. Spam classification).

	\subsubsection{Sparsity}
	\label{Sec_sparseness}

	It is well known  that the degree distribution of many networks that arise in practice  generally follows a 
	power law \cite{Faloutsos:1999:PRI:316194.316229}. This leads to a sparse 
	adjacency matrix. Furthermore, in~\cite{Pandurangan} it was observed that such 
	link distributions cause the probability values produced by PageRank to decay 
	according to a power law, making most of the pages unable to obtain reasonable 
	score~\cite{Xue:2005:EHS:1076034.1076068}. The latter is especially true for the \textit{Newly Added Pages} which usually 
	have too-few incoming links, and thus cannot receive reasonable ranking \cite{Cho:2004:ISE:988672.988676}. To show the performance of NCDawareRank in dealing with the problems caused by the low density of networks such as the Web-graph, we conduct the following two experiments. 

	\subsubsection{Newly Added Pages}
	In the first experiment, we test the performance of our method in dealing with the newly added pages problem.  Adopting the methodology of  Xue \textit{et al.}~\cite{Xue:2005:EHS:1076034.1076068}, we simulate the phenomenon by extracting 90\% of the incoming links of a set of randomly chosen pages. The altered graph then represents an ``earlier version'' of the Web, where these pages were new, and hence, the number of their incoming links was smaller.	
	
\begin{figure}
		\setlength\figureheight{0.18\linewidth} 
		\setlength\figurewidth{0.87\linewidth} 
		\centering
		\includegraphics[width = 0.99\columnwidth]{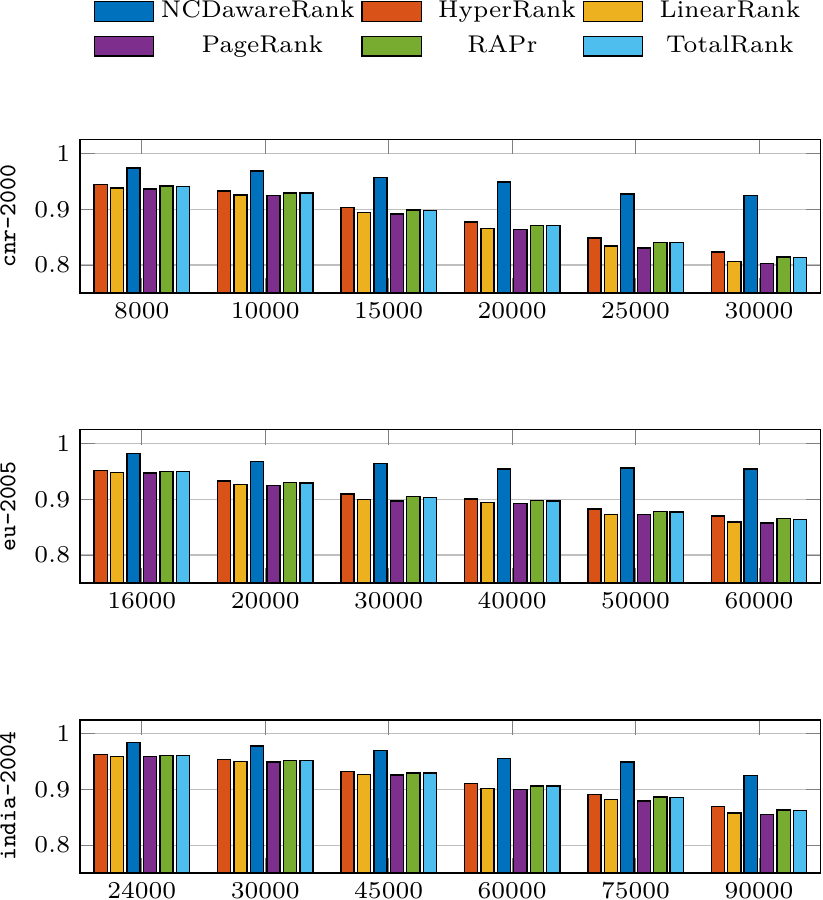}
		\caption[Newly added pages tests.]{Newly added pages tests.}
		\label{fig:NAP}
	\end{figure}

	In particular: 
	\begin{itemize}
		\item[-] First, we run all the algorithms on the complete graph, and we obtain a reference ranking for each method.
		\item[-] Then, we randomly choose a number of $n$ pages (for several values of $n$) and we randomly remove 90\% of their incoming links.
		\item[-] We rebuild the new factor matrices $\mathbf{A}$ and $\mathbf{R}$ using the  modified hyperlink matrix.
		\item[-] We re-run the algorithms and we compare how different is the ordering induced on the new graph from the original.
	\end{itemize}
	The measure used for the ranking comparison step is Kendall’s $\tau$ correlation coefficient. High value of this metric means that the new ordering is very close to the ordering on the original graph, where all the links were included.	
	We repeat the above procedure 10 times; each time for all the different values of newly added pages and we present the average results in Fig.~\ref{fig:NAP}. 

		 We see that both variants of NCDawareRank outperform all other algorithms, allowing the newly added pages to get a ranking more similar to the one arising using the complete set of incoming links; furthermore, their advantage becomes bigger as the number of newly added pages becomes larger. 
		 
		 The results are consistent with the way NCDawareRank views the Web. In NCDawareRank the importance of a page is not exclusively determined by its incoming links;  its ``neighborhood'' also matters, since the inter-level proximity matrix ensures that every link confers a small amount of rank to the corresponding NCD block of the target node. So, because the importance of a page is usually correlated with the importance of the website that contains it, even with fewer incoming links, newly emerging pages inherit some of the importance of the corresponding block, gathering a relative ranking score closer to that arising from the complete network.

		\subsection*{Ranking Stability in the Presence of Sparsity}
\begin{figure}
			\centering
			\includegraphics[width = 0.99\columnwidth]{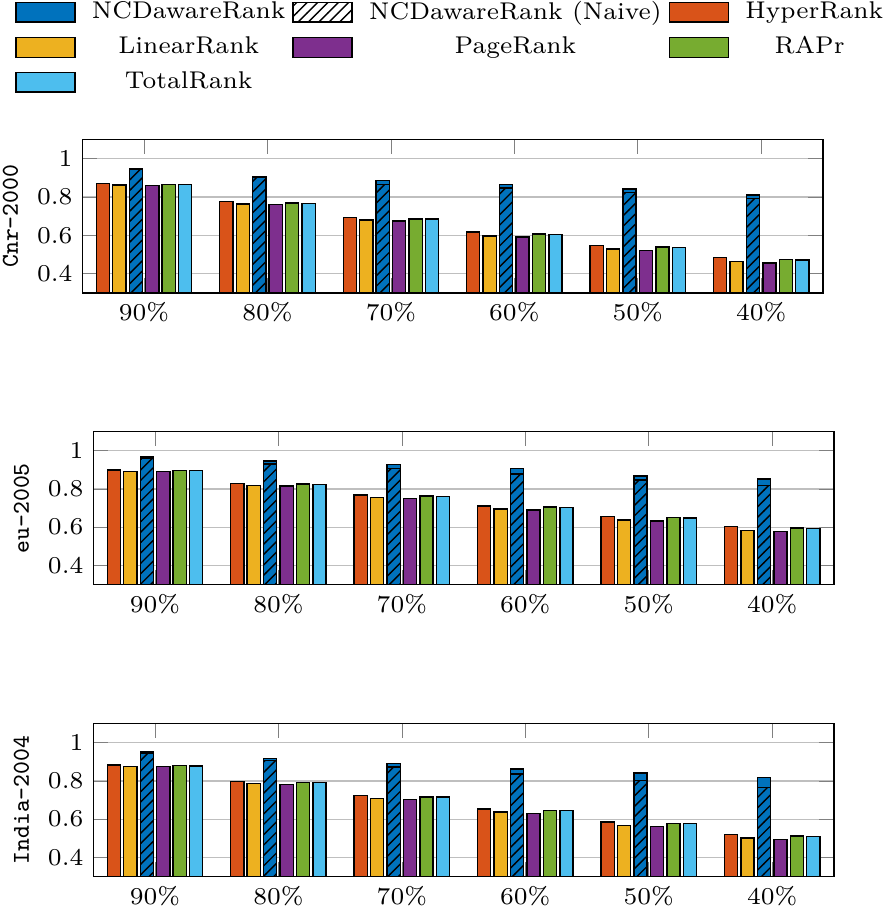}
			\caption{Ranking stability under sparseness.}
			\label{fig:Sparsity}
		\end{figure}

		In our second experiment, following the same methodology as before, we simulate the sparseness of the hyperlink network by randomly selecting to include 90\% - 40\% of the links on the altered network and we compare the ranking results of the algorithms against their corresponding original rankings. 
		Notice that in this case the sparsity is observed throughout the network, instead of being concentrated in a particular set of pages. In Fig.~\ref{fig:Sparsity} we see that while the network is still relatively dense (i.e. 90\% of the links are included) all algorithms tend to produce orderings very similar to those produced for the complete network. However, as the link  structure becomes sparser, the orderings begin to differ more and  more. The ranking vector produced by NCDawareRank is more resistant to this effect, compared to the other ranking methods with the standard version of our algorithm performing marginally better than its naive counterpart. Moreover, we clearly see that the advantage of our method becomes bigger as the network becomes sparser. 
		
		These results verify the intuition behind NCDawareRank; even though the direct link structure of the network collapses with the exclusion of such many links, the inter-level proximity captured by our novel teleportation model---and the corresponding matrix $\mathbf{M}$---decays harder and thus preserves longer the coarser structure of the network. This, results in a ranking vector that proves to be less sensitive to small changes of the underlying link structure.
		
		\nocite{nikolakopoulos2019recwalk,nikolakopoulos2020boosting,nikolakopoulos2014use,nikolakopoulos2019eigenrec,nikolakopoulos2016multi,nikolakopoulos2016multi}
	
\acknow{This work was supported by NSF 1901134. I would like to thank Prof. John D. Garofalakis;  co-author of the preliminary conference paper that first proposed the NCDawareRank framework. I would also like to thank Prof. Efstratios Gallopoulos for many insightful discussions.}

\showacknow{} 

% Bibliography
\bibliography{pnas-sample}	

\newpage

\onecolumn
\appendix

\section*{\LARGE Supporting Information Appendix }
\section{Nearly Completely Decomposable Systems: Theoretical Background and Preliminaries}
\label{Ch2:Preliminaries}	

\textit{Decomposability} is the leitmotif of this paper; it is exploited conceptually, qualitatively as well as computationally throughout our work. In this section, we introduce the basic assumptions supporting the property of Near-Complete Decomposability (NCD); we present formally the two fundamental existence theorems of Simon and Ando~\cite{simon1961aggregation}; and we explore intuitively their implications to the analysis of stochastic systems. Furthermore, we discuss how one can exploit the intuition behind Simon and Ando's results to get a \textit{Decompositional Procedure} for the approximation of the stationary distribution of an NCD Markov chain.
Finally, we discuss Meyer’s \textit{Stochastic Complementation Theory}~\cite{Meyer:1989:SCU:75568.75571} which provides a way for the above approach to yield exact results. 

The discussion of Simon and Ando’s results follows the original presentation of the authors’ in their seminal paper~\cite{simon1961aggregation}, as well as that of Courtois in his classic monograph~\cite{courtois1977decomposability}. The organization and exposition of the purely computational sections draws from Stewart~\cite{stewart:1994introduction} and, finally, the discussion of stochastic complementation theory adopts the notation introduced by Meyer in his original paper~\cite{Meyer:1989:SCU:75568.75571}. 
The reader is assumed to be familiar with elementary Linear Algebra, Discrete Time Markov Chains as well as some fundamental definitions and theorems of the theory of Non-Negative Matrices. 	For a detailed discussion of the notions introduced in this section the interested reader can refer to~\cite{courtois1977decomposability,Meyer:1989:SCU:75568.75571,stewart:1994introduction} and the references therein. 

	 \subsection{Nearly Completely Decomposable Systems}
	 \label{Subsec:NCDMarkovChains}
	 The pioneering work on NCD systems was done by Simon and Ando~\cite{simon1961aggregation},  who reported on state aggregation in linear models of economic systems, but the universality and the versatility of Simon’s idea have permitted the theory to be used in many complex problems from diverse disciplines ranging from economics, cognitive theory and social sciences, to computer systems performance evaluation, data mining and information retrieval~\cite{courtois1977decomposability,blockrank,MeyerRV13,meyer2012stochastic,yin2013continuous}. The intuition behind Simon and Ando’s approach is founded in the idea that \textit{large systems often have the property that their states can be clustered into groups, such that the interactions among the states of a group may be studied as if interactions among groups do not exist, and then the group level interactions can be carried out without reference to the interactions that take place within the groups}.
	 In the following section we formulate this intuitive idea rigorously, limiting our attention to stochastic systems. 
	 
	 \subsubsection{The Simon-Ando Theorems}
	 Let $\mathbf{P}$ be an $n\times n$ primitive stochastic matrix, denoting the transition probability matrix of an ergodic Markov chain.  Note that $\mathbf{P}$ can be written as follows:
	 \begin{equation}
	 \mathbf{P} = \mathbf{\tilde{P}} + \varepsilon \mathbf{Q},
	 \label{synistvses}
	 \end{equation}
	 where $\mathbf{\tilde{P}}$ is given by
	 \begin{equation}
	 \mathbf{\tilde{P}} \triangleq
	 \begin{pmatrix}
	 \mathbf{\tilde{P}_1} & \mathbf{0} & \mathbf{\cdots} & \mathbf{0} \\
	 \mathbf{0}    & \mathbf{\tilde{P}_2} & \mathbf{\ddots} & \mathbf{\vdots} \\
	 \mathbf{\vdots} & \mathbf{\ddots} & \mathbf{\ddots}  & \mathbf{0} \\
	 \mathbf{0}    &  \mathbf{\cdots} &  \mathbf{0}   & \mathbf{\tilde{P}_N} 
	 \end{pmatrix}.
	 \end{equation}
	 Matrices $\mathbf{\tilde{P}_I}$, $I = 1,\dots, N$, are irreducible stochastic matrices of order $n(I)$, therefore 
	 \begin{displaymath}
	 n=\sum_{I=1}^{N}n(I)
	 \end{displaymath}
	 and the row-sums of matrix $\mathbf{Q}$ are all zero. We choose $\varepsilon$ and  $\mathbf{Q}$ such that for each row $m_I , I = 1,\dots, N$, $m = 1,\dots, n(I)$ it holds:
	 
	 \begin{eqnarray}
	 \varepsilon \sum_{J\ne I}\sum_{l=1}^{n(J)}Q_{m_Il_J} & = & \sum_{J \ne I}\sum_{l=1}^{n(J)}P_{m_Il_J} \\ 
	 \varepsilon & \triangleq &\underset{{}}{\mathop{\underset{{{m}_{I}}}{\mathop{\max }}\,\left( \sum\limits_{J\ne I}{\sum\limits_{l=1}^{n(J)}{{{P}_{{{m}_{I}}{{l}_{J}}}}}} \right)}},
	 \end{eqnarray}
	 where $m_I$ denotes the $m^\textit{th}$ element of the $I^{\textit{th}}$ block. Parameter  $\varepsilon$ is called the \textit{maximum degree of coupling} between the subsystems $\mathbf{\tilde{P}_I}, I=1,\dots,N$. We use $\tilde{\lambda}_{m_I}$, $m=1,\dots,n(I)$ to denote the eigenvalues of $\mathbf{\tilde{P}_I}$, and we consider them ordered so that,
	 \begin{displaymath}
	 \tilde{\lambda}_{1_I}=1>|\tilde{\lambda}_{2_I}| \leq |\tilde{\lambda}_{3_I}| \leq \cdots \leq |\tilde{\lambda}_{n(I)_I}|.
	 \end{displaymath}
	 Adopting the notational conventions of Courtois~\cite{courtois1977decomposability}, we use $\tilde{\delta}$ to define the minimum of the absolute values of the differences between unity and all eigenvalues of $\mathbf{\tilde{P}}$ that are not equal to unity. We have
	 \begin{displaymath}
	 |1-\tilde{\lambda}_{m_I}| \geq \tilde{\delta} > 0.
	 \end{displaymath}
	 Since the eigenvalues of a matrix are continuous functions of its elements~\cite{meyer2000matrix,strang2003introduction} for every positive real number  $\delta$ we can define a small enough $\varepsilon$ so that, for every eigenvalue $\tilde{\lambda}_{m_I}$ of $\mathbf{\tilde{P}}$ there exists an eigenvalue $\lambda_{m_I}$ of $\mathbf{P}$ such that for all $m_I$
	 \begin{equation}
	 |\tilde{\lambda}_{m_I}-\lambda_{m_I}|< \delta.
	 \end{equation}
	 Hence, we can classify the eigenvalues of $\mathbf{P}$ into two categories:
	 \begin{eqnarray}
	 |1-\lambda_{1_I}| & < & \delta, \qquad \qquad I = 1,\dots,N, \nonumber\\
	 |1-\lambda_{m_I}| & > & \tilde{\delta} - \delta,\text{ } \qquad I = 1,\dots,N, \quad m = 2,\dots,n(I), \nonumber
	 \end{eqnarray}
	 where $\delta \to 0$ with $\varepsilon$.
	 
	 Assuming that $\mathbf{P}$ and $\mathbf{\tilde{P}}$ have linear elementary divisors (and thus there exist complete sets of not necessarily unique left and right eigenvectors), the spectral decomposition of matrix $\mathbf{P}^t$ can be written as
	 \begin{eqnarray}
	 \mathbf{P}^t & = & \sum_{I=1}^{N}\sum_{m=1}^{n(I)} \lambda^t_{m_I}\mathbf{Z}_{m_I},\text{ with } \\
	 \mathbf{Z}_{m_I} & = & s_{m_I}^{-1}\mathbf{v}_{m_I}\mathbf{u}_{m_I}^\intercal, \nonumber 
	 \label{t-step}
	 \end{eqnarray}
	 where $\mathbf{v}_{m_I},\mathbf{u}_{m_I} $ denote the left and right eigenvectors\footnote{The eigenvectors are assumed to be normalized to one using the norm  $\rVert \cdotp \lVert_1$} that correspond to eigenvalue $\lambda_{m_I}$, and $s_{m_I}$ is defined by%the condition number 
	 \begin{displaymath}
	 s_{m_I} \triangleq \mathbf{v}_{m_I}^\intercal \mathbf{u}_{m_I}.
	 \end{displaymath}
	 Taking into account the stochasticity of $\mathbf{P}$ we can write
	 \begin{displaymath}
	 \mathbf{P}^t  =  \mathbf{Z}_{1_1} + \sum_{I=2}^{N}\lambda^t_{1_I}\mathbf{Z}_{1_I} + \sum_{I=1}^{N}\sum_{m=2}^{n(I)} \lambda^t_{m_I}\mathbf{Z}_{m_I}.
	 \end{displaymath}
	 If we define for each block of $\mathbf{\tilde{P}}$, left and right normalized eigenvectors $\mathbf{\tilde{v} }_{m_I},\mathbf{\tilde{u} }_{m_I}$; scalars   $\tilde{s} _{m_I}$;  and the related matrices $\mathbf{\tilde{Z}}_{m_I}$, as before,  we can write for $\mathbf{\tilde{P}}$:
	 \begin{displaymath}
	 \mathbf{\tilde{P} }^t  =  \sum_{I=1}^{N}\mathbf{\tilde{Z}}_{1_I} + \sum_{I=1}^{N}\sum_{m=2}^{n(I)} (\tilde{\lambda} )^t_{m_I}\mathbf{\tilde{Z}} _{m_I}.
	 \end{displaymath}	 
	 Let us now consider the dynamic behavior of the processes $\mathbf{y}_{(t)}$ and  $\mathbf{\tilde{y}}_{(t)}$ defined by 
	 \begin{eqnarray}
	 \mathbf{y}_{(t)}^\intercal & = &\mathbf{y}_{(t-1)}^\intercal \mathbf{P} = \mathbf{y}_{(0)}^\intercal \mathbf{P}^t, \nonumber \\
	 \mathbf{\tilde{y}}_{(t)}^\intercal& = &\mathbf{\tilde{y}}_{(t-1)}^\intercal \mathbf{\tilde{P}} =  \mathbf{\tilde{y}}_{(0)}^\intercal \mathbf{\tilde{P}}^t. \nonumber
	 \end{eqnarray} 
	 The comparison of these two processes is made possible using the following two theorems of Simon and Ando~\cite{simon1961aggregation} which are presented here without proof.
	 
	 \begin{theorem}[Simon \& Ando~\cite{simon1961aggregation}]
	 	For an arbitrary positive real number $\xi$, there exists a number $\varepsilon_\xi$ such that for $\varepsilon < \varepsilon _\xi$,
	 	\begin{displaymath}\max_{i,j}\lvert[\mathbf{Z}_{m_I}]_{ij} - [\mathbf{\tilde{Z}}_{m_I}]_{ij}\rvert < \xi
	 	\end{displaymath} 
	 	with
	 	\begin{displaymath}
	 	2 \leq m \leq n(I), \qquad 1 \leq I \leq N, \qquad 1 \leq i,j \leq n.
	 	\end{displaymath}   
	 	\label{SA1}
	 \end{theorem}
	 
	 \begin{theorem}[Simon \& Ando~\cite{simon1961aggregation}]
	 	For an arbitrary positive real number $\omega$, there exists a number $\varepsilon_\omega$ such that for $\varepsilon < \varepsilon_\omega$,
	 	\begin{displaymath}
	 	\max_{m,l}\lvert[\mathbf{Z}_{1_K}]_{m_Il_J} -  [\mathbf{\tilde{v}}_{1_J}]_{l_J}\alpha_{IJ}(1_K)\rvert < \omega
	 	\end{displaymath}
		with 
		\begin{displaymath}
		1 \leq K, I, J \leq N, \qquad 1 \leq m \leq n(I), \qquad 1 \leq l \leq n(J),
		\end{displaymath}
	 	and where $\alpha_{IJ}(1_K)$ is given by 
	 	\begin{displaymath}
	 	\alpha_{IJ}(1_K)=\sum_{m=1}^{n(I)}\sum_{l=1}^{n(J)}[\mathbf{\tilde{v}}_{1_I}]_{m_I}[\mathbf{Z}_{1_K}]_{m_Il_J} .
	 	\end{displaymath}
	 	\label{SA2}
	 \end{theorem}
	 
	 \subsubsection{The Implications of Simon-Ando Theorems on Intuitive Grounds}
	 Let us restate here for clarity the spectral decompositions of $\mathbf{P}^t$ and $\mathbf{\tilde{P}}^t$
	  \begin{eqnarray}
	  \mathbf{P}^t & = & \underbrace{\mathbf{Z}_{1_1}}_{\text{Term }\mathrm{A}} + \underbrace{\sum_{I=2}^{N}\lambda^t_{1_I}\mathbf{Z}_{1_I}}_{\text{Term }\mathrm{B}} + \underbrace{\sum_{I=1}^{N}\sum_{m=2}^{n(I)} \lambda^t_{m_I}\mathbf{Z}_{m_I}}_{\text{Term }\mathrm{C}},
	  \label{t-phasma} \\
 	 \mathbf{\tilde{P} }^t & = & \underbrace{\sum_{I=1}^{N}\mathbf{\tilde{Z}}_{1_I}}_{\text{Term }\tilde{\mathrm{A}}} + \underbrace{\sum_{I=1}^{N}\sum_{m=2}^{n(I)} (\tilde{\lambda} )^t_{m_I}\mathbf{\tilde{Z}} _{m_I}}_{\text{Term }\tilde{\mathrm{C}}}.
 	 \label{t-step-star}
	  \end{eqnarray}
	 In NCD systems it holds that for each $I = 1,\dots, N$, the eigenvalue $\lambda_{1_I}$ is close to unity which means that $\lambda^t_{1_I}$ will also be close to unity for small values of $t$. Therefore, Terms $\mathrm{A}$ and $\mathrm{B}$ of the RHS of~(\ref{t-phasma}) will not differ significantly for $t < T_2$ (for some $T_2 > 0$), while Term $\tilde{\mathrm{A}}$ of~(\ref{t-step-star}) does not change at all. Thus, for $t < T_2$ the dynamic behavior of  $\mathbf{y}_{(t)}$ and $\mathbf{\tilde{y}}_{(t)}$ is determined by Terms $\mathrm{C}$ and $\mathrm{\tilde{C}}$ respectively.  However, as $\varepsilon \to 0$ we have
	 \begin{math}
	 \lambda_{m_I} \to \tilde{\lambda}_{m_I},
	 \end{math} 
	 and from Theorem~\ref{SA1} it follows that, 
	 \begin{displaymath}
	 \mathbf{Z}_{m_I} \to \mathbf{\tilde{Z}} _{m_I},
	 \end{displaymath}for every $ m = 2,\dots, n(I)$ and $I = 1,\dots, N$. This means that for sufficiently small $\varepsilon$ and $t < T_2$ the time paths of $\mathbf{y}_{(t)}$ and $\mathbf{\tilde{y}}_{(t)}$ are close. 
	 
	 Since the moduli of eigenvalues $\tilde{\lambda}_{m_I}$ are all away from unity for every $m =2,\dots , n(I)$, and $I = 1,\dots , N$, for each positive real number $\xi_1$ we can define the smallest time interval $T^\star_1$ such that
	 \begin{displaymath}
	 \max_{1 \leq i,j \leq n}\bigg| \sum_{I=1}^{N}\sum_{m=2}^{n(I)}(\tilde{\lambda} )^t_{m_I}[\mathbf{\tilde{Z}}_{m_I}]_{ij}\bigg|<\xi_1, \quad \text{ for }t>T^\star_1.
	 \end{displaymath}
	 Similarly, we can define the smallest interval $T_1$ such that
	\begin{displaymath}
	 \max_{1 \leq i,j \leq n}\bigg| \sum_{I=1}^{N}\sum_{m=2}^{n(I)}\lambda^t_{m_I}[\mathbf{Z}_{m_I}]_{ij}\bigg|<\xi_1, \quad \text{ for }t>T_1.
	 \end{displaymath}
	 Theorem~\ref{SA1} together with the fact that eigenvalues $\tilde{\lambda}_{m_I}$ converge to $\lambda_{m_I}$ with $\varepsilon$ ensures that 
	 \begin{displaymath}
	 T_1 \to T^\star_1, \qquad \text{ as\ \  } \varepsilon \to 0.
	 \end{displaymath} Now, since we can make $T_2$ as large as we want by choosing a sufficiently small $\varepsilon$, we can choose a small enough $\varepsilon$ so that $T_2>T_1$ holds. 
	 
	 Furthermore, given that $\varepsilon\neq0$, $\lambda_{1_I}\neq1, I=2,\dots,N$ also holds\footnote{Notice that if $\varepsilon=0$, every block $\mathbf{P_I}$ will be irreducible and $\lambda_{1_I}=\tilde{\lambda}_{1_I}$  will trivially hold for each $I$.}, and there will come a time $T_3>T_2$ such that for a small enough real number $\xi_3$,
	 \begin{displaymath}
	 \max_{1 \leq i,j \leq n}\bigg| \sum_{I=1}^{N}\lambda^t_{1_I}[\mathbf{Z}_{1_I}]_{ij}\bigg|<\xi_3, \quad \text{ for\ \ }t>T_3.
	 \end{displaymath}
	 Therefore, for $T_2 < t < T_3$, Term $\mathrm{C}$ of $\mathbf{P}^t$ is negligible and the time path of $\mathbf{y}_{(t)}$ is determined by the Terms $\mathrm{A,B}$ of $\mathbf{P}^t$. However, Theorem~\ref{SA2} specifies that for any $I$ and $J$, the entries of $\mathbf{Z}_{1_K}$:
	 \begin{displaymath}
	 [\mathbf{Z}_{1_K}]_{i_I1_J}, \dots , [\mathbf{Z}_{1_K}]_{i_Ij_J},\dots, [\mathbf{Z}_{1_K}]_{i_In(J)_J},
	 \end{displaymath}
	 depend upon $I,J$ and $j$, being almost independent of $i$, i.e. for any $I,J$ they are proportional to the elements of the unique stationary distribution of $\mathbf{\tilde{P}_J}$,
	 \begin{equation}
	  [\mathbf{\tilde{v}}_{1_J}]_{1_J}, \dots , [\mathbf{\tilde{v}}_{1_J}]_{j_J},\dots, [\mathbf{\tilde{v}}_{1_J}]_{n(J)_J},
	  \label{eq:BlockShortTermEquilibrium}
	 \end{equation}
	 being approximately the same for $i=1,\dots ,n(I)$.   Thereby, for $ T_2 < t < T_3$, the process $\mathbf{y}_{(t)}$ will vary with $t$, keeping among the elements of $[\mathbf{y}_{(t)}]_{j_J}$ of every subset $J$ an approximately constant ratio which is identical to the ratio between the elements of~(\ref{eq:BlockShortTermEquilibrium}). 	   
	   Finally, for $t > T_3$ the behavior of $\mathbf{y}_{(t)}$ is determined by Term $\mathrm{A}$, and $\mathbf{P}$ moves towards its long-term equilibrium defined by $\mathbf{v}_{1_1}$.

\subsubsection{NCD Decomposition Approximation}
\label{Ch:Preliminaries:NCDApproximation}
Let us consider a Markov chain $\mathbf{P}$, that is nearly completely decomposable into $L$ blocks:
\begin{displaymath}
\{\mathcal{D}_1,\mathcal{D}_2,\dots,\mathcal{D}_L\}.
\end{displaymath} 
 Strong interactions among the states within a block and weak interactions between the blocks themselves imply that the state space of a nearly completely decomposable Markov chain can be ordered so that the transition probability stochastic matrix has the form
  \begin{equation}
  \mathbf{P} = \begin{pmatrix}
  \mathbf{P_{11}} & \mathbf{P_{12}} & \dots & \mathbf{P_{1L}} \\
  \mathbf{P_{21}} & \mathbf{P_{22}} & \dots & \mathbf{P_{2L}} \\
  \vdots	& \vdots & \ddots & \vdots \\
  \mathbf{P_{L1}} & \mathbf{P_{L2}} & \dots & \mathbf{P_{LL}} 
  \end{pmatrix},
  \end{equation}
  with the nonzero elements of the off-diagonal blocks, being small compared to the elements of the diagonal blocks, i.e. we will assume that
  \begin{eqnarray}
  \lVert\mathbf{P_{II}}\rVert & = & \mathrm{O}(1),\qquad I=1,2,\dots,L , \nonumber \\
  \lVert\mathbf{P_{IJ}}\rVert & = & \mathrm{O}(\varepsilon),\qquad I\neq J ,\nonumber
  \end{eqnarray}
 where $\varepsilon $ is a sufficiently small positive real number and $\lVert\cdot\rVert$ denotes the spectral norm of a matrix. 
 
 A direct consequence of the Simon and Ando’s theorems is a natural procedure for the solution of a Markov chain that satisfies the above assumptions: Informally, the first theorem gives us the grounds to consider each substochastic diagonal block $\mathbf{P_{II}}$ is isolation; make it stochastic, in order to get a well-defined independent stochastic system $\mathbf{\tilde{P}_{I}}$; and then find its stationary distribution which can be thought as a good approximation to the probability distribution of the states in block $I$ (conditioned on being in block $I$). The second theorem asserts that the solutions found in this way \textit{are maintained} as the systems moves towards equilibrium under the weak interactions between the blocks, therefore, we can proceed to find the long-term probability of being in each block, considering exclusively the way these blocks interact with each other, and get an approximation of the global solution by weighting the individual solutions of the independent blocks by the equilibrium probabilities of being in each block.  

To clarify the above procedure we give an example that makes use of the following NCD matrix (which is sometimes called in the literature the Courtois matrix~\cite{stewart:1994introduction,stewart2009probability} since it was used as an example by Courtois in his classic monograph~\cite{courtois1977decomposability}),   
\begin{displaymath}
\mathbf{P}=
\begin{footnotesize}
\left(\begin{array}{ccc;{3pt/2pt}cc;{3pt/2pt}ccc}
0.85 & 0 & 0.149 & 0.0009 & 0 & 5\times 10^{-5} & 0 & 5\times 10^{-5} \\
0.1 & 0.65 & 0.249 & 0 & 0.0009 & 5\times 10^{-5} & 0 & 5\times 10^{-5} \\
0.1 & 0.8 & 0.0996 & 0.0003 & 0 & 0 & 0.0001 & 0 \\ \hdashline[3pt/2pt]
0 & 0.0004 & 0 & 0.7 & 0.2995 & 0 & 0.0001 & 0 \\
0.0005 & 0 & 0.0004 & 0.399 & 0.6 & 0.0001 & 0 & 0 \\\hdashline[3pt/2pt]
0 & 5\times 10^{-5} & 0 & 0 & 5\times 10^{-5} & 0.6 & 0.2499 & 0.15 \\
3\times 10^{-5} & 0 & 3\times 10^{-5} & 4\times 10^{-5} & 0 & 0.1 & 0.8 & 0.0999 \\
0 & 5\times 10^{-5} & 0 & 0 & 5\times 10^{-5} & 0.1999 & 0.25 & 0.55
\end{array}\right).
\end{footnotesize}
\end{displaymath}
 
 \begin{description}
 	\item[Solving the Diagonal Blocks as if they are Independent:] The first step is to assume that the system is completely decomposable into subsystems defined by stochastic matrices $\mathbf{\tilde{P}_I}$ and then to find the stationary distribution of each subsystem separately. Matrices $\mathbf{\tilde{P}_I}$ are formed by some sort of stochasticity adjustment of the strictly substochastic block diagonal matrices $\mathbf{P_{II}}$. The way these substochastic matrices are made stochastic  is known to have an effect to the degree of the approximation one gets in the end~\cite{stewart:1994introduction}. In fact, as we will discuss in the following section, one can define these matrices in such a way that the above procedure gives exact results. However, the computational burden of performing such ideal stochasticity adjustment is--- very often---forbidding for real-world applications of the method\footnote{Furthermore, when one uses the above decompositional approach solely as a way to find the stationary distribution faster---as opposed to obtaining a well-defined subsystem for further study---he can simply use the block diagonal substochastic matrices by themselves, and find the normalized left eigenvector that corresponds to the Perron root of the substochastic matrix~\cite{stewart2009probability}. Iterative Aggregation/Disaggregation Algorithms~\cite{DBLP:journals/jacm/CaoS85,KMS} following this approach are known to converge very fast to the exact solution for NCD matrices.}. 
 	
 	Returning to our example we see that the Courtois matrix is clearly NCD into 3 blocks: 
 	\begin{displaymath}
 	\{\{1,2,3\}, \{4,5\}, \{6,7,8\} \}.
 	\end{displaymath}
 	The first block has order 3, the second order 2, and the third block has order 3. Below we give one choice for the corresponding completely decomposable stochastic subsystems, and their stationary distributions:
 	\begin{eqnarray}
 	\mathbf{\tilde{P}_1} & = & \begin{pmatrix}
 	0.85 & 0 & 0.15\\
 	0.1 & 0.65 & 0.25 \\
 	0.1 & 0.8 & 0.1  
 	\end{pmatrix},\nonumber\\
 	\boldsymbol{\tilde{\pi}^\intercal_1} &=& \begin{pmatrix}
 	0.4 & 0.417391 & 0.182609
 	\end{pmatrix}. \nonumber \\[2em]
 	\mathbf{\tilde{P}_{2}} & = &
 	\begin{pmatrix}
 	0.7 & 0.3 \\
 	0.4 & 0.6   
 	\end{pmatrix},\nonumber\\
 	\boldsymbol{\tilde{\pi}^\intercal_2} & = & \begin{pmatrix}
 	0.511429 & 0.428571 
 	\end{pmatrix}. \nonumber \\[2em]
 	\mathbf{\tilde{P}_3} & = &
 	\begin{pmatrix}
 	0.6 & 0.25 & 0.150 \\
 	0.1 & 0.8 & 0.1 \\
 	0.2 & 0.25 & 0.55   
 	\end{pmatrix},\nonumber\\
 	\boldsymbol{\tilde{\pi}^\intercal_3} & = & \begin{pmatrix}
 	 	0.240741 & 0.555555 & 0.203704 
 	\end{pmatrix}. \nonumber
 	\end{eqnarray}
 	\item[Find the Long-Term Probability of Being in a Particular Block:] 	In order to compute the probability of being in a given block in the long run, we have to construct a $3\times 3$ stochastic matrix whose $IJ^{\textit{th}}$ element denotes the probability of a transition from block $I$ to block $J$. This can be done in two steps: First we replace each row of each block $\mathbf{P_{IJ}}$ of matrix $\mathbf{P}$ by the sum of its elements, $\mathbf{P_{IJ}}\mathbf{1}$, and then, we reduce each column subvector  $\mathbf{P_{IJ}}\mathbf{1}$ to a scalar which will represent the total probability of leaving any state of block $I$ to enter any state of block $J$. To determine the latter probability we need to sum the elements of $\mathbf{P_{IJ}}\mathbf{1}$, after each of these elements have been weighted by the conditional probability of being in a particular state of block $I$ given that we are in block $I$. 	Returning to our example and summing along each row of every block of $\mathbf{P}$ we get
 	\begin{displaymath}
 	\begin{pmatrix}
 	0.999 & 0.0009 & 0.0001 \\
 	0.999 & 0.0009 & 0.0001 \\
 	0.9996 & 0.0003 & 0.0001 \\
 	0.0004 & 0.9995 & 0.0001 \\
 	0.0009 & 0.999 & 0.0001 \\
 	5\times 10^{-5} & 5\times 10^{-5} & 0.9999 \\
 	6\times 10^{-5} & 4\times 10^{-5} & 0.9999 \\
 	5\times 10^{-5} & 5\times 10^{-5} & 0.9999   
 	\end{pmatrix}.
 	\end{displaymath} 
	Then reducing each column subvector to a scalar using the stationary distributions found in the first step, yields:
		\begin{equation}
		\mathbf{C}  =   
		\begin{pmatrix}
		\mathbf{\tilde{\pi}^{\intercal}_1} & \mathbf{\tilde{\pi}^{\intercal}_2} & \mathbf{\tilde{\pi}^{\intercal}_3}
		\end{pmatrix}\begin{pmatrix}
		0.999 & 0.0009 & 0.0001 \\
		0.999 & 0.0009 & 0.0001 \\
		0.9996 & 0.0003 & 0.0001 \\
		0.0004 & 0.9995 & 0.0001 \\
		0.0009 & 0.999 & 0.0001 \\
		5\times 10^{-5} & 5\times 10^{-5} & 0.9999 \\
		6\times 10^{-5} & 4\times 10^{-5} & 0.9999 \\
		5\times 10^{-5} & 5\times 10^{-5} & 0.9999   
		\end{pmatrix}  =  \begin{pmatrix}
		0.99911 & 0.00079 & 10^{-4} \\
		0.00061 & 0.99929 & 10^{-4} \\
		5.55\times 10^{-5} & 4.45\times 10^{-5} & 0.9999
		\end{pmatrix}. \nonumber
		\end{equation}
		It can be proved that when $\mathbf{P}$ is an irreducible stochastic matrix, the same thing holds for matrix $\mathbf{C}$, which is many times called the \textit{coupling matrix}. 

		The stationary distribution	of matrix $\mathbf{C}$ in our example is
 	\begin{displaymath}
 	\boldsymbol{\xi^\intercal} = \begin{pmatrix}
 	0.22573 & 0.277427 & 0.5
 	\end{pmatrix}.
 	\end{displaymath}
 	\item[Estimate the Global Solution:] We are now ready to get an approximation to the steady-state distribution of the complete Markov chain: 
 	\begin{displaymath}
 	\boldsymbol{\pi}^\intercal \approx \begin{pmatrix}
 	\xi_1\boldsymbol{\tilde{\pi}^\intercal_1} & \xi_2\boldsymbol{\tilde{\pi}^\intercal_2} & \xi_3\boldsymbol{\tilde{\pi}^\intercal_3}
 	\end{pmatrix}.
 	\end{displaymath}
 \end{description}
 	For our example, the approximation of the global solution is:
 	\begin{displaymath}
 	{\normalsize \boldsymbol{\tilde{\pi}}^\intercal} = {\footnotesize \begin{pmatrix}
 		0.089029 & 0.0929 & 0.040644 & 0.15853 & 0.118897 & 0.12037 & 0.277777 & 0.101852
 		\end{pmatrix}}
 	\end{displaymath}
 	which gives a good approximation of the exact solution:
 	\begin{equation}
 	{\normalsize \boldsymbol{\pi}^\intercal} = {\footnotesize \begin{pmatrix}
 		0.0893 & 0.0928 & 0.0405 & 0.1585 & 0.1189 & 0.1204 & 0.2778 & 0.1018
 		\end{pmatrix}}.
 	\label{eq:ExactSolutionofCourtoisChain}
 	\end{equation}
The complete procedure is summed up in Algorithm~\ref{alg1}. 
	\begin{algorithm}[h!]
		\caption{NCD Decomposition Approximation}
		\label{alg1}
		\begin{algorithmic}[1]
			\Require An irreducible NCD matrix $\mathbf{P}$.
			\Ensure Approximation of its stationary distribution $\mathbf{\tilde{\pi}}$. 
			\State  Perform a stochasticity adjustment of each diagonal block $\mathbf{P_{II}}$, that results in a well-defined irreducible stochastic matrix $\mathbf{\tilde{P}_{I}}$.
			\State Solve independently the systems
			\begin{eqnarray}
			\boldsymbol{\tilde{\pi}^\intercal_I}\mathbf{\tilde{P}_I} & = & \boldsymbol{\tilde{\pi}^\intercal_I}, \nonumber\\
			\boldsymbol{\tilde{\pi}^\intercal_I} \mathbf{1} & = & 1, \nonumber
			\end{eqnarray}
			for each $I=1,2,\dots,L$.
			\State Form the coupling matrix 
			\begin{equation}
			\mathbf{C} = \begin{pmatrix}
			\boldsymbol{\tilde{\pi}^\intercal_1}\mathbf{P_{11}}\mathbf{1} & \boldsymbol{\tilde{\pi}^\intercal_1}\mathbf{P_{12}}\mathbf{1} & \dots & \boldsymbol{\tilde{\pi}^\intercal_1}\mathbf{P_{1L}}\mathbf{1} \\[.3em]
			\boldsymbol{\tilde{\pi}^\intercal_2}\mathbf{P_{21}}\mathbf{1} & \boldsymbol{\tilde{\pi}^\intercal_2}\mathbf{P_{22}}\mathbf{1} & \dots & \boldsymbol{\tilde{\pi}^\intercal_2}\mathbf{P_{2L}}\mathbf{1} \\[.3em]
			\vdots	& \vdots & \ddots & \vdots \\
			\boldsymbol{\tilde{\pi}^\intercal_L}\mathbf{P_{L1}}\mathbf{1} & \boldsymbol{\tilde{\pi}^\intercal_L}\mathbf{P_{L2}}\mathbf{1} & \dots & \boldsymbol{\tilde{\pi}^\intercal_L}\mathbf{P_{LL}} \mathbf{1}
			\end{pmatrix}.\nonumber
			\end{equation}
			\State Solve the system
			\begin{eqnarray}
			\boldsymbol{\xi^\intercal}\mathbf{C} & = & \boldsymbol{\xi^\intercal}, \nonumber\\
			\boldsymbol{\xi^\intercal}\mathbf{1} & = & 1. \nonumber
			\end{eqnarray}
			\State Put together the approximate solution
			\begin{equation}
			\boldsymbol{\tilde{\pi}^\intercal} = \begin{pmatrix}
			\xi_1 \boldsymbol{\tilde{\pi}^\intercal_1} & \xi_2 \boldsymbol{\tilde{\pi}^\intercal_2} & \cdots & \xi_L \boldsymbol{\tilde{\pi}^\intercal_L}
			\end{pmatrix}.\nonumber
			\end{equation}
		\end{algorithmic}
	\end{algorithm}

\begin{remark}
	Notice that the coupling matrix $\mathbf{C}$ that arises from NCD Markov chains, will many times be ill-conditioned, however one can exploit the fact that for any irreducible stochastic matrix $\mathbf{A}$, the matrix 
		\begin{displaymath}
		\mathbf{A}(\alpha) \triangleq \mathbf{I} - \alpha(\mathbf{I}-\mathbf{A}),
		\end{displaymath}
		where $\alpha \in \mathfrak{R}\backslash\{0\}$, has a simple eigenvalue equal to 1 that is associated with a uniquely defined positive left-hand eigenvector, of unit 1-norm, which coincides with the stationary distribution of $\mathbf{A}$ (see~\cite{stewart:1994introduction} for a proof). This way one can ``engineer''  eigenvalues that are more conveniently distributed for iterative solution methods.
\end{remark}

	 \subsubsection{Iterative Aggregation/Disaggregation Methods}
	 Based on the approach presented above, one can develop an algorithm to compute iteratively the exact solution, by incorporating the approximation of each step back into the decomposition procedure. In particular, it was found that applying a power step to the obtained approximation before plugging it back into the decomposition, had a very beneficial effect. Later, this power step was replaced by a block Gauss-Seidel step, which is referred to by Stewart as a \textit{disaggregation step}; with the formation and solution of the coupling matrix $\mathbf{C}$ being the \textit{aggregation step}~\cite{stewart:1994introduction}.  The overall procedure became known as \textit{Iterative Aggregation/Disaggregation} (IAD). 
	 The earliest work along these lines can be traced back to Takahashi~\cite{takahashi1975lumping} and the standard two-level IAD have been studied extensively ever since~\cite{DBLP:journals/jacm/CaoS85,COURTOIS198659,IADHaviv,KMS,Schweitzer1986313}. Convergence proofs for two-level aggregation/disaggregation methods are given in~\cite{IADConvergence,IADMarek2003177}. Extensions to multiple levels of aggregation/disaggregation were first explored in~\cite{IADMultiLEvelHorton,IADMultiLEvelHorton2} and later in~\cite{Pultarová2011354,IADMultiLevelRAMA}.

	 \subsubsection{Stochastic Complementation}
	 \label{subsec:PreliminariesStochasticComplementation}
	 
	 In our discussion in the previous section we mentioned that the stochasticity adjustment of the strictly substochastic diagonal blocks of an NCD matrix can be done in such a way that the results we obtain from the NCD approximation procedure are exact. The completely decomposable blocks obtained by this approach are called \textit{Stochastic Complements}, and even though in most cases their formation requires prohibitive amount of computation\footnote{For a detailed discussion of the computational implications of stochastic complementation see Section 6.2.5 of~\cite{stewart:1994introduction}.}, it is useful to briefly discuss the topic here, firstly for the insight it provides into the theoretical aspects of nearly completely decomposable systems, and secondly because, as we  show in Section~\ref{Ch4:NCDawareRank:Sec:Algorithm}, in some cases careful exploitation of system's symmetries may provide an ``analytical shortcut'' that justifies their use.  	 
	 Below we present the basic definitions and theorems behind stochastic complementation adopting the notation used by Meyer in~\cite{Meyer:1989:SCU:75568.75571}. For proofs of the theorems and further discussion the interested reader may refer there.
	 
	 \paragraph{Definitions}
	 Let us consider an irreducible stochastic matrix $\mathbf{P}$ with an  $L$-level partition 
	 \begin{equation}
	 \mathbf{P} = \begin{pmatrix}
	 \mathbf{P_{11}} & \mathbf{P_{12}} & \dots & \mathbf{P_{1L}} \\
	 \mathbf{P_{21}} & \mathbf{P_{22}} & \dots & \mathbf{P_{2L}} \\
	 \vdots	& \vdots & \ddots & \vdots \\
	 \mathbf{P_{L1}} & \mathbf{P_{L2}} & \dots & \mathbf{P_{LL}} 
	 \end{pmatrix},
	 \label{eq:GeneralPwithPartitioned}
	 \end{equation}
	 in which all the diagonal blocks are square. Let $\mathbf{P_{\star I}}$ denote the $I^{\textit{th}}$ column of blocks from which $\mathbf{P_{II}}$ is excluded,
	 \begin{displaymath}
	 \mathbf{P_{\star I}} = \begin{pmatrix}
	 \mathbf{P_{1I}}\\
	 \vdots \\
	 \mathbf{P_{I-1,I}}\\
	 \mathbf{P_{I+1,I}}\\
	 \vdots\\
	 \mathbf{P_{L1}}
	 \end{pmatrix},
	 \end{displaymath}
	  and $\mathbf{P_{I\star}}$, the $I^{\textit{th}}$  row of blocks from which $\mathbf{P_{II}}$ is excluded, 
	  	 \begin{displaymath}
	  	 \mathbf{P_{I\star}} = \begin{pmatrix}
	  	 \mathbf{P_{I1}} &
	  	 \cdots &
	  	 \mathbf{P_{I,I-1}}&
	  	 \mathbf{P_{I,I+1}}&
	  	 \cdots&
	  	 \mathbf{P_{1L}}
	  	 \end{pmatrix}.
	  	 \end{displaymath}
	   Furthermore, let us use $\mathbf{P^\star_I}$ to denote the principal block submatrix of $\mathbf{P}$ obtained by deleting the $I^{\textit{th}}$ row and $I^{\textit{th}}$ column of blocks from $\mathbf{P}$. 
	 \begin{definition}[Stochastic Complement]  
	 The stochastic complement  $\mathbf{S_I}$ of $\mathbf{P_{II}}$ in $\mathbf{P}$ is defined to be the matrix
	 \begin{eqnarray}
	 \mathbf{S_I} \triangleq \mathbf{P_{II}} + \mathbf{P_{I\star}}(\mathbf{I} - \mathbf{P^\star_I})^{-1}\mathbf{P_{\star I}}	.
	 \end{eqnarray}
	 \end{definition}
	 
	 It has been shown in~\cite{Meyer:1989:SCU:75568.75571}, that every stochastic complement in $\mathbf{P}$ is also an irreducible stochastic matrix. In particular, Meyer showed the following theorems: 
	 \begin{theorem}[Stochasticity of the Complements~\cite{Meyer:1989:SCU:75568.75571}]
	 	Let $\mathbf{P}$ be an irreducible stochastic matrix partitioned as in~(\ref{eq:GeneralPwithPartitioned}). Each stochastic complement,
	 	   	 	\begin{displaymath}
	 	   	 	\mathbf{S_I} = \mathbf{P_{II}} + \mathbf{P_{I\star}}(\mathbf{I} - \mathbf{P^\star_I})^{-1}\mathbf{P_{\star I}},
	 	   	 	\end{displaymath}
	 	   	 	is also a stochastic matrix.
	 \end{theorem}	
	 \begin{theorem}[\cite{Meyer:1989:SCU:75568.75571}]
	 	Let $\mathbf{P}$ be an irreducible stochastic matrix partitioned as in~(\ref{eq:GeneralPwithPartitioned}), and let
	 	\begin{displaymath}
	 	\boldsymbol{\pi^\intercal} = \begin{pmatrix}
	 	\boldsymbol{\pi^\intercal_1} & \boldsymbol{\pi^\intercal_2} & \cdots & \boldsymbol{\pi^\intercal_L}
	 	\end{pmatrix}
	 	\end{displaymath}
	 	be the stationary distribution of $\mathbf{P}$ partitioned according to~(\ref{eq:GeneralPwithPartitioned}). Then the normalized version of each $\mathbf{\pi_I}$, is a stationary distribution of the stochastic complement $\mathbf{S_I}$.
	 \end{theorem}
	 \begin{theorem}[Irreducibility of the Complements~\cite{Meyer:1989:SCU:75568.75571}]
	 		Let $\mathbf{P}$ be an irreducible stochastic matrix partitioned as in~(\ref{eq:GeneralPwithPartitioned}). Then each stochastic complement,
	 		\begin{displaymath}
	 		\mathbf{S_I} = \mathbf{P_{II}} + \mathbf{P_{I\star}}(\mathbf{I} - \mathbf{P^\star_I})^{-1}\mathbf{P_{\star I}},
	 		\end{displaymath}
	 		is also an irreducible stochastic matrix.
	 \end{theorem}
	 \paragraph{Computing Stationary Distributions by Stochastic Complementation}
	 When the stochastic complements have been formed, one can compute the exact stationary distribution of the complete stochastic matrix $\mathbf{P}$ following the procedure outlined in the previous section. The fact that the final solution yielded by Algorithm~\ref{alg1} is exact, follows directly from the following two theorems.
	 \begin{theorem}[Aggregation~\cite{Meyer:1989:SCU:75568.75571}]
	 	An irreducible Markov chain whose states can be partitioned into $L$ clusters
	 	\begin{displaymath}
	 	\{1,2,\dots,n\} = \mathcal{S}_1\cup \mathcal{S}_2\cup \cdots \cup \mathcal{S}_L,
	 	\end{displaymath}
	 	can be compressed in a smaller $L$-state aggregated chain whose states are the individual clusters $\mathcal{S}_I$. The transition probability matrix $\mathbf{C}$ of the aggregated chain is called the \textbf{coupling matrix}\index{Coupling Matrix} and it is defined by
	 		\begin{equation}
	 		\mathbf{C} \triangleq \begin{pmatrix}
	 		\mathbf{s^\intercal_1}\mathbf{P_{11}}\mathbf{1} & \mathbf{s^\intercal_1}\mathbf{P_{12}}\mathbf{1} & \dots & \mathbf{s^\intercal_1}\mathbf{P_{1L}}\mathbf{1} \\[.3em]
	 		\mathbf{s^\intercal_2}\mathbf{P_{21}}\mathbf{1} & \mathbf{s^\intercal_2}\mathbf{P_{22}}\mathbf{1} & \dots & \mathbf{s^\intercal_2}\mathbf{P_{2L}}\mathbf{1} \\[.3em]
	 		\vdots	& \vdots & \ddots & \vdots \\
	 		\mathbf{s^\intercal_L}\mathbf{P_{L1}}\mathbf{1} & \mathbf{s^\intercal_L}\mathbf{P_{L2}}\mathbf{1} & \dots & \mathbf{s^\intercal_L}\mathbf{P_{LL}} \mathbf{1}
	 		\end{pmatrix},
	 		\end{equation}
	 		where $\mathbf{s_I}$ is the stationary distribution of the stochastic complement $\mathbf{S_I}$. \\ Furthermore, if $Y_t$ is the cluster occupied by the original chain at time $t$, then for ergodic chains, the aggregated transition probability $C_{IJ} = \mathbf{s^\intercal_I}\mathbf{P_{IJ}}\mathbf{1} $ can be expressed as 
	 		\begin{displaymath}
	 		C_{IJ} = \lim\limits_{t \to \infty} \Pr \{Y_{t+1} = J|Y_t = I\}.
	 		\end{displaymath}
	 \end{theorem}
	 \begin{theorem}[Disaggregation~\cite{Meyer:1989:SCU:75568.75571}] \label{Theorem:Dissagregation}
	 	If $\mathbf{P}$ is an irreducible stochastic matrix with an  $L$-level partition 
	 	\begin{displaymath}
	 	\mathbf{P} = \begin{pmatrix}
	 	\mathbf{P_{11}} & \mathbf{P_{12}} & \dots & \mathbf{P_{1L}} \\
	 	\mathbf{P_{21}} & \mathbf{P_{22}} & \dots & \mathbf{P_{2L}} \\
	 	\vdots	& \vdots & \ddots & \vdots \\
	 	\mathbf{P_{L1}} & \mathbf{P_{L2}} & \dots & \mathbf{P_{LL}} 
	 	\end{pmatrix},
	 	\end{displaymath}
	 	with square diagonal blocks, then the stationary distribution vector for $\mathbf{P}$ is given by
	 	\begin{displaymath}
	 	{\boldsymbol{\pi}^\intercal} = \begin{pmatrix}
	 	\xi_1 \mathbf{s^\intercal_1} & \xi_2 \mathbf{s^\intercal_2}  & \cdots & \xi_L \mathbf{s^\intercal_L}
	 	\end{pmatrix},
	 	\end{displaymath}
	 	where $\mathbf{s_I}$ is the unique stationary distribution of the stochastic complement
	 	\begin{displaymath}
	 	\mathbf{S_I} = \mathbf{P_{II}} + \mathbf{P_{I\star}}(\mathbf{I} - \mathbf{P^\star_I})^{-1}\mathbf{P_{\star I}},
	 	\end{displaymath}
	 	and where 
	 	\begin{displaymath}
	 	\boldsymbol{\xi^\intercal} = \begin{pmatrix}
	 	\xi_1 & \xi_2  & \cdots & \xi_L \end{pmatrix},
	 	\end{displaymath}
	 	is the unique stationary distribution vector for the $L\times L$ irreducible coupling matrix $\mathbf{C}$. 
	 \end{theorem}
	 
	 \paragraph{Example}
	 For the sake of example, the stochastic complements of the blocks for the Courtois matrix and the corresponding stationary distributions are:
	 \begin{eqnarray}
	 \mathbf{S_1} & = & \begin{pmatrix}
     0.8503 &    0.0004 &    0.1493 \\
     0.1003 &    0.6504 &   0.2493 \\
     0.1001 &    0.8002 &   0.0997
	 \end{pmatrix},\nonumber\\
	 \mathbf{s^\intercal_1} & = & \begin{pmatrix}
	 0.4012  &  0.4168  &  0.1819
	 \end{pmatrix} .\nonumber \\[2em]
	 \mathbf{S_{2}} & = &
	 \begin{pmatrix}
      0.7003 &    0.2997 \\
	  0.3995 &   0.6005 
	 \end{pmatrix},\nonumber\\
	 \mathbf{s^\intercal_2} & = & \begin{pmatrix}
	 0.5713  &  0.4287
	 \end{pmatrix}. \nonumber \\[2em]
	 \mathbf{S_3} & = &
	 \begin{pmatrix}
    0.6000  & 0.2499 &  0.1500 \\
    0.1000  & 0.8000 &  0.0999 \\
    0.1999  & 0.2500 &  0.5500
	 \end{pmatrix},\nonumber\\
	 \mathbf{s^\intercal_3} & = &\begin{pmatrix}
	     0.2408  &  0.5556 &   0.2036
	 \end{pmatrix}. \nonumber
	 \end{eqnarray}

	 The corresponding coupling matrix is
	 \begin{eqnarray}
	 \mathbf{C} & = & \begin{pmatrix}
	 \mathbf{s^\intercal_1}\mathbf{P_{11}}\mathbf{1} & \mathbf{s^\intercal_1}\mathbf{P_{12}}\mathbf{1} &  \mathbf{s^\intercal_1}\mathbf{P_{13}}\mathbf{1} \\[.3em]
	 \mathbf{s^\intercal_2}\mathbf{P_{21}}\mathbf{1} & \mathbf{s^\intercal_2}\mathbf{P_{22}}\mathbf{1} &  \mathbf{s^\intercal_2}\mathbf{P_{23}}\mathbf{1} \\[.3em]
	 \mathbf{s^\intercal_3}\mathbf{P_{31}}\mathbf{1} & \mathbf{s^\intercal_3}\mathbf{P_{32}}\mathbf{1} &
	 \mathbf{s^\intercal_3}\mathbf{P_{33}} \mathbf{1}
	 \end{pmatrix} \nonumber\\[1em]
	 & = & \begin{pmatrix}
	     0.9991  &  0.0008 &   0.0001 \\
	     0.0006  &  0.9993 &   0.0001 \\
	     0.0001  &  0.0000 &   0.9999
	 \end{pmatrix},
	 \end{eqnarray}
	 and its stationary distribution
	 \begin{displaymath}
	 \boldsymbol{\xi^\intercal} = \begin{pmatrix}
	 0.2225  &  0.2775 &   0.5000
	 \end{pmatrix},
	 \end{displaymath}
	  yielding a final solution
	  \begin{eqnarray}
	  	 	{\boldsymbol{\pi}^\intercal} & = & \begin{pmatrix}
	  	 	\xi_1 \mathbf{s^\intercal_1} & \xi_2 \mathbf{s^\intercal_2}  & \cdots & \xi_L \mathbf{s^\intercal_L}
	  	 	\end{pmatrix} \nonumber \\
	  	 	& = & \begin{pmatrix}	  
	  0.0893  &  0.0928  &  0.0405  &  0.1585 &   0.1189 &   0.1204 &    0.2778 &    0.1018
	  \end{pmatrix}.\nonumber
	  \end{eqnarray}
	  
	  \subsubsection{Closing Remarks}
	  In this section, we have presented briefly some basic theory regarding nearly completely decomposable stochastic systems. For a proof of the fundamental Simon and Ando’s theorems, the interested reader is referred to~\cite{simon1961aggregation} (or to~\cite{ModernProofofSimonAndoTheorems} for a more modern proof). For a rigorous analysis of the degree of approximation one gets following the simple aggregation procedure presented intuitively in ~Section \ref{Ch:Preliminaries:NCDApproximation}, the reader is referred to the second chapter of Courtois’ monograph~\cite{courtois1977decomposability}, which represents a classic and in our opinion beautiful treatment of the subject. For a thorough discussion of the computational implications of NCD stochastic matrices, maybe the best starting point is~\cite{stewart:1994introduction}, which also covers the basics of Meyer’s stochastic complementation theory, as well as a nice discussion of its relationship with block Gaussian elimination.  

\newpage

	\section{Characteristics of NCDaware Handling of Dangling Nodels}
\label{SI:dangling}

 Below we discuss several desirable properties of the NCDaware dangling node patching strategy, using the Web ranking application as the primary vehicle of exposition. 
\begin{description}
	\item[More Realistic Modeling.] The proposed strategy provides different random surfing behavior depending on the origin dangling node. In particular, when the random surfer leaves a particular dangling node he jumps to the nodes of the  NCD block (or blocks), that contains it. Assuming, for example Web-ranking as an application, and that the criterion behind the definition of the NCD blocks is the partition of Web-pages into websites, this strategy is intuitively closer to the idea of a back button in a browser, without any of the mathematical complications~\cite{fagin2000random,fagin2001}. 
	\item[Importance Propagation.] Under the strategy we propose, the importance of the dangling nodes \textit{is propagated back to its affiliated nodes}. For example, if a high-ranking document has a large number of incoming links due to the quality of its content, its relatively big importance will be propagated to the NCD block (e.g. the website) that contains it, rather than scattered throughout the Web in a simplistic and meaningless manner. In this way, its affiliated pages are being acknowledged in a tangible and quantifiable way, as they should.  
	\item[Lower Susceptibility to Link Spamming.] Typically the patching of the dangling nodes is approached in a uniform manner. 
	This can be exploited by spamming groups of nodes putting them in a position to be able to achieve disproportionately big ranking scores simply by creating enough artificial nodes (e.g., crawlable pages for web ranking applications) designed to funnel all the rank towards particular sets of nodes. Of course the very existence of the traditional teleportation matrix increases link-spamming susceptibility as well, however in case of strongly preferential patching the random surfer is left with no alternatives than to teleport with probability 1. Taking into account the fact that in certain applications dangling nodes constitute a large fraction of the overall nodes of the network (see e.g., \cite{Eiron:2004:RWF:988672.988714}) we see that the problem of direct ranking manipulation through link-spamming becomes even worse. 
	
	In our strategy the dangling nodes funnel their rank to their affiliated nodes instead. Hence, spamming nodes can only hope to gain rank through the teleportation model, the probability of following which, is controlled by our model. As a result, our strategy alleviates this problem, making the overall ranking measure less susceptible to link-spamming. In our experiments, we will see that this effect is confirmed by a number of tests using real-world datasets as well. 
	\item[Confined Handling.] Many real-world directed networks are reducible (e.g., the  Web-graph~\cite{LangvilleMeyer06,pagerank}); therefore the traditional patching of the dangling nodes in a sense ``artificially connects'' parts of the graph that are actually disconnected. With our strategy, and for reasonable decompositions of the underlying network, this is not case. For example, if one uses the partitioning of the Web into websites as a criterion of decomposition, the NCD blocks typically correspond to weakly connected subnetwork, therefore, our patching strategy ``respects'' the connectivity properties of the underlying network. As we explore in depth in Section \ref{Ch4:NCDawareRank:Sec:Algorithm}, this property entails a wealth of mathematically elegant and computationally useful implications, that can lead to an efficient parallel algorithm for the computation of the NCDawareRank vector, with our results being directly applicable to the computation of standard PageRank as well (subject to the adoption of a similarly confined strategy for handling the dangling nodes). 
\end{description}

\newpage
		
	\section{Example Network Decompositions and Corresponding NCDawareRank Matrices}
	\label{App:ExampleInter-Level}

\begin{figure}[h!]
			\centering
			\includegraphics[height=5.5cm]{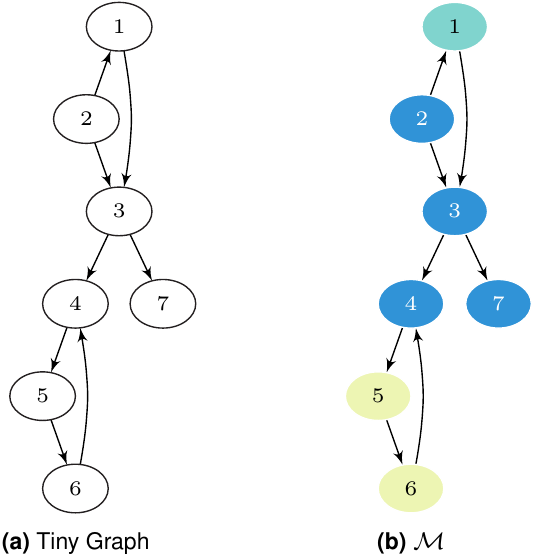}
			\caption[Example: Decomposition leading to a reducible indicator matrix.]{In the left figure we see a tiny graph that admits a decomposition $\mathcal{M}$ highlighted in the right subfigure. Same colored nodes  belong to the same block and are considered related according to a given criterion.}
			\label{Example:SinglePrimitive}
		\end{figure}

	To illustrate the above discussion, we give the following examples. First consider the graph of Fig.~\ref{Example:SinglePrimitive}(a) that can be decomposed as seen in Fig.~\ref{Example:SinglePrimitive}(b).
	
	The corresponding matrices $\mathbf{R,A}$ are the following:
 \begin{eqnarray}
		\mathbf{R} & = & 	{\small\begin{pmatrix}
		1/2 & 1/2 & 0 \\
		1/2 & 1/2 & 0 \\
		0 & 1 & 0\\
		0 & 1/2 & 1/2 \\
		0 & 0 & 1 \\
		0 & 1/2 & 1/2 \\
		0 & 1 & 0  
		\end{pmatrix} }, \qquad
		\mathbf{A}   =  {\small \begin{pmatrix}
		1& 0 & 0 & 0 & 0 & 0 & 0 \\
		0 & 1/4 & 1/4 & 1/4 & 0 & 0 & 1/4  \\
		0 & 0 & 0 & 0 & 1/2 & 1/2 & 0  
		\end{pmatrix}},\nonumber
		\end{eqnarray}
	and the indicator matrix $\mathbf{W}$ is
	\begin{eqnarray}
	\mathbf{W} &=& \mathbf{AR}
	= \begin{pmatrix}
	1/2 & 1/2 & 0\\
	1/8 & 3/4 & 1/8 \\
	0 & 1/4 & 3/4
	\end{pmatrix} ,
	\end{eqnarray}
	which is an irreducible matrix. Thus, the proposed decomposition  satisfies the criterion of Theorem~\ref{thm:PrimitivitySinleDecomposition}, and a well-defined random surfer model is produced without resorting to the uniform teleportation matrix.
	
\begin{figure}[h!]
		\centering
		\includegraphics[height=5.5cm]{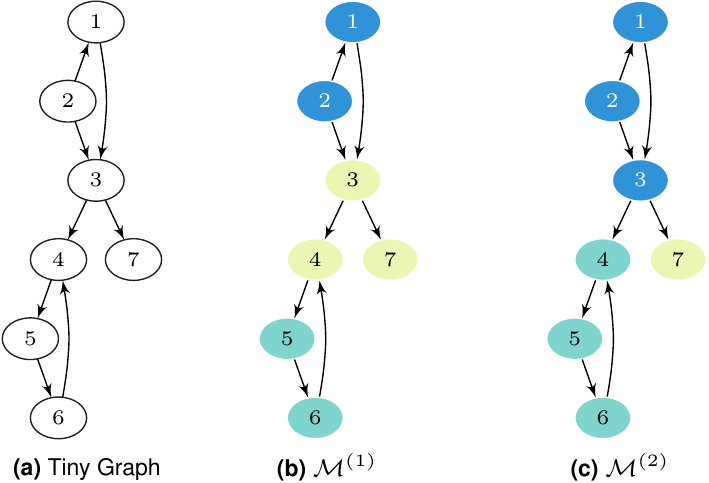}
		\caption[Example: Achieving primitivity through multiple decompositions]{In the left figure we see the tiny graph of our previous example that can be decomposed as seen in the subfigures (b) and (c).}
		\label{Example:MultiplePrimitive}
	\end{figure}

	\bigskip
	Let us now consider an example where multiple decompositions can be defined. In Fig.~\ref{Example:MultiplePrimitive}, we consider the same graph of our previous example, that admits two new decompositions $\mathcal{M}^{(1)},\mathcal{M}^{(2)}$.

	For the first decomposition, we have
	\begin{eqnarray}
		\mathbf{R_1} &=& 
	{\small 	\begin{pmatrix}
		1/2 & 1/2 & 0 \\
		1/2 & 1/2 & 0 \\
		0 & 1 & 0 \\
		0 & 1/2 & 1/2 \\
		0 & 0 & 1  \\
		0 & 1/2 & 1/2  \\
		0 & 1 & 0    
		\end{pmatrix}},
		\quad
		\mathbf{A_1}= 
	{\small 	\begin{pmatrix} 
		1/2 & 1/2 & 0 & 0 & 0 & 0 & 0 \\
		0 & 0 & 1/3 & 1/3 & 0 & 0 & 1/3 \\ 
		0 & 0 & 0 & 0 & 1/2 & 1/2 & 0 
		\end{pmatrix} },\nonumber
		\end{eqnarray}
	with indicator matrix,% $\mathbf{W_1}$
	\begin{eqnarray}
	\mathbf{W_1} & = & \begin{pmatrix}
	1/2 & 1/2 & 0 \\
	0 & 5/6 & 1/6 \\
	0 & 1/4 & 3/4   
	\end{pmatrix},
	\end{eqnarray}
	which is reducible. Thus, the first decomposition alone does not satisfy Theorem~\ref{thm:PrimitivitySinleDecomposition}. For the second decomposition we have
	\begin{eqnarray}
		\mathbf{R_2} &=& 
		{\small \begin{pmatrix} 
		1 & 0 & 0 \\
		1 & 0 & 0 \\
		1/3 & 1/3 & 1/3 \\
		0 & 1 & 0 \\
		0 & 1 & 0 \\
		0 & 1 & 0 \\
		0 & 0 & 1    
		\end{pmatrix}},
		\quad
		\mathbf{A_2}= 
	{\small 	\begin{pmatrix} 
		1/3 & 1/3 & 1/3 & 0 & 0 & 0 & 0 \\
		0 & 0 & 0 & 1/3 & 1/3 & 1/3 & 0 \\ 
		0 & 0 & 0 & 0 & 0 & 0 & 1 
		\end{pmatrix}},\nonumber
		\end{eqnarray}
	with indicator matrix %$\mathbf{W_2}$
	\begin{eqnarray}
	\mathbf{W_2} & = & \begin{pmatrix}
	7/9 & 1/9 & 1/9 \\
	0 & 1 & 0 \\
	0 & 0 & 1   
	\end{pmatrix},
	\end{eqnarray}	
	which is again reducible. Therefore, neither decomposition $\mathcal{M}^{(2)}$ alone could ensure the primitivity of the final stochastic matrix $\mathbf{P}$. 
	
	However, if one applies these two decompositions together, the final stochastic matrix that corresponds to the random surfing model
	\begin{displaymath}
	\mathbf{P} = \eta \mathbf{H} + \mu_1\mathbf{M_1} + \mu_2\mathbf{M_2} 
	\end{displaymath}
	becomes primitive, since matrix $\mathbf{W'}$:
	\begin{eqnarray}
	\mathbf{W'} &=& 
	\begin{pmatrix}
	\mathbf{A_1}\\
	\mathbf{A_2}
	\end{pmatrix}
	\begin{pmatrix}
	\mathbf{R_1} & \mathbf{R_2}
	\end{pmatrix} 
	\nonumber\\
	& = &
{\small 	\begin{pmatrix}
	1/2 & 1/2 & 0 & 0 & 0 & 0 & 0 \\
	0 & 0 & 1/3 & 1/3 & 0 & 0 & 1/3 \\ 
	0 & 0 & 0 & 0 & 1/2 & 1/2 & 0 \\
	1/3 & 1/3 & 1/3 & 0 & 0 & 0 & 0 \\
	0 & 0 & 0 & 1/3 & 1/3 & 1/3 & 0 \\ 
	0 & 0 & 0 & 0 & 0 & 0 & 1 
	\end{pmatrix}\begin{pmatrix}
	1/2 & 1/2 & 0 & 1 & 0 & 0 \\
	1/2 & 1/2 & 0 & 1 & 0 & 0 \\
	0 & 1 & 0 & 1/3 & 1/3 & 1/3 \\
	0 & 1/2 & 1/2 & 0 & 1 & 0 \\
	0 & 0 & 1  & 0 & 1 & 0 \\
	0 & 1/2 & 1/2  & 0 & 1 & 0 \\
	0 & 1 & 0  & 0 & 0 & 1   
	\end{pmatrix} }\nonumber \\
	& = & {\small \begin{pmatrix}
	1/2 & 1/2 & 0 & 1 & 0 & 0 \\
	0 & 5/6 & 1/6 & 1/9 & 4/9 & 4/9 \\
	0 & 1/4 & 3/4 & 0 & 1 & 0 \\
	1/2 & 2/3 & 0 & 7/9 & 1/9 & 1/9\\
	0 & 1/3 & 2/3 & 0 & 1 & 0\\
	0 & 1 & 0 & 0 & 0 & 1
	\end{pmatrix}},
	\end{eqnarray}
	is irreducible, and therefore it satisfies the criterion of Theorem~\ref{thm:PrimitivityMultipleDecomposition}.

	\begin{remark}
		Notice here that the conditions of Theorem~\ref{Theorem:PrimitivityConditions} are not necessary for the primitivity of the final stochastic matrix. For example for decompositions given in Fig.~\ref{Example:MultiplePrimitive} none of the conditions of Theorem~\ref{Theorem:PrimitivityConditions} is verified since: 
		\begin{eqnarray}
		\mathbf{A_1R_2} & = & \begin{pmatrix}
		1 & 0 & 0 \\
		1/9 & 4/9 & 4/9 \\
		0 & 1 & 0  
		\end{pmatrix}\ngtr \mathbf{0}, \nonumber \\
		\mathbf{A_2R_1} & = & \begin{pmatrix}
		1/2 & 2/3 & 0 \\
		0 & 1/3 & 2/3 \\
		0 & 1 & 0  
		\end{pmatrix}\ngtr \mathbf{0}, \nonumber
		\end{eqnarray} and we have seen that $\mathbf{W_1}$ and $\mathbf{W_2}$ are both reducible. Therefore, even if these conditions are not met, the criterion of Theorem~\ref{thm:PrimitivityMultipleDecomposition} should be checked.
	\end{remark}

	  \newpage 
	  
	\section{Numerical Example of NCDawareRank Centrality Computation}
	\label{App:NCDawareRankComputation}
	To illustrate the above algorithm we give the following numerical example. Let us consider a tiny Web-graph with adjacency matrix 
		 \begin{displaymath}
		 {\small \begin{pmatrix}
		 	0 & 1 & 0 & 0 & 0 & 0 & 0 & 0\\
		 	0 & 0 & 1 & 1 & 0 & 0 & 0 & 0\\
		 	0 & 1 & 0 & 1 & 0 & 0 & 0 & 0\\
		 	0 & 0 & 0 & 0  & 0 & 0 & 0 & 0\\
		 	0 & 0 & 0 & 0 & 0 & 1 & 1 & 1\\
		 	0 & 0 & 0 & 0 & 0 & 0 & 0 & 0 \\
		 	0 & 0 & 0 & 0 & 0 & 0 & 0 & 0 \\
		 	0 & 0 & 0 & 0 & 1 & 0 & 0 & 0 \\
		 	\end{pmatrix}},
		 \end{displaymath}
		together with decomposition into 4 NCD blocks 
		\begin{displaymath}
		\{\mathcal{A}_1,\mathcal{A}_2,\mathcal{A}_3,\mathcal{A}_4,\} \equiv \{\{v_1,v_2\},\{v_3,v_4\},\{v_5,v_6,v_7\},\{v_8\}\}.
		\end{displaymath} 
		The graph has three dangling nodes, that correspond to zero rows of the adjacency matrix, therefore a stochasticity adjustment is needed and it will be done using our dangling strategy. In particular, we have 
	 \begin{eqnarray}
		\mathbf{H} & = & {\scriptsize \begin{pmatrix}
		0 & 1 & 0 & 0 & 0 & 0 & 0 & 0\\
		0 & 0 & 1/2 & 1/2 & 0 & 0 & 0 & 0\\
		0 & 1/2 & 0 & 1/2 & 0 & 0 & 0 & 0\\
		0 & 0 & 0 & 0  & 0 & 0 & 0 & 0\\
		0 & 0 & 0 & 0 & 0 & 1/3 & 1/3 & 1/3\\
		0 & 0 & 0 & 0 & 0 & 0 & 0 & 0 \\
		0 & 0 & 0 & 0 & 0 & 0 & 0 & 0 \\
		0 & 0 & 0 & 0 & 1 & 0 & 0 & 0 \\
		\end{pmatrix}} +  %\\[1em]
%	\mathbf{H_{D}} & = & 
%	 & & 
	 {\scriptsize \begin{pmatrix}
		0 & 0 & 0 & 0 & 0 & 0 & 0 & 0\\
		0 & 0 & 0 & 0 & 0 & 0 & 0 & 0\\
		0 & 0 & 0 & 0 & 0 & 0 & 0 & 0\\
		0 & 0 & 1/2 & 1/2  & 0 & 0 & 0 & 0\\
		0 & 0 & 0 & 0 & 0 & 0 & 0 & 0\\
		0 & 0 & 0 & 0 & 1/3 & 1/3 & 1/3 & 0 \\
		0 & 0 & 0 & 0 & 1/3 & 1/3 & 1/3 & 0 \\
		0 & 0 & 0 & 0 & 0 & 0 & 0 & 0 \\
		\end{pmatrix}} \nonumber \\[1em]
	& = & {\scriptsize \begin{pmatrix}
		0 & 1 & 0 & 0 & 0 & 0 & 0 & 0\\
		0 & 0 & 1/2 & 1/2 & 0 & 0 & 0 & 0\\
		0 & 1/2 & 0 & 1/2 & 0 & 0 & 0 & 0\\
		0 & 0 & 1/2 & 1/2  & 0 & 0 & 0 & 0\\
		0 & 0 & 0 & 0 & 0 & 1/3 & 1/3 & 1/3\\
		0 & 0 & 0 & 0 & 1/3 & 1/3 & 1/3 & 0 \\
		0 & 0 & 0 & 0 & 1/3 & 1/3 & 1/3 & 0 \\
		0 & 0 & 0 & 0 & 1 & 0 & 0 & 0 \\
		\end{pmatrix}}.
		\end{eqnarray}
 The factors of the inter-level proximity matrix $\mathbf{M}$, defined using the above decomposition into 4 NCD blocks are 
	 \begin{displaymath}
		\mathbf{R} = {\small\begin{pmatrix}
		1 & 0 & 0  & 0\\
		1/2 & 1/2 & 0 & 0\\
		1/2 & 1/2 & 0 & 0\\
		0 & 1 & 0 & 0\\
		0 & 0 & 1/2 & 1/2\\
		0 & 0 & 1 & 0\\
		0 & 0 & 1 & 0\\
		0 & 0 & 0 & 1\\
		\end{pmatrix}},
		\qquad
		\mathbf{A} = {\small \begin{pmatrix}
		1/2 & 1/2 & 0 & 0 & 0 & 0 & 0 & 0\\
		0 & 0 & 1/2 & 1/2 & 0 & 0 & 0 & 0 \\
		0 & 0 & 0 & 0 & 1/3 & 1/3 & 1/3 & 0 \\
		0 & 0 & 0 & 0 & 0 & 0 & 0 & 1\\
		\end{pmatrix} }.
		\end{displaymath}
	Choosing  $\eta = 0.85, \mu = 0.1$ and teleportation vector $\mathbf{v} = \tfrac{1}{8}\mathbf{1}$, the final NCDawareRank matrix $\mathbf{P}$ will be given by
	 \begin{eqnarray}
		\mathbf{P} & = & 0.85\mathbf{H} + 0.1\mathbf{RA} + 0.05 \frac{1}{8}	\mathbf{1}\mathbf{1^\intercal} \nonumber \\
		& = & {\small\begin{pmatrix}
		0.05625&0.90625&0.00625&0.00625&0.00625&0.00625&0.00625&0.00625\\
		0.03125&0.03125&0.45625&0.45625&0.00625&0.00625&0.00625&0.00625\\
		0.03125&0.45625&0.03125&0.45625&0.00625&0.00625&0.00625&0.00625\\
		0.00625&0.00625&0.48125&0.48125&0.00625&0.00625&0.00625&0.00625\\
		0.00625&0.00625&0.00625&0.00625&0.02292&0.30625&0.30625&0.33958\\
		0.00625&0.00625&0.00625&0.00625&0.32292&0.32292&0.32292&0.00625\\
		0.00625&0.00625&0.00625&0.00625&0.32292&0.32292&0.32292&0.00625\\
		0.00625&0.00625&0.00625&0.00625&0.85625&0.00625&0.00625&0.10625\\
		\end{pmatrix}}. \nonumber
		\end{eqnarray}
	Notice that the model is block-level separable with respect to the partition of the NCD blocks into two aggregates: 
	\begin{displaymath}
	\mathcal{B} = \{\{\mathcal{A}_1,\mathcal{A}_2\},\{\mathcal{A}_3,\mathcal{A}_4\}\} = \{\{v_1,v_2,v_3,v_4\}\{v_5,v_6,v_7,v_8\}\} = \{\mathcal{B}_1,\mathcal{B}_2\}.
	\end{displaymath} 
	Therefore, we have
	 \begin{displaymath}
	 \mathbf{P} = {\small\left(\begin{array}{cccc;{3pt/2pt}cccc} 
	 	0.05625&0.90625&0.00625&0.00625&0.00625&0.00625&0.00625&0.00625\\
	 	0.03125&0.03125&0.45625&0.45625&0.00625&0.00625&0.00625&0.00625\\
	 	0.03125&0.45625&0.03125&0.45625&0.00625&0.00625&0.00625&0.00625\\
	 	0.00625&0.00625&0.48125&0.48125&0.00625&0.00625&0.00625&0.00625\\
	 	\hdashline[3pt/2pt]
	 	\addlinespace[.3em]
	 	0.00625&0.00625&0.00625&0.00625&0.02292&0.30625&0.30625&0.33958\\
	 	0.00625&0.00625&0.00625&0.00625&0.32292&0.32292&0.32292&0.00625\\
	 	0.00625&0.00625&0.00625&0.00625&0.32292&0.32292&0.32292&0.00625\\
	 	0.00625&0.00625&0.00625&0.00625&0.85625&0.00625&0.00625&0.10625\\
	 	\end{array}\right)},
	 \end{displaymath}
	and the maximum degree of coupling between the aggregates is 
	\begin{math}
	\varepsilon  = 	  0.025 .
	\end{math}
	The stochastic complements $\mathbf{S_1,S_2}$ using their definition are  
	 \begin{displaymath}
		\mathbf{S_1}  =   \mathbf{P_{11}} + \mathbf{P_{1\star}}(\mathbf{I} - \mathbf{P^\star_1})^{-1}\mathbf{P_{\star 1}}  = {\small \begin{pmatrix}
		0.0625&0.9125&0.0125&0.0125\\
		0.0375&0.0375&0.4625&0.4625\\
		0.0375&0.4625&0.0375&0.4625\\
		0.0125&0.0125&0.4875&0.4875\\
		\end{pmatrix}} , 
		\end{displaymath}
	and
	 \begin{displaymath}
		\mathbf{S_2}  =   \mathbf{P_{22}} + \mathbf{P_{2\star}}(\mathbf{I} - \mathbf{P^\star_2})^{-1}\mathbf{P_{\star 2}} 
		 = {\small \begin{pmatrix}
		0.029167&0.3125&0.3125&0.34583\\
		0.32917&0.32917&0.32917&0.0125\\
		0.32917&0.32917&0.32917&0.0125\\
		0.8625&0.0125&0.0125&0.1125\\
		\end{pmatrix}}.  
	\end{displaymath}
	Theorem~\ref{thm:stochasticcomplements} predicts that these stochastic complements can be expressed as NCDawareRank models of the subgraphs defined by the aggregates
	\begin{displaymath}
	\{v_1,v_2,v_3,v_4\} \quad \text{and} \quad \{v_5,v_6,v_7,v_8\},
	\end{displaymath}
   using the same parameters $\eta,\mu$ and the normalized corresponding part of the initial teleportation vector $\mathbf{v}$. Indeed,
	\begin{eqnarray}
	\mathbf{P_1} & = & \eta\mathbf{H_{11}} + \mu\mathbf{M_{11}} + (1-\eta-\mu)\dfrac{1}{\mathbf{v^\intercal_1}\mathbf{1}}\mathbf{1}\mathbf{v^\intercal_1} \nonumber \\
	& = &  0.85 {\small \begin{pmatrix}
	        0  &  1  & 0   &  0 \\
	        0  &  0  & 1/2 & 1/2 \\
	        0  & 1/2 & 0   & 1/2 \\
	        0  &  0  & 1/2 & 1/2
	\end{pmatrix}} + 0.1 {\small \begin{pmatrix}
	    1/2 & 1/2 & 0 & 0 \\
	    1/4 & 1/4 & 1/4 & 1/4 \\
	    1/4 & 1/4 & 1/4 & 1/4 \\
	    0   &   0 & 1/2 & 1/2  
	\end{pmatrix} }\nonumber \\
	& & + 0.05\frac{1}{1/2}{\small \begin{pmatrix}
	1 \\ 1 \\ 1 \\ 1
	\end{pmatrix} \begin{pmatrix}
	1/4 & 1/4 & 1/4 & 1/4
	\end{pmatrix}}\nonumber \\
	& = & {\small \begin{pmatrix}
		0.0625&0.9125&0.0125&0.0125\\
		0.0375&0.0375&0.4625&0.4625\\
		0.0375&0.4625&0.0375&0.4625\\
		0.0125&0.0125&0.4875&0.4875\\
		\end{pmatrix}}  \nonumber \\
	& \equiv & \mathbf{S_1},
	\end{eqnarray}
	and
	\begin{eqnarray}
	\mathbf{P_2} & = & \eta\mathbf{H_{22}} + \mu\mathbf{M_{22}} + (1-\eta-\mu)\dfrac{1}{\mathbf{v^\intercal_2}\mathbf{1}}\mathbf{1}\mathbf{v^\intercal_2} \nonumber \\
	& = &  0.85 {\small \begin{pmatrix}
	         0   & 1/3 & 1/3 & 1/3 \\
	         1/3 & 1/3 & 1/3 &  0  \\
	         1/3 & 1/3 & 1/3 &  0  \\
	         1   &   0 &   0 &  0
	        \end{pmatrix}} + 0.1 {\small \begin{pmatrix}
	       1/6   &  1/6  &  1/6  &   1/2 \\    
	       1/3   &  1/3  &  1/3  &     0 \\      
	       1/3   &  1/3  &  1/3  &     0 \\      
	       0     &    0  &    0  &     1    
		\end{pmatrix} }\nonumber \\
	& & + 0.05\frac{1}{1/2}{\small \begin{pmatrix}
		1 \\ 1 \\ 1 \\ 1
		\end{pmatrix} \begin{pmatrix}
		1/4 & 1/4 & 1/4 & 1/4
		\end{pmatrix}}\nonumber \\
	& = & {\small \begin{pmatrix}
		0.029167&0.3125&0.3125&0.34583\\
		0.32917&0.32917&0.32917&0.0125\\
		0.32917&0.32917&0.32917&0.0125\\
		0.8625&0.0125&0.0125&0.1125\\
		\end{pmatrix}} \nonumber \\
	& \equiv & \mathbf{S_2},\nonumber
	\end{eqnarray}
	as predicted.
	
	The unique stationary distributions of $\mathbf{S_1, S_2}$ are
	\begin{eqnarray}
	\mathbf{s^\intercal_1} & = & \begin{pmatrix}
	0.0266 & 0.1870 & 0.3243 & 0.4621
	\end{pmatrix}, \nonumber \\
	\mathbf{s^\intercal_2} & = & \begin{pmatrix}
	0.3053 & 0.2839 & 0.2839 & 0.1270
	\end{pmatrix}. \nonumber
	\end{eqnarray} 
	The corresponding coupling matrix will be given by
	\begin{equation}
	\mathbf{C}  =  \begin{pmatrix}
	\mathbf{s^\intercal_1}\mathbf{P_{11}}\mathbf{1} & \mathbf{s^\intercal_1}\mathbf{P_{12}}\mathbf{1} \\[0.3em]
	\mathbf{s^\intercal_2}\mathbf{P_{21}}\mathbf{1} & \mathbf{s^\intercal_2}\mathbf{P_{22}}\mathbf{1}  
	\end{pmatrix}  =  \begin{pmatrix}
	0.95 & 0.05 \\
	0.05 & 0.95
	\end{pmatrix},
	\end{equation}
	resulting to a steady state distribution 
	\begin{displaymath}
	\mathbf{\xi}^\intercal = \begin{pmatrix}
	0.5 & 0.5
	\end{pmatrix},
	\end{displaymath}
	which is equal to
	\begin{displaymath}
	\begin{pmatrix}
	\mathbf{v_1^\intercal}\mathbf{1} & \mathbf{v_2^\intercal}\mathbf{1}
	\end{pmatrix},
	\end{displaymath}
	as predicted by Theorem~\ref{thm:couplingsolution}. Therefore the final ranking vector is
	\begin{displaymath}
	{\boldsymbol{\pi}}^\intercal = \begin{pmatrix}
	0.0133  &  0.0935 &    0.1621 &    0.2310 &   0.1526  &   0.1419 &   0.1419 &    0.0635
	\end{pmatrix},
	\end{displaymath}
	which of course coincides with the direct solution of the stochastic matrix $\mathbf{P}$, that can be computed easily for such tiny graph by 
	\begin{displaymath}
	\boldsymbol{\pi^\intercal} = \mathbf{1^\intercal}\left(\mathbf{P}+\mathbf{11^\intercal}-\mathbf{I}\right)^{-1}.
	\end{displaymath}

	 \newpage 
	\section{Datasets}
	\label{App:Datasets}

	The snapshots of the Web used for our experiments are presented in Table~\ref{table:WebGraphs}, below. 
	
	\begin{table}[!hpbt]
		\centering
		\caption{Datasets Used in Section~\ref{Sec:experiments}}
		{\resizebox{0.7\textwidth}{!}{ \begin{tabular}{rrrrr} 
					\toprule
					\toprule
					Network &  \#Nodes & \#Edges & Dangling Nodes & Description \\
					\midrule
					\texttt{cnr-2000} & 325557 & 3216152 & 23.98\% & A small crawl of the CNR domain \\
					\texttt{eu-2005} & 862664 & 19235140 & 8.31\% & Crawl of the .eu domain in 2005 \\
					\texttt{india-2004} & 1382908 & 16917053 & 20.41\% & Crawl of the .in domain \\
					\texttt{uk-2002} & 18520486 & 298113762 & 14.91\% & Crawl of the .uk domain in 2002 \\
					\bottomrule
					\bottomrule
		\end{tabular}}}
		\label{table:WebGraphs}
	\end{table}
	More information about the networks---including details about the crawling and storing procedure---can be found in the Website of the \textit{Laboratory of Web Algorithmics} (see also~\cite{UbiCrawler,BRSLLP,BoVWFI}).

\newpage

\section{Supplementary Figures}
\label{Figs:Supplementary}

\begin{figure*}[h!]
	\centering
	\includegraphics[width = 0.92\textwidth]{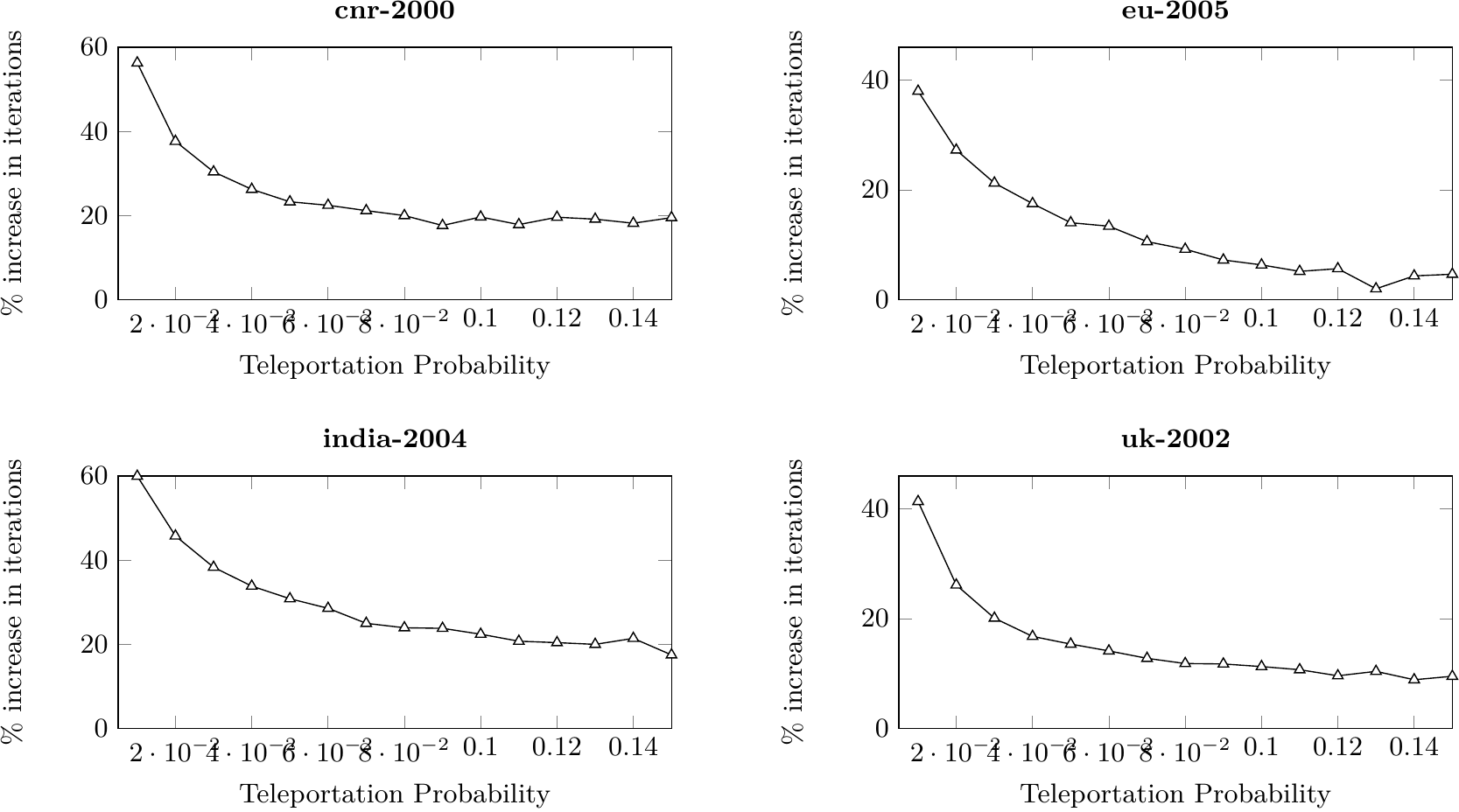}
	\caption{NCDawareRank vs PageRank. Relative convergence behavior for different values of the teleportation probability. The figure reports the percentage increase of the number of iterations needed by PageRank to converge with respect to NCDawareRank, for teleportation probabilities in the range $[0.01,0.15]$.}
	\label{fig:PercentageIncrease}
\end{figure*}

\showmatmethods{}

\end{document}